\documentclass[floatfix,aps,reprint,superscriptaddress,amsmath,amssymb,longbibliography,showpacs]{revtex4-1}
\pdfoutput=1
\usepackage[english]{babel}
\usepackage{amssymb}
\usepackage{graphicx}
\usepackage{color}
\usepackage{bbm}
\usepackage{hyperref}
\hypersetup{pdfauthor={L. Fiedler, P. Naaijkens, T.J. Osborne}}
\hypersetup{pdftitle={Jones index, secret sharing and total quantum dimension}}

\newcommand{\alg}[1]{\mathfrak{#1}}
\newcommand{\ket}[1]{\left|#1\right\rangle}
\newcommand{\bra}[1]{\left\langle#1\right|}
\newcommand{\mc}[1]{\mathcal{#1}}
\newcommand{\mf}[1]{\mathfrak{#1}}
\newcommand{\mr}[1]{\mathrm{#1}}
\newcommand{\bonds}{\textbf{B}}
\newcommand{\mb}[1]{\mathbbm}

\begin{document}
  \title{Jones index, secret sharing and total quantum dimension}

  \author{Leander Fiedler}
  \email[]{leander.fiedler@itp.uni-hannover.de}
  \affiliation{Institut f\"ur Theoretische Physik, Leibniz Universit\"at Hannover, Germany}
  \author{Pieter Naaijkens}
  \email[]{pnaaijkens@math.ucdavis.edu}
  \affiliation{Department of Mathematics, University of California, Davis, USA}
  \affiliation{JARA Institute for Quantum Information, RWTH Aachen University, Germany}
  \author{Tobias J. Osborne}
  \email[]{tobias.osborne@itp.uni-hannover.de}
  \affiliation{Institut f\"ur Theoretische Physik, Leibniz Universit\"at Hannover, Germany}
  \date{\today}

\begin{abstract}
	We study the total quantum dimension in the thermodynamic limit of topologically ordered systems.
  In particular, using the anyons (or superselection sectors) of such models, we define a secret sharing scheme, storing information invisible to a malicious party, and argue that the total quantum dimension quantifies how well we can perform this task.
  We then argue that this can be made mathematically rigorous using the index theory of subfactors, originally due to Jones and later extended by Kosaki and Longo.
  This theory provides us with a ``relative entropy'' of two von Neumann algebras and a quantum channel, and we argue how these can be used to quantify how much classical information two parties can hide form an adversary.

  We also review the total quantum dimension in finite systems, in particular how it relates to topological entanglement entropy.
  It is known that the latter also has an interpretation in terms of secret sharing schemes, although this is shown by completely different methods from ours.
  Our work provides a different and independent take on this, which at the same time is completely mathematically rigorous.
  This complementary point of view might be beneficial, for example, when studying the stability of the total quantum dimension when the system is perturbed.
\end{abstract}
\maketitle
\section{Introduction}
  Quantum phases can be understood as equivalence classes of ground states of quantum many body systems~\cite{PhysRevB.82.155138}. In this paper we are particularly interested in \emph{gapped} quantum phases, up to adiabatic evolution~\cite{PhysRevB.72.045141,2011arXiv1102.0842B}. A particularly interesting set of phases is that of topological ordered phases, i.e.\ classes of ground states that exhibit long range entanglement. There are several different ways of setting up an equivalence of phases~\cite{PhysRevB.82.155138,PhysRevB.72.045141,BraHaMi:2010,MR2842961}, but in general they are expected to give rise to the same equivalence relation. It is believed that topological order is a property of states alone \cite{PhysRevLett.96.110405}.
  While defining the equivalence relation from physical principles is a task in itself, the characterisation of all possible equivalence classes is a much more subtle endeavour.
  One way of tackling this problem is to find invariants for the equivalence classes which can be computed locally and which allow one to distinguish different phases.

  A possible candidate for an invariant is the topological entanglement entropy (TEE) \cite{PhysRevLett.96.110404,PhysRevLett.96.110405}, which is believed to be a strong indicator of topological order.
  It is motivated by systems where the ground state satisfies an area law. In states with long-range entanglement, where this area law is expected to hold, the TEE is a correction of order $\mc O(1)$ to the von Neumann entropy of the reduced density matrix of the ground state on a disk shaped region.
  Furthermore, for the usual examples of anyonic systems, such as the toric code model \cite{MR1951039} and the string-net models \cite{LeWe:2005}, it is proportional to $\mr{log}(\mc D)$, where $\mc D$ is the total quantum dimension of the modular tensor category describing the anyons.
  The proportionality factor depends on the geometry of the bipartition of the system.
  The total quantum dimension itself characterises to some degree the anyonic nature of the local excitations of the ground state, as it is given by the quantum dimensions $d_a$ of the different types of anyons via $\mc D^2=\sum_{a}d_a^2$ \cite{Preskill:1999}.
  A total quantum dimension that is larger than the number of distinct particles signifies non-abelian anyons \cite{RowelWang:2015}, since an anyon $a$ is abelian if and only if $d_a = 1$.
  The quantum dimension $d_a$ of an anyon of type $a$ can be understood as the asymptotic growth of the Hilbert space that encodes $n$ anyons $a$ placed on a plane and conditioned on global vacuum~\cite{Preskill:1999}.

  In the thermodynamic limit of topologically ordered systems the total quantum dimension can be related to the Jones-Kosaki-Longo (JKL) index of certain inclusions of algebras of observables localised in cones \cite{klindex}. Under precise (and natural) technical assumptions this index coincides with $\mc D^2$.
  The reason is that the JKL index gives us a way to compare the size of two (infinite dimensional!) algebras. As we shall see later, in our setting the big algebra is related to the smaller one precisely through ``charge transporters'', which in turn are in correspondence with the different types of anyons.
  This suggests that there should be a connection between the JKL index and the topological entanglement entropy. However, it is a priori not clear how these very different concepts are related. Investigating this relation is one of the main goals of this paper.

  In particular, we show with the example of the toric code how a secret sharing scheme for classical information between two parties naturally arises, and how we can relate it to the inclusion of algebras mentioned above.
  The amount of classical information that can be hidden with this scheme is then given by the JKL index.
  We compare this to a similar result in finite dimensions \cite{KaFuMu:2015}, where the topological entanglement entropy was shown to coincide with the optimal achievable rate of a (different) secret sharing scheme via the irreducible correlation.
  Based on this we argue that the JKL index is indeed closely related to the topological entanglement entropy.
  This picture is strengthened by the observation that the index is in a sense optimal and that it is related to a relative entropy between the corresponding von Neumann algebras.
  This is a generalisation of the relative entropy known from finite dimensional systems. Using this relative entropy and its relation to the index, we can interpret the index as a bound on the amount of classical information that can be encoded in the above secret sharing scheme.

\subsection{Total quantum dimension and the TEE}
  An anyon model can be specified in terms of a set of \emph{particle types}, together with a set of \emph{fusion rules}, certain matrices describing the interchange of two anyons, i.e.\ the \emph{braiding}, and tensors relating the different orders in which one can fuse $n$ anyons. These rules have to satisfy certain compatibility conditions. Mathematically, this means an anyon model is described by a \emph{modular tensor category}~\cite{MR2200691,*Wang}. To each anyon type one can associate a quantum dimension $d_i$. One way to interpret this dimension is as a ``scaling factor'' describing the asymptotic growth of the state space of $n$ anyons of that type. It also describes the growth in ground-state degeneracy of a model when it is placed on an $n$-torus~\cite{PhysRevB.41.9377,*PhysRevB.85.075107}. The \emph{total quantum dimension} is defined as $\mathcal{D}^2 = \sum_i d_i^2$, where the sum is over all anyon types. In the language of tensor categories, $\mathcal{D}^2$ is called the (global) dimension of the category~\cite{MR1966524}.

  Based on arguments involving topological quantum field theory, Kitaev and Preskill \cite{PhysRevLett.96.110404} introduced a way to calculate the total quantum dimension: they defined an entropic quantity $S_{top}^{KP}$ and argued that it is equal to $\log \mc{D}$. Levin and Wen also defined a similar entropic quantity $S_{top}^{LW}$, and showed that $S_{top}^{LW} = \log \mc{D}^2$ for so-called quantum double models~\cite{PhysRevLett.96.110405}. The difference of a factor of two between the two definitions can be attributed to the different shapes of the regions used in their definition.

  The TEE has become a key tool in the study of topological order because it allows for a fairly practical approximation of the total quantum dimension: one only needs to solve the model on a torus large enough that the entropies for the various regions involved in its definition can be non-trivial \cite{Isakov:2011fk,*PhysRevB.76.184442,*2013arXiv1303.4455B}. Unfortunately the TEE does suffer from some shortcomings: it is far from clear how to extend it to higher dimensional systems (see, however~\cite{PhysRevB.84.195120} for recent progress) and situations involving symmetry protection~\cite{2013arXiv1303.4190H} and it is also a deeply non-trivial task to show that it is stable under adiabatic equivalence (for some partial progress see \cite{PhysRevB.86.245116}). Another issue is that one can construct examples of states that appear to have a universal TEE term in their entanglement entropy but which are topologically trivial. One such example is due to Bravyi (cf.\ Section II.C of~\cite{ZouHaah:2016} for a description).

\subsection{Secret Sharing}
  Secret sharing schemes can be seen as an instance of error correction codes. They are based on the idea that, given a set of states of the system, one needs access to a certain ``minimal'' set of observables on the system in order to distinguish states in this set.
  This becomes particularly interesting when considering settings where information should be encoded in such a way that only observables that act on sufficiently large parts of the system are able to decode the hidden information.
  Classical secret sharing schemes where discussed in~\cite{Shamir:1979,*Blakley:1979} and later generalised to the quantum setting in~\cite{ClevGotLo:1999}.
  There are certain bounds on the amount of information that can be encoded in such schemes~\cite{ClevGotLo:1999,Gottesman:2000,*KretKriSpe:2008}, that is, bounds on the size of the regions (also called shares) and the minimal number required to decode the information, given the total system size.
  Here we consider secret sharing schemes in the context of topologically ordered states.

  For topologically ordered systems, such as the toric code, the ground states of the Hamiltonian are locally indistinguishable~\cite{BraHaMi:2010}. That is, with access to observables that act on a few sites of the system only, it is not possible to distinguish the ground states. In order to do so one needs observables that act non-locally, that is, on a part of the system that is large compared to the system size.
  Note that this is exactly an error correction condition on the ground state space:
  local perturbations of the ground state can be detected and afterwards corrected.
  Hence we can regard the ground state space as a quantum code, where the resulting size of the code space is determined by the total quantum dimension $\mc D$ of the anyon model and the genus of the manifold in which the system is embedded~\cite{MR1951039}.

  In the thermodynamic limit, however, locally indistinguishable states converge to the \emph{same} state (in the weak$*$-topology), since their expectation values on local operators coincide as soon as the system size is big enough.
  Hence in that setting we cannot directly appeal to the degeneracy of the ground state space.
  Nevertheless, it should still be possible to use the topological charges to secretly share information between two parties in the system, if one restricts the corresponding regions in which the excitations are distributed accordingly.
  The intuition for this comes from the observation that in two dimensions excitations above the ground state always occur in conjugate pairs at the endpoints of a string, where the excitations do not depend on the exact geometry of the string but only on it's endpoints.
  As long as one can ``keep'' these endpoints ``away'' from a possibly malicious third party by restricting their observables it should be impossible for them to determine which pair of excitation was created.
  The expected size of the code space is then again given by the total quantum dimension $\mc D$ of the anyon model.
  This forms the basis for our secret sharing scheme in the thermodynamic limit.

\subsection{Content of this work}
\begin{figure}
    \includegraphics[width=0.95\columnwidth]{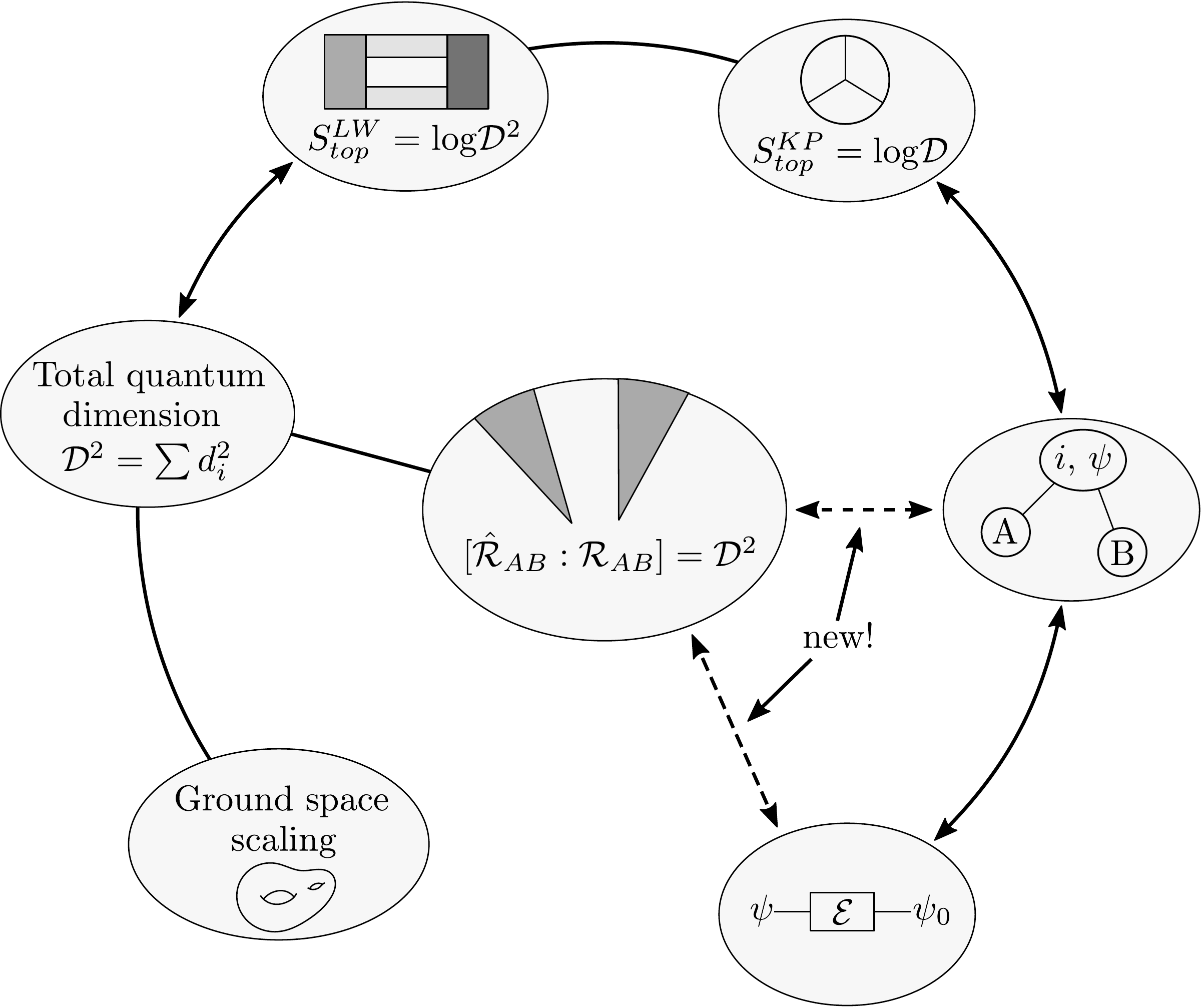}
	\caption{An anyon model has a total quantum dimension $\mathcal{D}^2$. The existence of anyons in models is related to long range entanglement in the system. By judiciously choosing combinations of entanglement entropies of distinct regions, the entanglement due to topological order can be isolated. This leads to topological entanglement entropies $S_{top}^{KL}$ and $S_{top}^{LW}$, the differences being mainly due to the choice of regions. These entropies are argued to be related to the total quantum dimension $\mc D^2$~\cite{PhysRevLett.96.110404,*PhysRevLett.96.110405}. The latter can be obtained as a Jones-Kosaki-Longo index associated to cone algebras~\cite{klindex}. In this paper we concentrate on the dashed arrows, arguing that by reinterpreting the topological entropy in terms of an information hiding task~\cite{KaFuMu:2015}, it can be naturally connected to the cone index. The log of the index tells us something about how much information we can hide, and the index theory provides us with a channel $\mc{E}$ which describes the limited powers of an adversary Eve.
    Finally, $\mc{D}$ is related to the ground space degeneracy as a function of the genus of the surface on which the model is defined~\cite{PhysRevB.41.9377,*PhysRevB.85.075107}.}
    \label{fig:mindmap}
\end{figure}

  In this paper we try to connect the different approaches to obtain the total quantum dimension $\mc D$ and related quantities. In particular, we advocate an (operator) algebraic way to obtain the total quantum dimension $\mathcal{D}$, which will allow for generalisations to different dimensions and symmetry protected cases. In addition, we interpret $\mathcal{D}$ in terms of a information hiding task, making a direct connection between the TEE and the JKL index possible. Although we do not claim that our approach solves the problems with the TEE mentioned above, we believe it offers additional insight to the nature of topological order. In particular, as we formulate $\mathcal{D}$ in terms of observable algebras it is easier to anticipate a proof of the stability of the index under adiabatic equivalence and to extend it to more exotic scenarios.
   There are many ways to think about $\mc D$, some of which are outlined in Fig.~\ref{fig:mindmap}.
   That figure also shows how our work fits into the big picture.

  We will start in Section~\ref{sec:finite} with explaining our intuition about the index at a finite dimensional variant, discuss drawbacks and problems that arise in the context of finite system sizes, and illustrate our intuition with the example of a chain of Fibonacci anyons.
  In Section~\ref{sec:thelimit} we recall the necessary notions and properties of two dimensional models in the thermodynamic limit that we want to consider and discuss how the Jones-Kosaki-Longo index appears in this context.
  Section~\ref{sec:secshare} then is devoted to constructing a secret sharing scheme in the example of the toric code on the infinite plane and to explaining how it relates to the index. Furthermore we discuss how this is connected to recent work~\cite{KaFuMu:2015}, where for finite two-dimensional lattice systems it was shown that there is a connection between the topological entanglement entropy and the irreducible correlation of certain secret sharing schemes. We also discuss the role of superselection sectors in our construction.

  One of the main contributions of our work is discussed in Section~\ref{sec:channels}. There we illustrate how index theory can be used to study the secret sharing scheme in the context of quantum information theory.
  In particular, one gets a quantum channel ``for free'', and it is possible to define a relative entropy for certain algebras. Using this relative entropy the (logarithm of the) index can be recovered, and we see how this provides us with bounds on the amount of information that can be hidden in the secret sharing scheme.
  As we are mainly working in an algebraic setting, in Section~\ref{sec:private} we shed some light on how one can reformulate the picture of secret sharing schemes in terms of private subsystems of a channel between the corresponding algebras of observables, and give some of the details for the example of the toric code.
  Finally we remark on the stability of the index under local perturbations.

  The goal of this work is to focus on the physical ideas and intuition behind our constructions.
  Many parts can be made mathematically rigorous, but this requires substantial mathematical machinery, in particular from the theory of von Neumann algebras. We refer to the relevant literature whenever this is the case.
  However, since we work in an operator-algebraic framework, some basic terminology of this field is unavoidable.
  The appendix contains a motivation on why we use this language to describe systems in the thermodynamic limit, as well as an introduction to the basic notions that we use in the course of this work.

\section{Finite dimensions}\label{sec:finite}
  To explain the main idea behind our index approach, we first consider a finite-dimensional variant. Although the main idea can be made clear in this case, it is a little surprising that a careful algebraic analysis seems infeasible, \emph{precisely} because of the finite dimensionality. We return to this point later. Although the finite dimensional case is perhaps somewhat naive in light of these limitations, it nevertheless provides some intuition for the approach we take in the thermodynamic limit.
  \subsection{Motivation: a secret sharing task}
  \begin{figure}
	  \includegraphics[width=0.75\columnwidth]{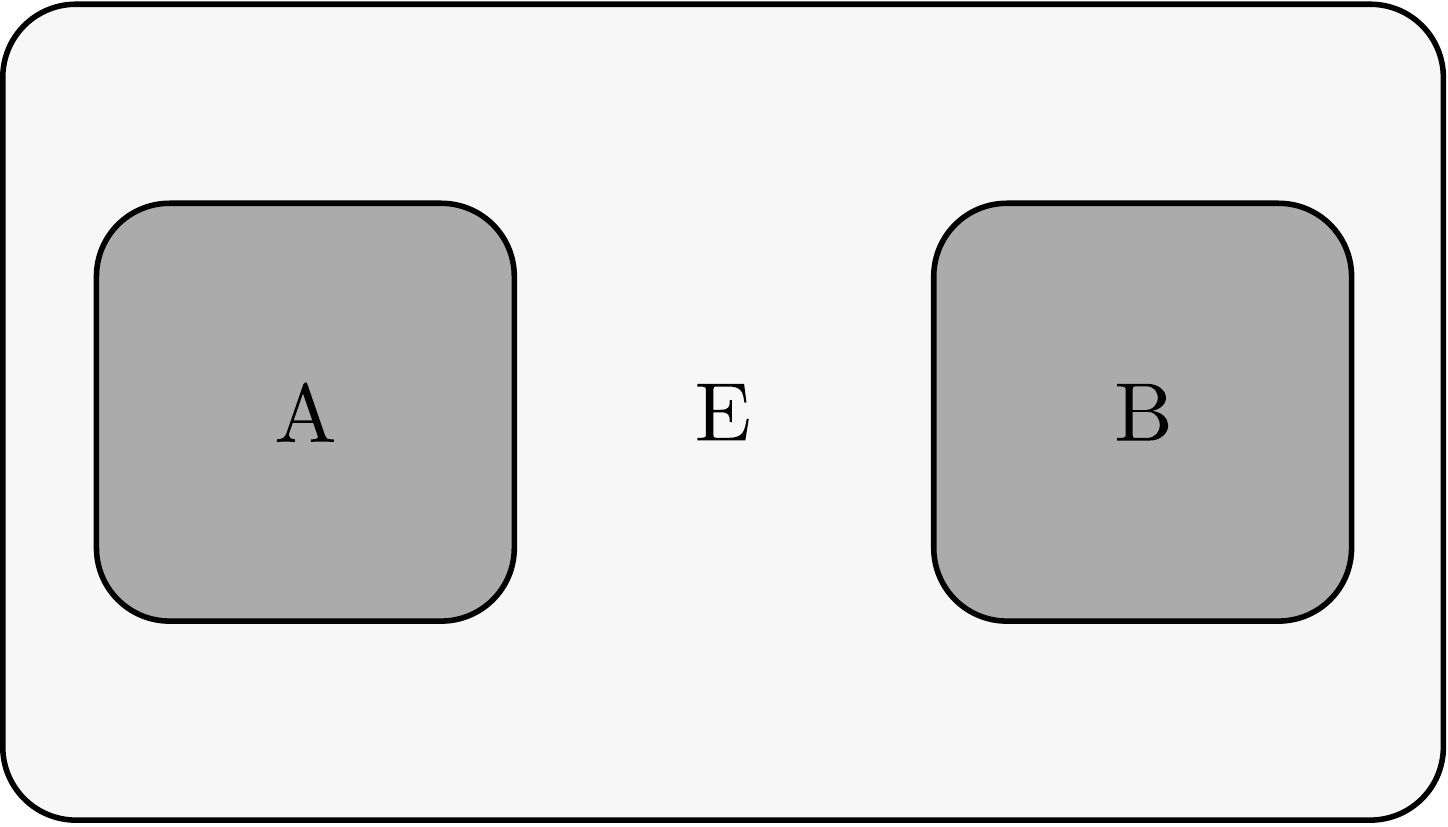}
	  \caption{The system is partitioned into three parts: regions A and B controlled by Alice and Bob, and the rest (E), controlled by Eve.}
	  \label{fig:abe}
  \end{figure}

\begin{figure}
    \includegraphics[width=0.7\columnwidth]{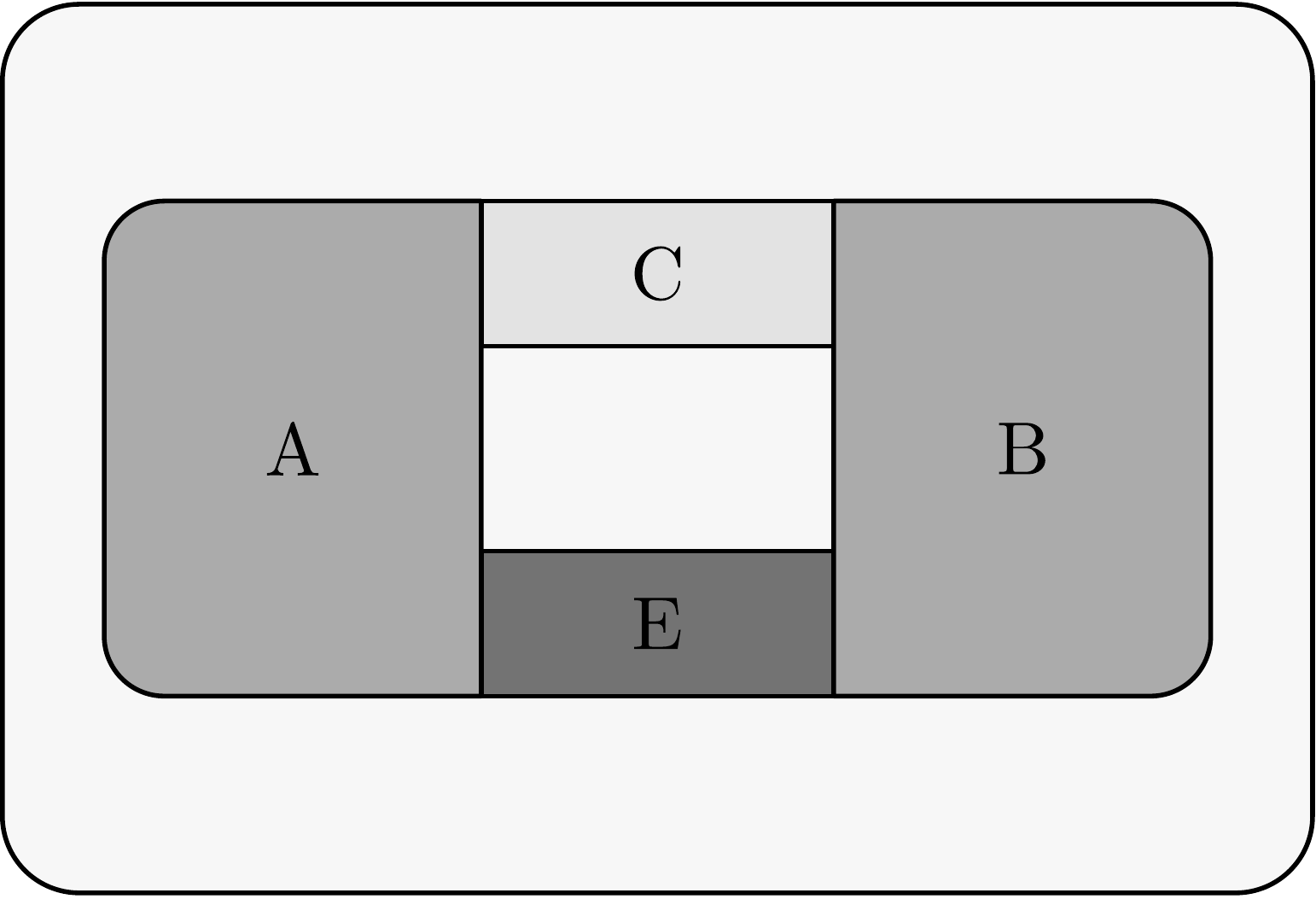}
    \caption{Alice and Bob control the regions A and B, but in addition have access to a region C, ``bridging'' their parts.
    Eve on her part has access to a disjoint region $E$, small compared to the system size. Alice and Bob can change their choice of region C, according to Eve's choice of the region E.}
    \label{fig:mutual}
  \end{figure}

  The results in this section are not completely rigorous, but are intended as motivation for the (rigorous!) results in the thermodynamic limit, which we describe later. In the finite-dimensional ``toy model'', the set-up is as follows. The system, defined on a lattice $L$, is divided into three parts $A$, $B$, and $E$, like in Figure~\ref{fig:abe}. Alice and Bob each control disjoint parts of the system, and Eve (perhaps some government agency) controls the rest. Suppose the system is initially in the state $|\Omega\rangle$. Alice and Bob have the task of storing a classical message in $|\Omega\rangle$; they want to use the system to set up something akin to a quantum I2P network which would be anonymous and immune from censorship. To achieve this task they are allowed to do \emph{any} joint quantum operation on their respective regions $A$ and $B$. In this case one can easily deduce that the configuration space for their anonymous and secret messages is given by
  \begin{align*}
  	\mathcal{V}_{AB} := \text{span}\{ U_{AB} |\Omega\rangle\,|\, \text{supp}(U_{AB}) \subset AB \}.
  \end{align*}
  However, in achieving their information-hiding goal there is no reason we should restrict Alice's and Bob's operations to act only on $AB$. Indeed, they are allowed to touch sites belonging to Eve, \emph{as long as Eve does not find out}. Clearly, in this finite-dimensional setting, this won't help if Eve is allowed to perform \emph{any} bona fide quantum operation on her part of the system. However, if for some reason Eve's capabilities are restricted (eg.\ perhaps government funding for building spying networks has been cut), there is a possibility for Alice and Bob to exploit this limitation and potentially hide more information in $|\Omega\rangle$. In this paper we postulate that Eve should only be able to do \emph{local} measurements. Here ``local'' means small compared to the system size, and compared to the regions that Alice and Bob control \footnote{Perhaps the best analogy here is that Alice and Bob control separate nation-size states that cannot be completely encircled by an antagonistic spying nation.}. In particular, we disallow measurements that act on all sites on a ring around either Alice's or Bob's region (see Fig.~\ref{fig:mutual}). It will become clear from the example given below why we impose this restriction. Eve can do an unlimited number of such operations in succession, so long as the resulting operation doesn't encircle $A$ or $B$ \footnote{Note that, in the finite case, the set $\mathcal{O}_E$ of Eve's observables \emph{does not} obviously form an algebra.}.

  Alice and Bob can do any joint operation on their part of the system and in that way can store a classical or quantum message that is inaccessible to Eve. Given that Eve has limited eavesdropping capabilities, the question is if this allows Alice and Bob to encode additional signals into $|\Omega\rangle$. This is the problem that we want to answer.  To this end, consider the space $\widehat{\mathcal{V}}_{AB}$ generated by states that Eve cannot distinguish from $|\Omega\rangle$ by the operations at her disposal. If we denote $\mathcal{O}_E$ for the set of operations that Eve is allowed to perform, it can be defined as
  \begin{equation}
  	\begin{split}
		\widehat{\mathcal{V}}_{AB}& := \operatorname{span}\{ |\Omega_{ABE}\rangle \, |\, \\
		&\langle\Omega_{ABE}| O_E |\Omega_{ABE}\rangle = \langle\Omega|O_E|\Omega\rangle, \forall O_E\in \mathcal{O}_E\}.
  	\end{split}
  \end{equation}
  Note that this is precisely the statement that the space $\widehat{\cal V}_{AB}$ forms an error correction code that corrects the errors caused by Eve's observables $\mc O_E$ \cite{KnillLafla:1995,*KnilLafVio:2000,*NieCavScBa:1998,*Gottesman:2009}.
  Clearly the Hilbert space $\mathcal{V}_{AB}$ is contained in $\widehat{\mathcal{V}}_{AB}$. Note that the condition on the expectation values above is non-linear in the vectors of the form $U_{ABE} |\Omega\rangle$, so that it is not quite natural to take the linear span. For the models we have in mind, however, this is \emph{does} make sense. A typical feature of these models is that the states corresponding to anyons of \emph{distinct} type cannot be converted into each other with local operations (if the compensating anyonic excitations, which are necessarily there because of charge conservation, are localised far away). In other words, the anyons belong to different superselection sectors. Taking a superposition of such states, one sees that for local observables $O_E$ the cross terms vanish when calculating the expectation value. This is essentially why this somewhat naive approach works in models such as the toric code.

    We now have enough information to explain the calculation of the index invariant. This is given by the \emph{ratio} of the dimensions of $\mathcal{V}_{AB}$ and $\widehat{\mathcal{V}}_{AB}$ of Alice's and Bob's regions:
    \begin{align}
    	\label{eq:dimfrac}
    	[\widehat{\mathcal{V}}_{AB}: \mathcal{V}_{AB}] := \frac{\dim \widehat{\mathcal{V}}_{ A  B}}{\dim\mathcal{V}_{ A B}}.
    \end{align}
  We will later consider a different (and less naive) definition for this index in an operator-algebraic setting.
  For now we note that an equivalent way to express the index in this toy model is as a difference of entropies: here Alice and Bob are comparing the rates of two  maximally mixed signal ensembles, one built from the Hilbert space $\mathcal{V}_{AB}$, namely $\rho_{AB} := \mathbb{I}/\dim\mathcal{V}_{ A B}$ and the one built from  $\widehat{\mathcal{V}}_{ A  B}$, namely $\widehat{\rho}_{AB} := \mathbb{I}/\dim \widehat{\mathcal{V}}_{ A  B}$:
  \begin{align*}
  	\log[\widehat{\mathcal{V}}_{AB}: \mathcal{V}_{AB}] := S(\widehat{\rho}_{ A  B})-S({\rho}_{ A  B}).
  \end{align*}
  At this point we illustrate the task above by an example. Consider Kitaev's toric code~\cite{MR1951039}. In this model, one can create pairs of (anyonic) excitations by acting with path operators on a ground state. These paths are either drawn on the lattice or on the dual lattice, or a combination of the two. Using such a path operator $F_\xi$, Alice and Bob can create a pair of excitations, where one excitation is in Alice's part, while the other one belongs to Bob. The claim is that Eve, with the operations at her disposal, cannot detect that such a pair of excitations was created. Indeed, it is well known that the state $F_\xi |\Omega\rangle$, where $|\Omega\rangle$ is a ground state, only depends on the endpoints of $\xi$. Hence, since Eve can only do local measurements, one can always choose a path that avoids the support of Eve's measurement, in which case it is clear that Eve cannot detect it. Note that the only way to detect the excitations is to measure the total charge in a region by measuring the path operator corresponding to a Wilson loop enclosing the region. This is precisely how Alice and Bob can detect the presence of a charge in their respective parts of the system. Since in the toric code charge addition is done modulo two, and there are two fundamental charges (electric and magnetic), they have access to a factor of four additional orthogonal states in $\widehat{\mathcal{V}}_{AB}$ relative to $\mathcal{V}_{AB}$ to hide information from Eve. Thus the index for this case is $4$, which is the total quantum dimension for the toric code. Since Alice and Bob can only measure charges locally in their region, relative phases between the different charged states get lost upon measurement. Hence they can only retrieve four \emph{classical} bits of information.

  \subsection{Problems with this approach}
  There are some drawbacks to this approach. They mainly stem from two causes: (i) the index quantity is not obviously independent of the regions $A$ and $B$; and (ii) there is no clear algebraic structure underlying the set of allowed operations for Eve. Her local operations do \emph{generate} an algebra, but this algebra is too big: it contains all operations on Eve's region. In some cases there \emph{is} a natural choice of algebra: for example in the toric code one can choose the abelian algebra generated by all star and plaquette operators acting on $E$. However, in general it seems to be difficult to get a good handle on Eve's operations, and consequently, it is difficult to find out what all the allowed operations for Alice and Bob are. We argue below that these difficulties can be overcome by passing to the thermodynamic limit. This is the starting point of our analysis.

  The naive analysis here can be refined by using techniques developed by Haah~\cite{Haah:2014}. He considers ground states of local commuting projection Hamiltonians for which examples include Kitaev's toric code and the Levin-Wen models. His main goal is to define an invariant for such Hamiltonians that is stable with respect to local perturbations that do not close the gap. Part of his construction is to identify the different types of (anyonic) excitations in the model. As discussed above, such excitations are precisely what allow Alice and Bob to share classical information. A key ingredient in his construction are algebras associated to annuli. These algebras are obtained by looking at the observables supported on the annulus, and dividing out the observables that commute with all terms of the Hamiltonian supported on the annulus, i.e., those operators that do not create excitations in the annulus. This quotient algebra can then be decomposed into smaller algebras using projections which correspond to the different particle types. This construction is -- in a sense -- dual to ours outlined above. His procedure allows one to detect a single charge sitting inside the annulus. Since the total charge should be zero on the ground state space, the compensating charge can be thought of as sitting inside a different annulus. By growing (and deforming) the annuli on the outside, we can make them fill the entire space outside of the parts in the interior. Hence, we again have divided the system into three regions -- Alice, Bob and Eve. The only difference is then that we are interested in which information can be hidden from Eve (i.e., which are invisible to her), while Haah considers all the charges that can be detected inside the annuli. These two notions are clearly related, but we will not pursue this connection any further in this paper.

\subsection{The Fibonacci chain}
  Before we discuss the thermodynamic limit, we consider another example which sheds some light on the relation to the algebraic properties of the anyons (for example, given as a modular tensor category, cf.~\cite{MR2200691,Wang}). For concreteness, suppose that we have $n$ anyons, fusing to the vacuum. A basis for such states can be conveniently represented in terms of fusion trees. The key point then is to define the appropriate notions of a local operation for Alice, Bob and Eve, given that they each control a fixed set of the anyons. To this end we follow the approach of \cite{2013arXiv1310.0373P,*2013arXiv1310.4140K}. Of the $n$ anyons, Alice (Bob) controls a group of $n_A\,(n_B)$ anyons, Eve the rest. These groups of anyons are assumed to have total charge given by labels $\rho_A, \rho_B$ and $\rho_E$. The local operations are then precisely those operations on the respective groups of anyons that leave this total charge unchanged. We can then construct the spaces $\widehat{\mc{V}}_{AB}$ and $\mc{V}_{AB}$.

\begin{figure}
  \includegraphics[width=0.8\columnwidth]{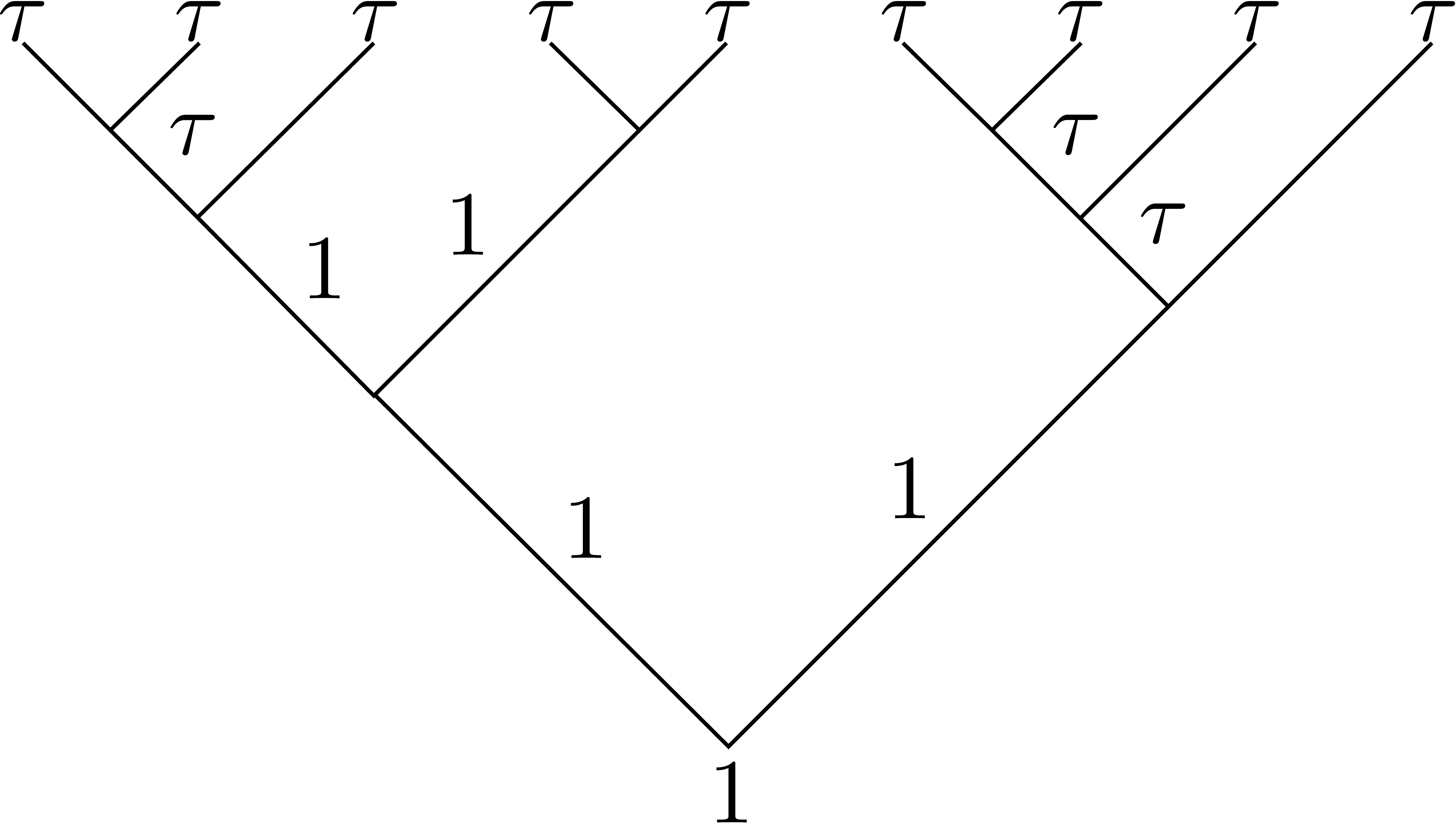}
  \caption{Consider a chain of $n$ $\tau$-anyons, grouped into three groups. Alice, Bob control the left and right group, Eve the ones in the middle. Since the fusion rules are $\tau \otimes \tau = 1 \oplus \tau$, these $n$ anyons can fuse to the trivial charge $1$. The states describing these different configurations span a vector space: the \emph{fusion space}. A basis can be described by considering fusion trees as above. For example, the two leftmost $\tau$ anyons fuse to $\tau$, while the two in the middle fuse to $1$. Fusing the anyons in a different order would give a different basis, which is related to the old one by a unitary transformation.}
  \label{fig:fibonacci}
\end{figure}

  As an example we consider the Fibonacci model~\cite{TreTroWaLu:2009} with anyons $1, \tau$ and assume that $\rho_A = \rho_B = \rho_E = 1$.
  The Hilbert space of the system is also called the \emph{fusion space}.
  Its states describe the different ways that the anyons can fuse.
  In the Fibonacci chain there is only one non-trivial fusion rule: $\tau \otimes \tau = 1 \oplus \tau$.
  That is, if we fuse two $\tau$ anyons, we either get a $\tau$ anyon again, or the trivial anyon $1$.
  A basis for the Hilbert space of the system can then be obtained by labeling all different ways $n$ distinct $\tau$ anyons can fuse to the trivial anyon.
  This can be done conveniently with the help of fusion trees, which label the outcome of the fusion operations.
  In Figure~\ref{fig:fibonacci} we illustrated this with an example of a handful of anyons.
  In this example, the two left-most anyons fuse to a $\tau$, while the fourth and fifth anyon fuse to the trivial charge.
  The order in which the fusion is performed should be the same for all basis elements, but is otherwise arbitrary.
  Choosing a different order amounts to a basis transformation~\cite{Wang}.

  Now choose a fusion tree where $\rho_A = \rho_B = \rho_E = 1$, that is, all Alice's anyons fuse to the trivial anyon, and the same is true for Bob and Eve.
  Write $|\Omega\rangle$ for the corresponding fusion state.
  If Alice and Bob act with local operations on the state $|\Omega\rangle$, they cannot change the total charge in their respective regions.
  That is, they can only make states such that Alice's anyons fuse to $1$, and the same is true for Bob.
  By finding all fusion trees subject to these constraints we find the space $\mathcal{V}_{AB}$.

  In contrast, if they are allowed to do non-local operations as well, there are additional possibilities: they can collude and make states such that the total charge in Alice's region and that in Bob's region is $\tau$, but in such a way that these two $\tau$'s fuse to $1$, so that the total charge of the system remains trivial.
  They can do this without changing the total charge of Eve (because two $\tau$'s can fuse to $1$), so she is not able to detect this.
  This gives a bigger space $\widehat{\mathcal{V}}_{AB}$.

  Finding the dimensions of $\mathcal{V}_{AB}$ and $\widehat{\mathcal{V}}_{AB}$ now reduces to the straightforward combinatorial task of counting all admissible fusion trees.
  If the number of anyons $n_A$ and $n_B$ tend to infinity, the ratio of the dimensions of these spaces tends to $1 + \varphi^2$, where $\varphi$ is the golden ratio. This is precisely the total quantum dimension of the Fibonacci model. Of course, this is not a useful way to find the total quantum dimension, since this immediately follows from the given data. However, by considering this abstract setting it does shed more light on the secret sharing task, giving support to the definition in equation~\eqref{eq:dimfrac}.

  These different examples show that the essential step is to find the appropriate notion of what a local operation should be, emphasising that the algebraic point of view is a natural one.

\section{Thermodynamic Limit}\label{sec:thelimit}
  To obtain a clear-cut, purely algebraic construction of the communication task described in the previous section we have to go to the thermodynamic limit.
  Instead of keeping track of the system size $N$, we start with infinitely many sites from the outset~\cite{MR887100,MR1441540}. The sites are labelled by a countable set $\bonds$. Typically, in the models we are interested in $\bonds$ is the set of edges (bonds) between nearest neighbours in a $\mathbb{Z}^2$ lattice or of a honeycomb lattice.
  For simplicity we assume that the local dimension is the same $d$ for each site, but this can easily be generalised.

  This setting is most conveniently described in the operator-algebraic framework, where the observables of the system are modelled by a $C^*$-algebra $\alg{A}$. This can be thought of as the algebra of all observables (or, more general, operations) that can be approximated arbitrarily well (in norm) by observables that only act on a finite number of sites. We refer to Appendix~\ref{app:cstar} for an overview of the main definitions.

  The results in this section are not new. Rather, we recall the main objects of interest in the operator-algebraic approach to topological phases, with a view towards our intended applications. Technical details can be found in~\cite{toricendo,klindex}.

\subsection{Alice, Bob and Eve again}
  \begin{figure}
  	\includegraphics[width=0.75\columnwidth]{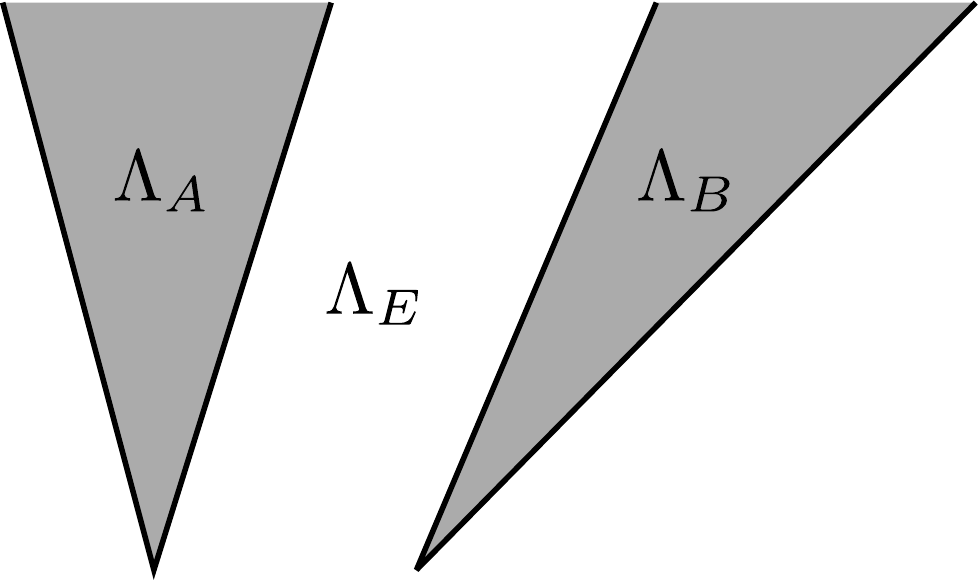}
  	\caption{Alice and Bob each control infinite disjoint cone-like regions $\Lambda_A$ and $\Lambda_B$, Eve controls the rest of the system.}
  	\label{fig:2cone}
  \end{figure}

  We can divide the system into three parts again. Alice and Bob both control (disjoint) cone-like regions (see Fig.~\ref{fig:2cone}), and
  Eve controls the complement. We write $\Lambda_E$ for the set of all sites that Eve controls, and similarly $\Lambda_A$ and $\Lambda_B$ for Alice's and Bob's cones.
  The corresponding observable algebras are denoted by $\alg{A}(\Lambda_i)$. These are the algebras of all observables that can be approximated arbitrarily well in norm by observables acting on only finitely many sites inside $\Lambda_i$. The specific shape of the cones is not that important, as long as they are without holes, disjoint, and extend to infinity. It will become clear below why this choice of regions is natural.

  We now suppose that $\alg{A}$ is represented on some Hilbert space $\mc{H}$ by an irreducible representation $\pi_0$, that is $\pi_0(\mf A)$ is a $C^*$-subalgebra of $\alg B(\mc H)$. Which representation to use (in general, there are many inequivalent choices) is dictated by physical principles; in our case it will come from a pure, translation invariant ground state $\omega_0$ via the GNS construction (see appendix for a short introduction). Motivated by the discussion above we postulate that Alice and Bob can perform every operation that commutes with all of Eve's local operations, hence this is given by $\widehat{\mc{R}}_{AB} := \pi_0(\alg{A}(\Lambda_E))'$.
  Recall that the prime denotes the commutant so that $\widehat{\mc{R}}_{AB}$ is the set of all bounded operators in $\alg{B}(\mc{H})$ that commute with each $\pi_0(A)$, $A \in \alg{A}(\Lambda_E)$.
  On the other hand we can consider all operations that Alice can implement on the cone she controls. These are given by the von Neumann algebra $\mc{R}_A := \pi_0(\alg{A}(\Lambda_A))''$.
  Taking the double commutant is natural here: it ensures that all the relevant spectral projections are in the algebra \cite{MR887100}. We define $\mc{R}_B$ similarly. The operations Alice and Bob can do together when only acting on their cones is then $\mc{R}_{AB} := \mc{R}_A \vee \mc{R}_B$, where the wedge denotes the von Neumann algebra generated by the two algebras.
  Note that by locality, $\pi_0(\alg{A}(\Lambda_A \cup \Lambda_B)) \subset \pi_0(\alg{A}(\Lambda_E))'$. Taking commutants twice it follows that we have an inclusion $\mc{R}_{AB} \subset \widehat{\mc{R}}_{AB}$ of von Neumann algebras.
  The algebras are in fact factors, if one assumes that the ground state representation $\pi_0$ is irreducible (equivalently, $\omega_0$ is a pure state), and $\mc{R}_{AB}' \cap \widehat{\mc{R}}_{AB} = \mathbb{C} I$~\cite[Lemma 3.2]{klindex}. Such an inclusion of von Neumann algebras is called an \emph{irreducible subfactor}.

  \subsection{Jones-Kosaki-Longo index}\label{sec:jkl}
  We are interested in the question of which extra operations Alice and Bob can perform. These extra operations are precisely those that are in $\widehat{\mc{R}}_{AB}$ but not in $\mc{R}_{AB}$. Therefore we would like to know how much ``bigger'' the algebra $\widehat{\mc{R}}_{AB}$ is compared to $\mc{R}_{AB}$.
  One way in which this can be quantified is by the Jones~\cite{MR696688} (or rather, in our case, Kosaki-Longo~\cite{MR829381,*MR1027496}) index $[\widehat{\mc{R}}_{AB}:\mc{R}_{AB}]$ of inclusions of von Neumann algebras. It can be thought of as a generalisation of the index of a subgroup $H$ in a group $G$. For our purposes, the technical details and properties behind this index (a subject on its own in operator algebra) play only a minor role. Rather, in the sequel we will focus on some properties that follow from the general theory, in particular the existence of a particular quantum channel and the Pimsner-Popa basis, a way to write elements of $\widehat{\mc R}_{AB}$ as a linear combination of a finite number of ``basis'' elements with coefficients in the smaller algebra $\mc{R}_{AB}$.

  If we assume two technical assumptions, the approximate split property (in \cite{haagdtoric} this property was referred to as the distal split property) and Haag duality~\cite{haagdtoric,haagdabelian}, it can be shown that the index does not decrease if one enlarges the cones. Haag duality (in a representation $\pi_0$) is a property of the commutants of cone algebras. In particular, it says that if $\Lambda$ is a cone, $\pi_0(\alg{A}(\Lambda))' = \pi_0(\alg{A}(\Lambda^c))''$, where $\alg{A}(\Lambda^c)$ is the algebra generated by all local observables outside of the cone.
  One inclusion follows easily from locality, but the other inclusion is non-trivial, and may fail in general. We will not give a precise definition here of the approximate split property (see~\cite{haagdtoric} for the details), but in the present setting it amounts to saying that the map $AB \mapsto A \otimes B$, with $A \in \mathcal{R}_A$ and $B \in \mathcal{R}_B$ extends to an isomorphism $\mathcal{R}_{AB} \to \mathcal{R}_A \overline{\otimes} \mathcal{R}_B$ of von Neumann algebras, if $A$ and $B$ are two separated cones. When one thinks of finite dimensional systems this looks like a trivial statement, but in the thermodynamic limit it is not, and this property is related to deep operator-algebraic questions. For example, Longo used the split property in his solution to the factorial Stone-Weierstrass conjecture, which at the time was open for a long time~\cite{DoLo:1984,*Longo:1984}. One consequence is that if $\alg{A} \subset \alg{B}$ is an inclusion of $C^*$-algebras, then any factor state (in the sense that its GNS representation is a factor) of $\alg{A}$ extends to a factor state of $\alg{B}$. It also has been important in understanding entanglement properties in algebraic quantum field theory~\cite{Werner:1987:1,*Summers:1996}.
  
  In general, we expect the index to be independent of the choice of cones (as long as their opening angles are big enough).
  In the next subsection it will become clear that the inclusion $\mc{R}_{AB} \subset \widehat{\mc{R}}_{AB}$ is related to the different charges of the model, and to operators that move them around.
  Keeping this in mind, independence of the index on the choice of cones can be interpreted physically by saying that excitations can be localised well enough. That is, as long as the opening angle of the cone is not too small, each anyon can be localised in such a cone (regardless of the orientation of the cone). This can be shown explicitly for the toric code~\cite{klindex}.

  Note that the index is a property of the \emph{state}, just like topological entanglement entropy. This is perhaps not apparent at first sight, but one should keep in mind that the respective algebras are taken in a specific representation $\pi_0$.
  This representation, in turn, usually is obtained from a state (in our case, generally a translation invariant ground state), and different states in general lead to different algebras (and possibly different values for the index).

  \subsection{Superselection sectors}\label{sec:ssect}
  In the finite-dimensional toy model we observed that the extra power that Alice and Bob have at their disposal is due to the existence of anyonic excitations which live in different superselection sectors. This is also true in the thermodynamic limit, where there is an elegant characterisation of such sectors. There they appear because there are inequivalent irreducible representations of $\mf A$. This is equivalent to saying that vector states corresponding to distinct representations are not superposable, i.e.\ a relative phase between such vectors cannot be observed with any observable in $\alg{A}$~\cite{MR1405610}.

  Not all representations of $\alg{A}$ are physically relevant. In the models we are interested in, charges are created by applying string-like operators. By moving one end of the string to infinity, we can obtain a state with a \emph{single} charge. In topologically ordered models states created by such string-like operators only depend on the endpoints of the string. Hence the direction in which the charge is moved to infinity is not observable. In fact, if we restrict to operations outside an \emph{arbitrary} cone containing the endpoint of the string, the charge cannot be detected at all and the system appears to be in the ground state. In other words, the charges can be localised in cones. Another natural condition is that we should be able to move the charges around.

  On the level of representations $\pi$, these features are encoded by demanding that they satisfy the following criterion for all cones $\Lambda$:
  \begin{align}
  	\label{eq:superselect}
  	\pi_0 \upharpoonright \alg{A}(\Lambda^c) \cong \pi \upharpoonright \alg{A}(\Lambda^c).
  \end{align}
  Here with $\upharpoonright$ we mean that we restrict the representation to the subalgebra $\alg{A}(\Lambda^c)$. That is, the criterion demands that if we restrict the representations $\pi$ and $\pi_0$ to observables \emph{outside} of a cone $\Lambda$, they become unitarily equivalent. Note that this restriction is important: for example, the representations $\pi_0$ and $\pi$ are \emph{not} equivalent if $\pi$ describes a single anyonic excitation. That is, in such case there is no unitary $V$ such that $\pi_0(A) = V \pi(A) V^*$ for all $A \in \alg{A}$, but if we only require this to hold for $A \in \alg{A}(\Lambda^c)$, such a unitary \emph{does} exist. In algebraic quantum field theory a similar criterion is used, and it is known that (under some additional technical assumptions), studying these equivalence classes of representations allows one to find all relevant properties of the charges in the theory, for example their statistics and fusion rules~\cite{MR1405610}. Using similar ideas this can also be done for quantum lattice models, such as Kitaev's quantum double~\cite{toricendo,haagdabelian}.

  How does this relate to the choice of the algebras $\mc{R}_{AB}$ and $\widehat{\mc{R}}_{AB}$? As in the finite dimensional setting, the idea is that Alice and Bob each control their own regions. The algebra $\mc{R}_{AB}$ describes the local operations they can perform. For example, it allows them measure the total charge in their region or move charges around. But it \emph{does not} give them the ability to move a charge from one cone to the other, or equivalently, create a pair of conjugate charges (one in each cone). One can however show that \emph{charge transporters} $V$ that can move a charge from one cone into the other \emph{are} contained in $\widehat{\mc{R}}_{AB}$. This shows that $\widehat{\mathcal{R}}_{AB}$ is bigger than $\mathcal{R}_{AB}$, and it is precisely the observation that it contains the charge transporters that will allow us to connect it to the quantum dimension.

\section{Secret sharing}\label{sec:secshare}
  We now have the technical tools to describe a version of the quantum information task of Section~\ref{sec:finite} in the thermodynamic limit. In particular, we will describe how we can use charges localised in cones to store data that is invisible to Eve, using the presence of superselection sectors, and argue how our procedure is related to the topological entanglement entropy.

  In Section~\ref{sec:finite} we described how an information hiding-task can be implemented for systems on a finite lattice in 2 dimensions, motivating our index approach. Although the naive method there works, this finite dimensional description suffers from drawbacks, such as the index described there not manifestly being independent of the regions $A$ and $B$ and that the set of allowed operations for Eve not carrying a nice algebraic structure.
  Here we describe an analogue setting in the thermodynamic limit of the toric code on the plane and show that it overcomes both drawbacks, while still resulting in an operationally sensible picture.

  As in the finite dimensional variant, the task for Alice and Bob is to share information encoded in some quantum state on the whole system in such a way that Eve cannot access this information with any local measurement on her system.
  This means that Alice and Bob should be capable of reconstructing the shared information encoded in the quantum state just by performing local operations on their respective part of the system, while Eve cannot cannot access this information by using operations on her part of the system~\footnote{If we speak of ``local'' we always mean that the observable acts on finitely many particles on the lattice. Furthermore, in this context ``local'' additionally means that the observable is localised in one of the cones.}.
  This is exactly the situation described by secret sharing schemes as treated in~\cite{Gottesman:2000}.
  In such schemes the parts of the system that are capable of reconstructing the shared information solely by performing local operations are usually referred to as \emph{authorised}, whereas those that cannot are called \emph{unauthorised}.
  In our setting Alice and Bob will comprise the authorised parts of the system and Eve is unauthorised.
  Secret sharing schemes are usually defined for systems described by a finite dimensional Hilbert space, where the partition of the system into subsystems is given by a tensor product structure.
  In the thermodynamic limit of the toric code the system's Hilbert space is clearly infinite dimensional and the we do not have an obvious partition into tensor factors.
  In fact, one can show that the ground state Hilbert space does not factor~\cite{haagdtoric} as $\mathcal{H}_\Lambda \otimes \mathcal{H}_{\Lambda^c}$, where $\mathcal{H}_\Lambda$ is the Hilbert space related to a cone~\footnote{Although we do not claim that this is the case here, this touches upon a more fundamental property of infinite dimensional systems. Recently Slofstra has found a counterexample to Tsirelson's problem~\cite{Slofstra:2016}, by showing that there are commuting operator models for two-party correlations that are not equivalent to a tensor product model.}.
  In \cite{Gottesman:2000} it was shown, however, that there exists a characterisation of secret sharing schemes by error correction conditions.
  We will not generalise this secret sharing scheme to infinite dimensions, but will use this characterisation to illustrate that we indeed find a secret sharing scheme in the thermodynamic limit of the toric code.
  This is motivated by the observation that error correction schemes can be formulated in terms of operators~\cite{KriLafPoLe:2006,*BenyKemKri:2007} and, more generally, for von Neumann algebras~\cite{Crann:2015wj}.

  We will briefly review the authorised and unauthorised sets comprising a secret sharing scheme in finite dimensions. Given a subspace $\mc C\subset\mc K$ of some $n$-partite Hilbert space $\mc K$, the authorised sets $A\subset\{1,\dots,n\}$ are characterised by the condition that $\mc C$ corrects errors on their complements $A^c$.
  That is, for all $\phi,\psi\in\mc C$ and for all $E\in\alg B(\mc K_{A^c})$ it holds that $\langle\phi,E\phi\rangle=\langle\psi,E\psi\rangle$.
  Unauthorised sets $U\subset\{1,\dots,n\}$ are characterised by the condition that $\mc C$ corrects errors on them, i.e. $\langle\phi,F\phi\rangle=\langle\psi,F\psi\rangle$ for all $\phi,\psi\in\mc C$ and $F\in\alg B(\mc K_{U})$.
  For such pure state quantum secret sharing schemes it is easy to see that the no-cloning theorem implies that the unauthorised sets must be the complements of authorised sets and vice versa~\cite{Gottesman:2000}.

\begin{figure}
    \includegraphics[width=0.75\columnwidth]{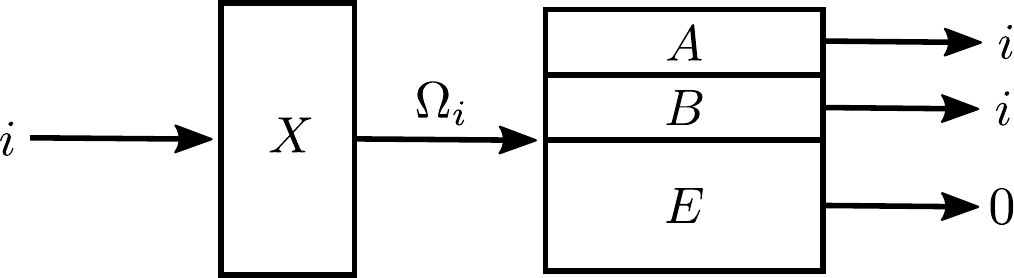}
    \caption{Example of a protocol for a secret sharing scheme: A sender $X$ chooses a state $\Omega_i$ from the set of code states $\{\Omega_0=\Omega,\Omega_X,\Omega_Z,\Omega_Y\}$ and sends it to the system consisting of authorised shares Alice ($A$) and ($B$) and unauthorised share Eve ($E$). The shares then operate locally on their part of the system to detect which state they received. The authorised shares Alice and Bob will be able to recover the information $i$ of which state they received, whereas Eve will always get the same answer ($0$) no matter which state was sent.}
    \label{fig:protocol}
  \end{figure}

  The setting we are considering here corresponds to the case where the shared information is classical.
  That is, the set of code states $\mc C$ consists of a choice of orthonormal vectors $\{\psi_i\}$.
  Then the conditions for unauthorised sets remain the same but the authorised sets $A$ are characterised by demanding that for each pair of indices $i,j$ and each operator $E\in\alg B(\mc K_{A^c})$ it holds $\langle\psi_i,E\psi_j\rangle=\delta_{i,j}\langle\psi_j,E\psi_j\rangle$ \cite{Gottesman:2000}. Here it is no longer true that unauthorised sets have to be complements of authorised sets, for classical information can be cloned.
  Figure~\ref{fig:protocol} shows what a protocol implementing a secret sharing scheme for classical information looks like.
  In the following section we describe how we can set up a secret sharing scheme in the thermodynamic limit of the toric code, specify a set of states which serve as code states, and check the above conditions.

\subsection{The use of cones}
  We now come back to the thermodynamic limit and start by considering two disjoint cones $\Lambda_A$ and $\Lambda_B$ which are separated sufficiently far enough from each other~\footnote{In \cite{haagdtoric} this is defined rigorously for the toric code, and in \cite{klindex} this is extended to more general models.}.
  For concreteness we describe the example of the toric code, but we believe that the method can be generalised to similar models; in particular Kitaev's quantum double models for abelian groups $G$ can be handled directly by using results from~\cite{haagdabelian}.
  These cones represent the regions to which Alice and Bob have access. The complement $\Lambda_E$ of the union of these two cones is considered to be controlled by a (possibly malicious) third party Eve.
  Eve cannot access $\Lambda_A$ or $\Lambda_B$.
  With the notations introduced in Section~\ref{sec:thelimit} we denote the von Neumann algebras of observables localised in the cones $\Lambda_A$ and $\Lambda_B$ by $\mc R_A$ and $\mc R_B$.
  The von Neumann algebra generated by the local observables on Eve's part is written as $\mc R_E$, while the algebra of observables commuting with $\mc R_E$ is denoted $\widehat{\mc R}_{AB}$.
   Here we are working in the translation invariant ground state representation of $\mf A$, that is, the cyclic representation $\pi_0$ associated to the (unique) translation invariant ground state $\omega_0$ of the toric code, on a Hilbert space $\mc H$, with $\omega_0$ represented as a unit vector $\Omega\in\mc H$.

  We are interested in ways to create states that Alice and Bob can distinguish, but Eve cannot.
  Of course, if Alice and Bob have access to both $\Lambda_A$ and $\Lambda_B$, they can just store their information by acting with local operators in one (or both) of the cones, and Eve will not be able to detect this.
  This scenario in itself is not that interesting, so we ask the question what they can do if they \emph{in addition} have access to operations that are not generated by the local observables in $\Lambda_A$ or $\Lambda_B$, but nevertheless invisible to Eve.
  Potentially, this gives them more power compared to the ``baseline'' scenario of local operations on their cones, and it are these additional capabilities that we want to investigate.
  The idea is to create a pair of charges, with one end of the pair in each cone.
  Since we are interested in the \emph{additional} power of Alice and Bob, we can disregard local modifications of these states that can be obtained by acting with observables in $\mathcal{R}_A$ or $\mathcal{R}_B$.
  Such operations include moving the charge around in the cone, or introducing pairs of charge and conjugate charge within a cone.
  We will come back to this point after we introduce the main idea in more detail.

\begin{figure}
    \centering
	\includegraphics[width=0.95\columnwidth]{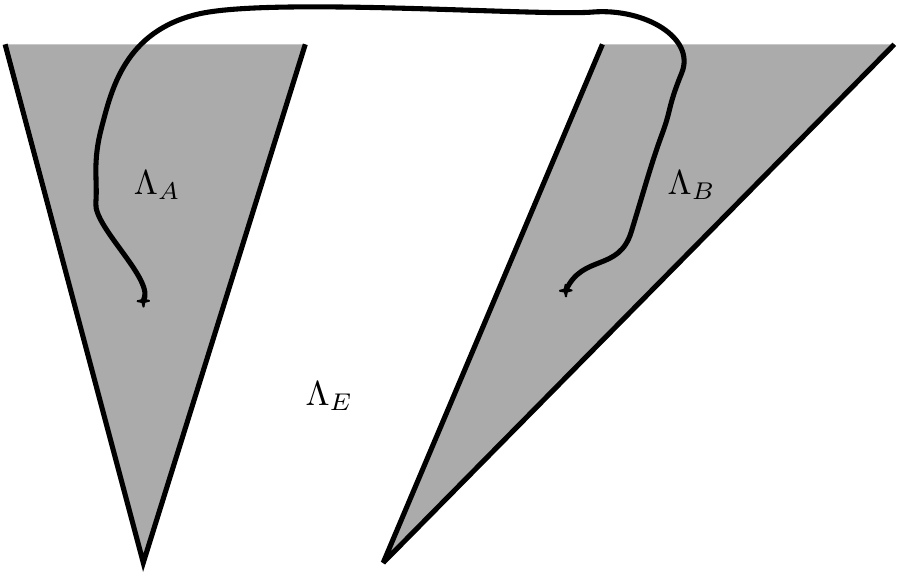}
    \caption{The cones $\Lambda_A$ and $\Lambda_B$ are the regions Alice and Bob have access to, and the remainder of the system $\Lambda_E$ is controlled by Eve. The charge transporters between $\Lambda_A$ and $\Lambda_B$ are constructed from sequences of ribbons with fixed endpoints (one in each cone) such that with growing length the ribbons avoids finite sets in $\Lambda_E$. The solid line indicates an infinite ribbon that stretches out to infinity and which we can think of representing a charge transporter.}
    \label{fig:ct}
  \end{figure}

  Note that the operations that Alice and Bob can perform in their respective cones commute with the observable Eve has at hand. This is the locality condition that is already built into the construction of the systems.
  We will show that we can use the charge transporters $V_X, V_Z$ and $V_Y$ that create pairs of excitations distributed over the cones $\Lambda_A$ and $\Lambda_B$ to construct states that the authorised parts can distinguish.
  They are unitaries on the Hilbert space $\mc H$ and one can think of them as creating correlations between the cones when applied to the ground state vector $\Omega$.
  Even though they are not localised in $\Lambda_A \cup \Lambda_B$, they still commute with all of Eve's observables, and hence are elements of $\widehat{\mc R}_{AB}$.
  The reason is that they can be obtained as weak operator limits of path operators.
  That is, one chooses a site in each cone, and connects them with a path (see Fig.~\ref{fig:ct}).
  Then, as $n$ grows, we let the path go to infinity (in the sense that it will avoid any finite subset of $\Lambda_E$ eventually, keeping the endpoints fixed).
  The corresponding path operators then converge to the charge transporter in the weak-operator topology.
  As a result the charge transporters commute with all of Eve's local observables, and hence are contained in $\widehat{\mc R}_{AB}$~\cite{toricendo}.
  From this it already follows that if $E\in\mc R_E$ is any operator on Eve's part of the system, its expectation values in the states $V_X\Omega, V_Z\Omega, V_Y\Omega$ and $\Omega$ coincide.
  That is, consider for example the state $V_i\Omega$ with $i=X,Z$ or $Y$, then $\langle V_i\Omega,EV_i\Omega\rangle=\langle\Omega,V_i^*EV_i\Omega\rangle$. Since $V_i\in\widehat{\mc R}_{AB}$ we have $[V_i,E]=0$, and therefore
  \begin{align*}
    \langle V_i\Omega,EV_i\Omega\rangle=\langle\Omega,E\Omega\rangle,\quad i=X,Z,Y.
  \end{align*}
  Hence Eve cannot distinguish between the states in $\mc C:=\{\Omega, V_X\Omega, V_Z\Omega, V_Y\Omega\}$ by acting with local operators on her system.
  On the other hand Alice and Bob can distinguish these states by acting locally on their cones:
  the construction of the charge transporters includes a specification of a site in each cone and the transporters create a pair of conjugate excitations from the ground states at these sites.
  Alice and Bob can now locally measure the flux through a (Wilson) loop around the respective sites to determine which excitation or whether at all was created.
  In other words, they do a charge measurement in their respective cones.
  The position of the site inside the cone itself is not so important here (that is why we did not specify it further) since there are always unitaries with support in either of the cones that can move the excitation around.

  This means, that Alice and Bob each possess a POVM that allows them to distinguish the states in $\mc C$.
  The choice of this POVM is not unique, since it depends on the loops around the site at which the excitation created by $V_i$ is localised. It can in principle be any loop of any size as long it surrounds the excitation. The flux measurement corresponds to a projection on the enclosed area onto the excitation one wants to measure (details descriptions of these projections can be found in \cite{MR1951039,BoMa:2008}). The POVM elements then simply consist of the projections on the different excitations that can occur, given that the loop is fixed.

  The above construction can be extended a bit. In the end, Alice and Bob are only interested in the \emph{total} charge in their region.
  Acting with local operators in either of the cones might move the existing charge around, or create pairs of conjugate charges, but the \emph{total} charge does not change.
  Hence, instead of just the states $V_i \Omega$, we can consider the four spaces $\mathcal{R}_{AB} V_i \Omega$ (or their closure).
  To Eve, all these states look the same, since $\mathcal{R}_{AB} \subset \mathcal{R}_E$, but Alice and Bob can in principle distinguish the four subspaces.
	Note that this is less practical, since the location of the charge of interest is not known, and Alice and Bob have to make sure their charge measurements encompass a big enough region.
	The projections that measure the \emph{total} charge in $\Lambda_A$ or $\Lambda_B$ are both in $\mathcal{R}_{AB}$, but these operations clearly are not local any more.
	For simplicity, we restrict to the set $\mathcal{C}$ of four states, and ignore the local modifications in the cones: the corresponding statements for that case can be obtained straightforwardly \footnote{The choice of taking $\mc C=\{\Omega,V_X\Omega,V_Z\Omega,V_XV_Z\Omega\}$ from the spaces $\mathcal{R}_{AB} V_i \Omega$ can be interpreted as choosing different implementations of secret sharing schemes with same information to be hidden and same access structure. Additionally Alice and Bob can act locally on the state within their cones without the corresponding other being able to notice.}.

  If we go back to the finite-dimensional description of secret sharing schemes where the Hilbert space is described by a tensor product of $n$ Hilbert spaces, it can be easily checked that for a set $A\subset\{1,\dots,n\}$ to be authorised with respect to a code space $\mc C$ is equivalent to requiring the existence of a POVM $\{E_i\}$ acting on $\mc H_{A}$ such that $E_i\psi_j=\delta_{i,j}\psi_j$ where $\psi_j\in\mc C$. It is necessary that the POVM elements commute with the operators acting on the complement of $A$.
  It makes sense then to rephrase the condition on $A$ to be authorised to the following: elements $E$ that commute with the observables $\alg B(\mc H_A)$ on $A$ such that it holds that $\langle\psi_i,E\psi_j\rangle=0$ if $i\neq j$. Note that this is equivalent to the original definition, since the Hilbert space decomposes as a tensor product of the authorised part $\mc H_A$ and its complement $\mc H_{A^c}$.

  In the infinite dimensional setting we do not necessarily have a decomposition of the Hilbert space into tensor products. But the algebraic view still allows for the characterisation of authorised sets. That is, we say that a subalgebra $\mc N$ of our system's algebra $\pi_0(\mf A)''$ is authorised with respect to the classical code space $\mc C$ if for all $E\in\mc N'$ and all $\psi_i,\psi_j\in\mc C$ with $i\neq j$ it holds that $\langle\psi_i,E\psi_j\rangle=0$.

  We can show that in our example of the toric code this is true for both Alice's and Bob's observables $\mc R_A$ and $\mc R_B$.
  In order to prove this it is crucial to understand what the structure of the commutants $\mc R_A'$ and $\mc R_B'$ is.
  In the following we only consider Alice's observables $\mc R_A$, since the case for Bob can be understood analogously.
  In fact it was shown~\cite{haagdtoric} that the von Neumann algebra $\mc R_A$ satisfies \emph{Haag duality}, that is, $\mc R_A=\mc R_{A^c}'$. Hence the commutant $\mc R_A'$ is exactly given by the observables $\mc R_{BE}$ on the complement $A^c=BE$ of Alice's cone.
  In the thermodynamic limit, this statement is non-trivial, the reason being that the algebras $\mc R_{BE}$ and $\mc R_A$ do not live on different tensor factors of the underlying Hilbert space. Moreover, it is in fact false for the algebra $\mc R_{AB}$ (in the sense that $\mc{R}_{AB}' \neq \mc{R}_{AB^c} = \mc{R}_E$) of observables localised in both cones $\Lambda_A$ and $\Lambda_B$, which is the whole reason that our construction works.
  In that case $\mc R_{AB}$ is properly included in the $\widehat{\mc R}_{AB}$, which as we recall is the algebra of observables that commute with all of Eve's observables.
  Hence Haag duality for cone algebras and its failure for the algebra $\mc R_{AB}$ are important for our setting.
  Even more, the identification of Alice's and Bob's part of the system, $\Lambda_A$ and $\Lambda_B$, as authorised sets by just the requirement that $\mc C$ corrects errors on their complements $\Lambda_A^c$ and $\Lambda_B^c$ still makes sense.

  Now let $\psi,\phi\in\mc C$ be distinct states and $E\in\mc R_A'=\mc R_{BE}$. Without loss of generality we can assume that $E$ is a product of ribbon operators $F_B$ and $F_E$ in $\mc R_B$ and $\mc R_E$, since the linear span of products of such forms a strongly dense subalgebra of $\mc R_{AB}$ \cite{haagdtoric}. Furthermore, recall that $\psi=V_i\Omega$ and $\phi=V_j\Omega$ for some $i\neq j$, including the possibility that $V_0= I$.
  Hence $\langle\psi,E\phi\rangle=\langle\Omega,V_i^*F_EF_BV_j\Omega\rangle$.
  Recall also that the action of the charge transporters $V_k$ on the ground state $\Omega$ create an excitation at some site $s_B\in\Lambda_B$ and its conjugate at some site $s_A\in\Lambda_A$.
  Thus, in order this scalar product be non-zero, the ribbons to which the operators $F_E$ and $F_B$ correspond, needed to connect the sites $s_B$ and $s_A$ with each other.
  But this is impossible, since this would require a ribbon in $\Lambda_A$ starting at $s_A$ and connecting to the boundary in $\Lambda_A$, and the corresponding ribbon operator would be in $\mc R_A$.
  Hence the product $V_i^*F_EF_BV_j$ creates non-trivial excitations above the ground state when acting on $\Omega$, and it follows $\langle\psi,E\phi\rangle=0$, since states that contain excitations above the ground state are orthogonal to the latter (see also \cite{BeShoWha:2011}).

  Summarising this, we consider the collection of orthonormal states $\mc C=\{\Omega,V_X\Omega,V_Z\Omega,V_Y\Omega\}$, or rather the classes of states $\mathcal{R}_{AB} V_i \Omega$ (recall that these are orthogonal spaces).
  Alice and Bob can distinguish these classes of states by doing charge measurements in the cones $\Lambda_A$ or $\Lambda_B$, hence $\Lambda_A$ and $\Lambda_B$ are authorised sets.
  On the other hand, all states look the same for local measurements in $\Lambda_E = (\Lambda_A \cup \Lambda_B)^c$, hence this can be regarded as an unauthorised set.
  The key properties used for this characterisation are Haag duality of cone algebras in the ground state representation and the detailed knowledge about the local excitation structure above the ground state.

  We would like to stress that this scheme cannot be extended to one for sharing quantum information. Given two different code states $\psi_i$ and $\psi_j$, $i\neq j$ (or, two states in different classes, in the general setting), we could in principle also prepare the superposition $\psi:= \frac{1}{\sqrt{2}}\left(\psi_i+e^{i \varphi} \psi_j\right)$, where $\varphi$ is a fixed complex phase. But we cannot distinguish this superposition from any other of these two states by any \emph{local} operation on $\Lambda_A$ or $\Lambda_B$.
  To see this, consider a local observable $A\in\mc R_A$. Then, since $A$ just acts in the cone $\Lambda_A$ the operator $V_i^*AV_j,i\neq j$ creates non-trivial excitations above the ground state, hence $\langle\psi_i,A\psi_j\rangle=0$, and therefore $\langle\psi,A\psi\rangle=\frac{1}{2} \langle\psi_i,A\psi_i\rangle+ \frac{1}{2}\langle\psi_j,A\psi_j\rangle$.
  So local charge measurements lose information about relative phases in the superposition of code states. That is, the code states obey a superselection rule.

  Another question is if there are more states which we can add to the already known code states to increase the amount of information Alice and Bob can share (where, as before, we are only interested in states up to acting with local operators in $\Lambda_A$ or $\Lambda_B$).
  That is, are there perhaps other operations in $\widehat{\mathcal{R}}_{AB}$ that lead us to a new class of states that can be distinguished with local operations in $\Lambda_A$ or $\Lambda_B$?
  This turns out not to be the case, which can be understood by the Jones-Kosaki-Longo index.
  The charge transporters which are used to create the code states are such that they commute with all observables at Eve's disposal, but are not localised in either cone. Hence the question whether this set of states is complete translates to the question whether we found all charge transporters, corresponding to charges that we may not have found yet.
  This question can be answered by computing the value of the Jones-Kosaki-Longo index:
  it turns out that all the observables $\widehat{\mc R}_{AB}$ that commute with Eve's allowed operations $\mc R_E$ are precisely generated by those in the cone and the unitary operators $V_X,V_Z,V_Y$, that is $\widehat{\mc R}_{AB}$ is generated as a von Neumann algebra by the charge transporters and $\mc R_{AB}$ \cite{klindex}.
  This result follows from two steps: first one proves the general result that the index provides a bound on the number of inequivalent charges (and hence, the number of ``inequivalent'' charge transporters).
  Then one can calculate the index itself, and see that the known charge transporters already saturate this bound.
  This argument shows that we can regard the Jones-Kosaki-Longo index then as the maximal number of code states for a secret sharing scheme of classical information.
  We come back to this point in more detail in Section~\ref{sec:channels}.

  There is one other point to discuss: is it really necessary to use the operations in $\widehat{\mc R}_{AB}$?
  Indeed, if $F_{\xi}$ is a path operator between the sites in cone $\Lambda_A$ and $\Lambda_B$, then $V_i \Omega = F_{\xi} \Omega$.
  The problem is that Alice and Bob at some point have to apply an operator to store the classical information.
  If they do this by creating a pair of excitations in one cone, and then move one of the excitations to the other cone, Eve could detect the excitation as it moves through her region.
  Even if Alice and Bob are able to do this so quickly that Eve has no chance of finding out, there is another method for Eve to detect this.
  So far we have assumed that Eve does not alter her part of the state.
  That is, the state we start with is $\Omega$.
  But nothing prevents Eve from doing any operation on her part of the system.
  For example, she could create a configuration of charges on her side.
  If the path operator that Alice and Bob apply to the system crosses one of the paths of Eve's operators, the state will acquire a phase because of the anyonic nature of the excitations, which opens up possibilities of detection.
  If Alice and Bob use the operations in $\widehat{\mc R}_{AB}$, this cannot happen, although it should be noted that in a laboratory setting this may not be very practical (or even possible).
  We briefly comment on this last point below.

\subsection{TEE and the irreducible correlation}\label{sec:irrcor}
  There exists an interesting interplay between the topological entanglement entropy \cite{PhysRevLett.96.110404, PhysRevLett.96.110405} and the irreducible correlation of the state, which provides the TEE with an operational interpretation as the achievable rate of a certain secret sharing scheme of classical information \cite{KaFuMu:2015}.

  The irreducible correlation is a measure of correlations on multipartite quantum systems. More generally, the \emph{$k$-th irreducible correlation} $C^{(k)}(\rho)$ of a state $\rho$ on an $n$-partite system quantifies how much correlations are contained in the $k$-th reduced density matrix (RDM) $\rho^{(k)}$ that are not contained in the $(k-1)$-RDM $\rho^{(k-1)}$.
  It is given by the expression $C^{(k)}(\rho)=S(\tilde\rho^{(k)})-S(\tilde\rho^{(k-1)})$, where $\tilde\rho^{(l)}$ is the state that maximises $S(\sigma)$ when optimising over all states that have the same $l$-RDM as $\rho$. Precise definitions can be found in \cite{ZhouYou:2007,KaFuMu:2015}.

  Consider now a state $\rho$ of a quantum many body system that satisfies an area law with a ``topological'' contribution, i.e.\ $S(A)=\beta|\partial A|-n_{\partial A}\gamma+\mc O(|\partial A|^{-1})$ for (large) regions $A$ with some constant  $\beta>0$ and $n_{\partial A}$ being the number of connected components of the boundary of $A$, and $\gamma$ the TEE.
Assume that the correlation length is finite, i.e.\ $\rho_{AC}=\rho_A\otimes\rho_C$ for disjoint regions $A,C$ that are far away from each other, and that the conditional mutual information between $A$ and $C$ conditioned on $B$ is zero, i.e. $I(A:C|B)=0$, if the regions $A$ and $C$ are connected through a third region $B$ such that $ABC$ has no holes.
  Under these conditions it was shown in \cite{KaFuMu:2015} that the topological entanglement entropy $S_{top}$ 
  coincides with the 3rd irreducible correlations of the reduced density matrix (RDM) on regions $ABC$ \cite{KaFuMu:2015}. Here $ABC$ is a partition of the system in a configuration as considered in Kitaev-Preskill \cite{PhysRevLett.96.110404} and similarly in Brown et al.~\cite{2013arXiv1303.4455B}, or as in the Levin-Wen definition of TEE \cite{PhysRevLett.96.110405}. Note that, borrowing the notation from the introduction, $S_{top}^{KP}=\gamma$ and $S_{top}^{LW}=2\gamma$.

  If we consider a finite region $ABC$ as in \cite{PhysRevLett.96.110404,PhysRevLett.96.110405} the 3rd irreducible correlation $C^{(3)}(\rho_{ABC})$ thus characterises the correlation in $\rho_{ABC}$ that are not contained in the RDM of any bipartition of the tripartite system $ABC$.
  Furthermore, in~\cite{KaFuMu:2015} it was shown that then $C^{(3)}(\rho_{ABC})$ carries an operational interpretation in terms of the \emph{maximal rate} $r_{ABC}$ of a certain secret sharing scheme of classical information.
  To be more precise, regarding $\rho_{ABC}$ as a tripartite state over $ABC$ it holds that $C^{(3)}(\rho_{ABC})$ is equal to the optimal sharing rate $r_{ABC}$ for a secret sharing scheme for classical information that encodes information in $\rho_{ABC}$ such that the information can only be decoded by having access to all three regions $A$, $B$ and $C$ \footnote{The optimal rate determines how many bits can at most be encoded in the state $\rho_{ABC}$ that it there exists a decoding channel that reliably can recover the information in asymptotic many uses of the scheme.}.

  That is, for optimal encoding and decoding channels the number bits that can be encoded is given by $r_{ABC}$.
  Now since $C^{(3)}(\rho_{ABC})=r_{ABC}$ this means that the number of bits that can be encoded using the tripartite correlations of $\rho_{ABC}$ is given by $S_{top}$.
  For the case of anyon models where the topological entanglement entropy of the ground state is given by $\gamma=\log\mc D$ and $\mc D$ is the total quantum dimension, the number of bits that can be encoded is $\mc D$ in case of $S_{top}^{KP}$ and $\mc D^2$ in case of $S_{top}^{LW}$. Therefore the total quantum dimension determines the maximal amount of information we can encode in $\rho_{ABC}$ by just using the tripartite correlations in this state.
  The difference between the two settings of \cite{PhysRevLett.96.110404,PhysRevLett.96.110405} is a result of the different topologies in the choice of $ABC$.
  Intuitively in the Levin/Wen type of regions there exist operators acting along non-contractible loops that leave the ground state invariant and that contribute to $C^{3}(\rho_{ABC})$ whereas in the Kitaev/Preskill setting such loops can be contracted.

  In the thermodynamic limit of the toric code we have, however, a different geometry of the regions $A$, $B$ and $C$, where we identify the regions $A$ and $B$ with the cones $\Lambda_A$ and $\Lambda_B$ controlled by Alice and Bob, respectively, and $C$ with Eve's part $\Lambda_E$. Also note that in this case we have that $ABC$ comprises the whole system, as opposed to the finite dimensional case, where $ABC$ just needs to be a sufficiently large region.
  As discussed in the previous sections, in this setting $\mc D^2$ is the dimension of the code space of a secret sharing scheme for classical information between the algebras over disjoint cones.
  More precisely, the number of equivalence classes $\mc R_{AB}V_i\Omega$ of states that differ only by local operators in the respective cones, is given by $\mc D^2$. The code space is maximal in so far as that the Jones-Kosaki-Longo index bounds the number of superselection sectors from above~\cite{klindex}, and is equal to $\mc D^2$.
  In this sense $\log{[\widehat{\mc R}_{AB} : {\mc R}_{AB}]}=\log{\mc D^2}$ is the optimal sharing rate of that scheme and we regard this as an infinite dimensional analogue of the results obtained in~\cite{KaFuMu:2015}.
  In the next section this equality is discussed in more detail.
  In the general case we expect that the index also carries a similar interpretation.

  The JKL index can also be related to a relative entropy for the inclusion $\mc R_{AB}\subset\widehat{\mc R}_{AB}$ which can be interpreted in terms of a Holevo quantity, giving a bound on how much better we can distinguish states using operations from $\widehat{\mc{R}}_{AB}$, compared to with just operations from $\mc{R}_{AB}$.
  The details can also be found in the next section.
  Consequently the thermodynamic limit exhibits a very similar structure as in the situation of finite lattices.

\subsection{Can we work around superselection sectors?}
  The secret sharing task we described depends on the presence of charge transporters, and hence of superselection sectors. These are modelled as equivalence classes of representations, satisfying the localisation criterion~\eqref{eq:superselect}. The idea to use superselection sectors to assist in quantum information tasks is not new, see for example~\cite{VerstraeteCirac}, where the authors apply the fact that local operators cannot distinguish the different superselection sectors to a data hiding protocol.

  It is natural to ask if superselection sectors can be used to circumvent certain no-go theorems in quantum information. Unfortunately, this turns out not to be the case, if we assume an adversary Eve has access to an auxiliary system to store compensating charges~\cite{Kitaev:2004p5175}. Hence the authors of~\cite{Kitaev:2004p5175} conclude that superselection sectors cannot be used to increase the security of quantum information protocols.

  This result appears to be at odds with our claim that Alice and Bob can share a secret securely with the help of superselection sectors (which is essential in our construction). This is not the case, since the two settings are fundamentally different.  In particular, Kitaev et al.\ consider the case where the superselection sectors are given by a compact group symmetry. The adversary Eve is then allowed to do \emph{any} operation that commutes with this symmetry. If she has an auxiliary system available in which she can store a compensating charge, she can implement arbitrary transformations without breaking the symmetry. In our setting, we know that the symmetry is not given by a group (since our anyons are describe by a modular tensor category), and Eve does not have an auxiliary system at her disposal. In addition, she can only do \emph{local} operations, which further limits her powers. In particular, such operations cannot interpolate between different superselection sectors, at least not in our setting, where the we describe infinite systems.

  We also do not need to assume that the total charge in the system is zero, it is enough to know that the total charge in both Alice's and Bob's cone is trivial, which they can check before starting the secret sharing protocol. If there are only \emph{abelian} sectors, even this assumption is not necessary: Alice and Bob can each measure the total charge in their cone before the protocol starts, and record the result. Since the fusion rules in that case give a \emph{unique} charge after fusing two anyons, they can compensate their measurements by computing the result of fusion with the conjugate charge. In the non-abelian case this is no longer true, since there are multiple fusion outcomes.

\section{Channels and Entropy}\label{sec:channels}
  The secret sharing task we described suggests a description in terms of quantum channels. In particular, we would like to have a quantum channel that compares the ``full'' operations available to Alice and Bob, described by $\widehat{\mc{R}}_{AB}$, to the strictly local operations $\mc{R}_{AB}$. Fortunately the index theory for subfactors provides such a map. This is the map that we investigate in this section. Moreover, it is possible to define a relative entropy for subfactors. This relative entropy is related to the index of the inclusion. Here we will argue that this relative entropy makes it possible to connect the index to the well-known Holevo $\chi$-quantity, which tells us how well we can distinguish states, and is related to the classical capacity of a quantum channel.

  We will again consider the toric code here, although the abstract constructions work for any subfactor with finite index.
  The toric code however has the advantage of being simple enough to allow a concrete analysis, and at the same time providing a clear physical interpretation. It will allow us to match the mathematical constructions to physical processes.

\subsection{Channels}\label{sec:channelex}
  The inclusion $\mc R_{AB}\subset\widehat{\mc R}_{AB}$ of the cone observables into the algebra of observables that commute with Eve's observables is accompanied by a \emph{conditional expectation} $\mc E:\widehat{\mc R}_{AB}\to\mc R_{AB}$ \cite{klindex,MR829381}, that is, a generalisation of the partial trace to the language of operator algebras.
  A conditional expectation is a (normal) completely positive map such that $\mathcal{E}(ABC) = A \mathcal{E}(B) C$ for all $A, C \in \mc{R}_{AB}$ and $B \in \widehat{\mc{R}}_{AB}$.
  The subalgebra ${\mc R}_{AB}$ hereby plays the role of the subsystem.
  In fact $\mc E$ is a channel; it is linear, completely positive, preserves the identity operator and is normal in the sense that it maps normal states to normal states.
  These are states that are represented by density matrices on the Hilbert space on which the algebra is represented.
  As mentioned earlier elements of $\widehat{\mc R}_{AB}$ can be expressed as linear combination of some ``basis'' with coefficients in $\mc R_{AB}$.
  Moreover the algebra $\widehat{\mc R}_{AB}$ is generated as a von Neumann algebra by $\mc R_{AB}$ and the charge transporters $\{V_X,V_Z\}$.
  With the notation $V_i, i=0,X,Z,Y$ with $V_0=I$ and $V_Y = V_X V_Z$, the basis expansion of elements $X\in\widehat{\mc R}_{AB}$ is then $X=\sum_iA_iV_i$ with $A_i\in\mc R_{AB}$ \cite{klindex}. Another way of saying this is that $\widehat{\mc R}_{AB}$ is a left module over $\mc R_{AB}$.
  In this case the operators $V_i$ are also called a ``Pimsner-Popa basis''.

  The channel $\mc E$ is then given by
\begin{equation}
	\label{eq:condexp}
	\mc E : \widehat{\mc{R}}_{AB} \to \mc{R}_{AB} : X \mapsto A_0.
\end{equation}
  In a sense it leaves the states $\Omega,V_X\Omega,V_Z\Omega$ and $V_Y\Omega$ invariant. This can be seen as follows.
  In the Schr{\"o}dinger picture the channel is given by the unique completely positive map $\mc E_*$ determined by $\mc E_*(\rho) := \rho\circ\mc E$ where $\rho$ is a normal state over $\mc R_{AB}$.
  We do not intend to give a full characterisation of $\mc E_*$ here. Instead we show how it acts the vector states $\Omega,V_X\Omega,V_Z\Omega$ and $V_Y\Omega$, where $\Omega$ is the ground state. Since these are vectors in the Hilbert space, they give rise to normal states on $\mc R_{AB}$ and on $\widehat{\mc R}_{AB}$.
  Let $\psi$ be any of these states and $X\in\widehat{\mc R}_{AB}$ as above, and let $\rho$ be the corresponding state on $\mc R_{AB}$.
  As shown in Section~\ref{sec:secshare}, $\bra\Omega V_iAV_j\ket\Omega=0$ for any $A\in\mc R_{AB}$ and $V_i\neq V_j$.
  In particular, this implies that $\bra{\psi} A V_i \ket{\psi} = 0$ if $i \neq 0$, and we find $(\mc E_*\rho)(X)=\rho(A_0)=\bra\psi A_0\ket\psi=\bra\psi X\ket\psi$.
  Hence the states corresponding to $\psi$ are invariant under the action of $\mc E_*$. For superpositions this is no longer true since $\mc E_*$ erases the off-diagonal elements of the density matrices in this basis.
  Of course the situation is much more complicated for general normal states on $\mc R_{AB}$ but this illustrates well the classical nature of the secret sharing scheme.
  Note that the argument still holds if we consider the states $U V_i \Omega$, with $U$ a unitary in $\mc{R}_{AB}$, so that we again have four classes of (vector) states.

  Before we come to the information-theoretical interpretation of the map $\mc{E}$, we first make another interesting observation.
  There is a canonical way to get a tower of inclusions of von Neumann algebras if we have a finite index subfactor.
  We here give an example of extending the tower downwards.
  Recall that the charge transporters constitute a unitary representation of the group $\mathbb{Z}_2\times \mathbb{Z}_2$ on the Hilbert space of the ground state representation. This representation induces an action on the operators by conjugation. This action maps $\mc R_{AB}$ into itself.
  Therefore the twirl $\mc E_1(A):=\frac14\sum_iV_iAV_i^*, A\in\mc R_{AB}$ is a conditional expectation from the cone algebra $\mc R_{AB}$ to the subalgebra $\mc R_0$ of fixed points of this action. The inclusion $\mc R_0\subset\mc R_{AB}$ then has index $[\widehat{\mc{R}}_{AB} : \mc{R}_{AB}]$. Furthermore, the channel $\mc E_1$ is implemented by the projection $P_0= \frac{1}{4} \sum_i V_i$ in the sense that $\mc E_1(A)P_0=P_0AP_0$.
  This subalgebra consists of these operations in the cones $\Lambda_A$ and $\Lambda_B$ which cannot distinguish the states $\Omega,V_X\Omega,V_Z\Omega$ and $V_Y\Omega$ from each other. In this sense the channel $\mc E_1$ can be interpreted as the completely depolarising channel on these states.

    \subsection{Relative entropies and classical information}
	The index can be connected to a relative entropy~\cite{MR860811}. Here we follow the work of Hiai~\cite{MR1096438,*MR1150623}, who discusses the case of general subfactors (not just Type II$_1$~\footnote{Von Neumann algebras which have trivial centres (in other words, factors), can be classified in types I, II$_1$, II$_\infty$ and Type III\@. Type I factors are precisely those that are isomorphic to $\alg{B}(\mathcal{H})$ for some Hilbert space $\mathcal{H}$. The type of the factors has important implications for the technical parts of the index theory, but the qualitative features are largely the same.}) and gives different characterisations of the index.
  We start with defining the relative entropy of a pair of von Neumann algebras $\alg{N} \subset \alg{M}$ with respect to a normal state $\varphi$ on $\alg{M}$. This is given by
  \begin{equation}
	  \label{eq:relent}
  H_\varphi(\alg{M} | \alg{N}) = \sup_{(\varphi_i)} \sum_i \left[ S(p_i \varphi_i, \varphi) - S(p_i \varphi_i \upharpoonright \alg{N}, \varphi \upharpoonright \alg{N}) \right],
  \end{equation}
  where again we use $\upharpoonright$ to denote restriction to a subalgebra. The supremum is over all \emph{finite} convex combinations such that $\varphi = \sum_i p_i \varphi_i$, with $\varphi_i$ a normal state.
  That is, we consider all different preparations of the state $\varphi$.
  The relative entropy $S(p_i \varphi_i, \varphi)$ is to be understood in the sense of Araki~\cite{MR0425631,*MR0454656} (see~\cite{MR1230389} for an introduction).
  Compared to these references we switched the order of the arguments to agree with the usual definition in quantum information.
  The definition of Araki reduces to the well-known formula for the quantum relative entropy of finite systems if the algebras are matrix algebras.
  We also note that the terms in square brackets are positive.
  This is perhaps not immediately clear, but essentially follows from the monotonicity of the relative entropy (restricting the states is like tracing out a part of the system).

  First we find it useful to find a physical interpretation of equation~\eqref{eq:relent}. Intuitively, it should capture how well we can distinguish states when we have all operations in $\alg{M}$ at our disposal, compared to when only measurements (or, POVM's) from $\alg{N}$ are allowed. To make this intuition more precise, consider the following scenario which is typical when trying to send classical information over a quantum channel. We largely follow Holevo~\cite{MR2986302} (but also see~\cite{MR3088659}), and for the moment consider finite dimensional systems. Let $\rho$ be a state on the system. If $\rho$ is a mixed state, there are different ways to prepare this state. In particular, consider a probability distribution $p_x$ and let $\rho_x$ be states such that $\rho = \sum_x p_x \rho_x$. That is, Alice picks a state according to the probability distribution $p_x$. The question then is if Alice sends this state to Bob, how well Bob can recover the distribution $p_x$. In general, even if Alice sends many copies, Bob cannot recover $p_x$ exactly, for example when the $\rho_x$ are pure but overlapping states. How well Bob is able to recover $p_x$ is governed by the Holevo $\chi$-quantity, defined as
  \begin{equation}
  	\label{eq:chiquantity}
  	\chi( \{p_x\}, \{ \rho_x \}) := S(\rho) - \sum_x p_x S(\rho_x) = \sum_{x} p_x S(\rho_x, \rho).
  \end{equation}
  This is a quantum generalisation of the Shannon information, and gives an upper bound on the amount of information Bob can recover. The equality follows from the definition of the relative entropy.

  In the infinite setting that we are interested in, the definition of the entropy $S(\rho)$ is problematic (since it typically scales with the dimension of the system), and it is better to stick to the relative entropy. We therefore take the right hand side of equation~\eqref{eq:chiquantity} as the definition of $\chi$. Using the identity $S(p_i \varphi_i, \varphi) = p_i S(\varphi_i,\varphi) + p_i \log(p_i)$, we can rewrite equation~\eqref{eq:relent} to
  \[
  	H_\varphi(\alg{M}|\alg{N}) = \sup_{(\varphi_i)} \chi( \{p_i\}, \{\varphi_i\} ) -  \chi( \{p_i\}, \{\varphi_i \upharpoonright \alg{N} \} ).
  \]
  By the previous paragraph, this tells us the maximum amount of extra information we can gain if we are allowed to use operations from $\alg{M}$, compared to when only operations from $\alg{N}$ are allowed, in case the state $\varphi$ is sent.
  Sometimes it is also called the ``quantum privacy'', since it tells us how much information is inaccessible for $\alg{N}$.

  We now come back to the inclusion $\mc{R}_{AB} \subset \widehat{\mc{R}}_{AB}$. The discussion of the secret sharing protocol shows that $\widehat{\mc{R}}_{AB}$ contains operators that are not in $\mc{R}_{AB}$, that make it possible to share classical information. Conversely, it is possible to discern more states using operations in $\widehat{\mc{R}}_{AB}$ compared to $\mc{R}_{AB}$. Hence we expect that there are states $\varphi$ such that $H_\varphi(\widehat{\mc{R}}_{AB}|\mc{R}_{AB}) > 0$. This is indeed the case. In fact, we will relate these relative entropies to the quantum dimension, by relating it to the Jones-Koski-Longo index of the inclusion.

  To do this, recall that if the subfactor $\mc{R}_{AB} \subset \widehat{\mc{R}}_{AB}$ has finite index, then there is a conditional expectation $\mc{E}: \widehat{\mc{R}}_{AB} \to \mc{R}_{AB}$ such that there is some $\lambda > 0$ with $\mc{E}(X) \geq \lambda X$ for all positive operators $X \in \widehat{\mc{R}}_{AB}$.
  In fact, there is a unique conditional expectation $\mathcal{E}$ maximising the constant $\lambda$~\cite{MR1027496}.
  In the example of the toric code it is the map $\mathcal{E}$ of equation~\eqref{eq:condexp}.
  The index is then equal to the inverse of the best such constant, with the convention that the index is infinite if there is no conditional expectation for which such a (positive) $\lambda$ exists. Conversely, the existence of such a conditional expectation implies that the index is finite, in particular there is a $\lambda > 0$.

  Consider then the conditional expectation $\mc{E}$ that maximes the bound.
  One can then define the relative entropy with respect to $\mathcal{E}$ by
  \begin{align*}
  	H_{\mathcal{E}}(\widehat{\mc{R}}_{AB}|\mc{R}_{AB}) &:= \sup  H_\varphi(\widehat{\mc{R}}_{AB}|\mc{R}_{AB})
\end{align*}
  The supremum is over all faithful normal states $\varphi$ on $\widehat{\mc{R}}_{AB}$ such that $\varphi \circ \mc{E} = \varphi$. 
  In general the relative entropy $H_{\mathcal{E}}(\widehat{\mc{R}}_{AB}|\mc{R}_{AB})$ is bounded from above by the logarithm of the index $[\widehat{\mc{R}}_{AB} : \mc{R}_{AB}]$ (see below for the argument in the easier Type II$_1$ case).
  By Corollary 7.2 of Ref.~\cite{MR1096438}, however, equality is attained if and only if the conditional expectation $\mc E$ maximises the bound in the previous paragraph. Hence we have
  \begin{align}
    H_{\mathcal{E}}(\widehat{\mc{R}}_{AB}|\mc{R}_{AB}) &= \log [ \widehat{\mc{R}}_{AB} : \mc{R}_{AB} ].\label{eq:indexrelent}
  \end{align}
  In the next subsection we will see that $[\widehat{\mc{R}}_{AB} : \mc{R}_{AB}]$ is related to the quantum dimension, so this relates the quantum dimension to a quantity that has a clear operational interpretation in terms of the amount of information that can be hidden.
  As an aside, for such $\varphi$ one can actually simplify the formula for $H_\varphi(\alg{M}|\alg{N})$ a bit:
  \[
  	H_\varphi(\alg{M}|\alg{N}) = \sup_{(\varphi_i)} \sum_i S(\varphi_i, \varphi_i \circ \mc{E}).
  \]
  The optimisation is again over all (finite) decompositions of $\varphi$.

  To get some intuition for the quantity $H_\mc{E}(\alg{M}|\alg{N})$ for some inclusion $\alg{N} \subset \alg{M}$ of von Neumann algebras, it is useful to consider the case where $\alg{M}$ and $\alg{N}$ are of Type II$_1$.
  It can be shown that this is not true in the case we are interested in~\cite{haagdtoric}, but the example is illustrative nonetheless.
  In the Type II$_1$ case, there is a (faithful) tracial state $\tau$ on $\alg{M}$, that is, a state such that $\tau(AB) = \tau(BA)$.
  It should be noted that Type II$_1$ factors are defined on infinite dimensional Hilbert spaces, so that $\tau$ is not the familiar trace of bounded (trace class) operators.
  If the index is finite, a trace preserving conditional expectation $\mc{E} : \alg{M} \to \alg{N}$ exists, with the index being equal to the inverse of the best constant $\lambda$ as above.
  Note that this further supports the notion of $\mc{E}$ as a quantum channel (since in the usual setting they are required to preserve the trace).
  In that case, it can be shown that the relative entropy can be rewritten as follows, where we set $\lambda_{x_i} := \tau(x_i)$~\cite{MR1096438}:
  \begin{align}
	  \label{eq:finitecase}
    H_\mc{E}(\alg{M}|\alg{N}) = \sup_{(x_i)} \sum_{i} \lambda_{x_i} \left[S(\mc{E}(\rho_{x_i})) - S(\rho_{x_i})\right].
  \end{align}
  Here $\rho_{x_i}$ is the density operator $x_i/\lambda_{x_i}$ and the entropy $S$ is defined with respect to $\tau$. The supremum is over all finite sets of positive operators $x_i$ such that $\sum_i x_i = 1$. In other words, it is an optimisation over all (finite) POVMs.
  Note that instead of looking at states, we now look at the possible operations we can use to distinguish states.
  The quantity between square brackets is called the \emph{entropy gain} in~\cite{Holevo.ChoiJamiolkowski}.

  Before we comment on the physical interpretation of equation~\eqref{eq:finitecase}, we come back to the claim on why it is equal to the Jones index.
  Again, we consider the Type II$_1$ case for simplicity, following Pimsner and Popa~\cite{MR860811}.
  The case of infinite factors that we need here is technically much more involved, but uses some similar ideas~\cite{MR1096438,*MR1150623}.
  Recall that there is a $\lambda > 0$ such that $\mathcal{E}(X) \geq \lambda X$ for all positive $X$.
  Then, since the logarithm is operator increasing, from equation~\eqref{eq:finitecase} one can show that $H_\mc{E}(\alg{M}|\alg{N}) \leq \log \lambda^{-1}$.
  Since one of the equivalent definitions of the index is that it is the inverse of the best of such constants $\lambda$, it follows that $H_{\mathcal{E}}(\alg{M}|\alg{N}) \leq \log [\alg{M}: \alg{N}]$.
  To complete the argument Pimsner and Popa find lower bounds for $H_\mathcal{E}(\alg{M}|\alg{N})$, and show that in the case of irreducible factors (such as we consider here), equality is in fact attained.
  The proof of this is more involved, and requires properties of subfactors that are outside of the scope of this paper.

  To understand equation~\eqref{eq:finitecase} a bit better, note that since $\mc{E}(I) = I$, we can add $S(I) - S(\mc{E}(I))$ to the right hand side of equation~\eqref{eq:finitecase}.
  But in that case, it simplifies to
  \[
  	\sup_{ I = \sum \lambda_x \rho_x} \chi( \{\lambda_x\}, \{ \rho_x \}) - \chi( \{\lambda_x\}, \{ \mathcal{E}(\rho_x) \}).
  \]
  Note that that the optimisation is \emph{only} over ensembles that sum up to the completely mixed state.
  This should be contrasted with the (Holevo) channel capacity $\chi_{\mc{E}} := \sup_{\{p_x\}, \{\rho_x\}} \chi(\{\lambda_x\}, \{\mc{E}(\rho_x) \})$, which gives the amount of classical information that can be transmitted using the channel~\cite{holevo1979capacity,*HolevoCapacity}.
  Note that here the optimisation is over \emph{all} ensembles.

  We also like to point out the similarity to wiretap channels.
  In a quantum wiretap channel, quantum information is sent form Alice to Bob, with an eavesdropper Eve.
  Such a channel maps density operators on $\mc{H}_A$ into $\mc{H}_B \otimes \mc{H}_E$ via a map $\rho \mapsto V \rho V^*$, where $V$ is an isometry.
  Note that any quantum channel can be written in this form by means of a Stinespring dilation.
  The point of the wiretap channel is that certain information is inherently private, in the sense that no measurement on $\mathcal{H}_E$ can recover it.
  This was first studied for quantum channels by Schumacher and Westmoreland~\cite{PhysRevLett.80.5695}.
  Later this analysis was extended, for example by allowing simultaneous use of multiple copies of the channel~\cite{Cai2004,*Devetak2005}.
  This for example leads to a proof that the (classical) private information is bounded from below by the (quantum) channel capacity.
  Although our setting is slightly different, the definition of what information is inaccessible or private is essentially the same.

  There is yet another description of essentially the same problem, in terms of a subfactor that is closer to the protocol outlined earlier.
  The inclusion $\mc{R}_{AB} \subset \widehat{\mc{R}}_{AB}$ could be understood by considering the charge transporters.
  The interpretation above however does not directly connect to the secret sharing scheme described earlier.
  A property of the index is that it is invariant under taking commutants:
  \[
  	[ \widehat{\mc{R}}_{AB} : \mc{R}_{AB} ] = [\mc{R}_{AB}' : \widehat{\mc{R}}_{AB}' ].
  \]
  Note that $\widehat{\mc{R}}_{AB}' = \pi(\alg{A}(\Lambda_E))''$, that is, the von Neumann algebra generated by all local observables accessible to Eve.
  In contrast, $\mc{R}_{AB}'$ contains \emph{more} operations.
  In particular, it contains projections that measure the total charge in one of the cones.
  These projections are not in Eve's algebra, hence she cannot use them.
  This is precisely what Alice and Bob use to hide information from her, and by a similar analysis as we have provided above, the amount of information that can be hidden in this way is quantified by the index.

\subsection{Total quantum dimension}
  The discussion above gives a relation between the index $[ \widehat{\mc{R}}_{AB} : \mc{R}_{AB} ]$ and the amount of inaccessible classical information.
  In particular, this can be quantified by equation~\eqref{eq:indexrelent}, so it would be good to have a better understanding of $[\widehat{\mc{R}}_{AB} : \mc{R}_{AB} ]$.
  From Section~\ref{sec:jkl} we see that this number tells us (in a sense) how much bigger $\widehat{\mc{R}}_{AB}$ is than $\mc{R}_{AB}$, while Section~\ref{sec:ssect} and the example in Section~\ref{sec:channelex} indicate that this is related to the superselection sectors (or anyons) of the theory.
  On the other hand, the topological entanglement entropy is related to the logarithm of the total quantum dimension, while also quantifying achievable rates in a secret sharing scheme, as discussed in Section~\ref{sec:irrcor}.
  Hence it would be reasonable to assume that there is a relation between the index and the total quantum dimension.

  This is indeed the case, and can be shown without any reference to any communication protocols.
  Already in 1989 Longo showed that the quantum dimension $d_i$ of a representative of a superselection sector can be obtained as the index of a certain inclusion of von Neumann algebras~\cite{MR1027496}.
  Later in 2001 it was shown that for the class of rational conformal field theories on the circle, the total quantum dimension is equal to the index of an inclusion $\mc{R} \subset \widehat{\mc{R}}$, very similar to the inclusion $\mc{R}_{AB} \subset \widehat{\mc{R}}_{AB}$~\cite{MR1838752}, and indeed our results are partially motivated by that paper.

  A similar strategy can be applied to the lattice models that we are interested in.
  If we assume (in addition to the technical conditions of Haag duality and the approximate split property mentioned above) that each charge has a corresponding conjugate charge (or show that they exist), it is always possible to define the quantum dimension of a charge.
  In that case, the relation $[\widehat{\mc{R}}_{AB} : \mc{R}_{AB}] = \sum_{i} d_i^2$ holds~\cite{MR1838752,klindex}, where the sum is over all distinct charges $\rho_i$, and $d_i$ is the corresponding statistical (quantum) dimension.
  If we do not assume existence of conjugate charges, the index still gives an upper bound on the number of them.

  It should be noted that this is more than abstract theory.
  For example, for the toric code one can explicitly show that Haag duality and the approximate split property hold~\cite{haagdtoric}.
  It is also possible to explicitly obtain representatives of different superselection sectors, and for example show that conjugates exist~\cite{toricendo}.
  Finally, independently from the superselection sector analysis, it can be shown that $[ \widehat{\mc{R}}_{AB} : \mc{R}_{AB} ] = 4$~\cite{klindex}.
  In fact, this result can be used to show that in fact any superselection sector of the model is equivalent to one of the explicit representatives that can be constructed.
  Hence for the toric code, the whole program can be carried out in full detail, and we see that also using the index method, we see that we can hide four classical bits.

  To summarise the discussion, we can conclude that the total quantum dimension gives tells us how much classical information can be hidden, in the setup described above.
  This provides an alternative interpretation way of thinking about the total quantum dimension.
  One of the advantages is that the argument is completely rigorous, and independent of any results on the finite dimensional models.
  In particular, we do not need to assume the relation between the topological entanglement entropy and the total quantum dimension.
  We also point out that the analysis is not restricted to the topologically ordered quantum spin systems that we have looked at so far.
  Rather, they can be applied to all models (once one makes appropriate technical assumptions) for which one can do a superselection structure analysis in terms of localised and transportable representations.
  This in particular applies to rational conformal field theories on the circle in the operator-algebraic approach~\cite{MR1838752}.

\section{Private quantum subsystems}\label{sec:private}
  We have discussed an operational interpretation of the Jones-Kosaki-Longo index in terms of a secret sharing task: the anyonic charges allow Alice and Bob to store (classical) bits which are not available to the adversary Eve. This is reminiscent of the theory of \emph{private quantum codes} or \emph{private subsystems} (see~\cite{PhysRevA.90.032305} and references therein). We argue that our construction can be interpreted in this way.

  Our description is stated in terms of observables, hence it is most natural to use the Heisenberg picture. Therefore in our setting a quantum channel will be a completely positive (cp) normal (i.e., continuous with respect to the weak-operator topology) map $\mathcal{E} : \alg{M} \to \alg{N}$ between two von Neumann algebras. Its dual is a normal cp map $\mathcal{E}_*: \alg{N}_* \to \alg{M}_*$, mapping normal states to normal states. Since we are dealing with infinite dimensional von Neumann algebras (and Hilbert spaces) it is necessary to go beyond the setting of~\cite{PhysRevA.90.032305}, and we will use the recent generalisation to von Neumann algebras by Crann et al.~\cite{Crann:2015wj}. Let $\mathcal{E} : \alg{M} \to \alg{B}(\mathcal{H})$ be a quantum channel, and $P$ a projection on the Hilbert space $\mathcal{H}$.
  Suppose moreover that $\mathcal{N}$ is a von Neumann algebra on $P \mathcal{H}$. Then $\alg{A}$ is called \emph{private} for $\mathcal{E}$ with respect to $P$ if $P \mathcal{E}(\alg{M}) P \subset \mathcal{N}'$.

  Our setup immediately leads to an example of a private quantum channel. The index theory gives us a normal conditional expectation $\mc{E}: \widehat{\mc{R}}_{AB} \to \mc{R}_{AB}$. Hence in particular, $\mathcal{E}$ is a normal cp map. We can choose $\alg{N} =  \left(\mathcal{R}_A \vee \mathcal{R}_B \right)' = \mathcal{R}_{AB}'$.
  Since $\mathcal{R}_A \vee \mathcal{R}_B$ is a von Neumann algebra, and therefore equal to its double commutant, it follows that $\alg{N}' = \mc{R}_{AB}$. Hence $\alg{N}$ is private for $\mathcal{E}$ with respect to $P = I$.
  Note also that $\mc R_E\subset\mc N$, that is, Eve's observables are private for $\mc E$.

  One can show that $\alg{N}$ is private for $\mc{E}$ if and only if it is correctable (in the sense of~\cite{MR2541204}) for any complementary channel $\mathcal{E}^c$ of $\mathcal{E}$~\cite[Thm. 4.7]{Crann:2015wj}. That is, there is some channel $\mc{R}$ such that $\mathcal{E}^c \circ \mc{R} = \operatorname{id}_{\alg{N}}$.
  Here $\mathcal{E}^c$ is a channel of the form $\mathcal{E}^c(X) = V^* X V$ for all $X \in \pi(\widehat{\mathcal{R}}_{AB})'$, where $(\pi, V, \mathcal{H})$ is a Stinespring triple for the channel $\mathcal{E}$.

  Consider again the example of the toric code. In that case we have an explicit description of $\widehat{\mathcal{R}}_{AB}$, which allows us to identify such a Stinespring triple.
  In particular, we know that $\widehat{\mathcal{R}}_{AB}$ is isomorphic to the crossed product $\mathcal{R}_{AB}\rtimes_\alpha(\mathbb{Z}_2\times\mathbb{Z}_2)$, where $\alpha_g(A)=V_gAV_g^*$ and $g\mapsto V_g$ is a unitary representation of $\mathbb{Z}_2\times\mathbb{Z}_2$ obtained by mapping $(1,0)\mapsto V_X$ and $(0,1) \mapsto V_Z$~\cite{klindex}.
  Concretely, denote $\mathcal{H}$ for the Hilbert space of the GNS representation of the translational invariant ground state of the toric code. Then we can define a representation $\pi$ of $\widehat{\mathcal{R}}_{AB}$ by sending $\sum_{k=0,X,Y,Z} A_k V_k$ (with $A_k \in \mathcal{R}_{AB}$) to the following operator, acting on $ \mathcal{H}_S := \mathcal{H} \oplus \mathcal{H} \oplus \mathcal{H} \oplus \mathcal{H}$:
  \[
  \pi\left(\sum_k A_k V_k\right) =
  	\begin{pmatrix}
  		A_0 & A_X V_X & A_Z V_Z & A_Y V_Y \\
  		A_X V_X & A_0 & A_Y V_Y & A_Z V_Z \\
  		A_Z V_Z & A_Y V_Y & A_0 & A_X V_X \\
  		A_Y V_Y & A_Z V_Z & A_X V_X & A_0
  	\end{pmatrix}.
  \]
  This can be shown to give an isomorphism of $\widehat{\mathcal{R}}_{AB}$ with the crossed product. Now define an isometry $V: \mathcal{H} \to \mathcal{H}_S$ by $V \psi = (\psi, 0, 0, 0)$. Then by a short calculation we check that
  \[
  	\mathcal{E}(X) = V^* \pi(X) V, \quad X \in \widehat{\mc{R}}_{AB},
  \]
  that is, $(\pi, V, \mathcal{H}_S)$ is a Stinespring triple for $\mathcal{E}$. Note that $\pi(\widehat{\mc{R}}_{AB}) V \mathcal{H}$ is dense in $\mathcal{H}_S$, hence the Stinespring dilation is minimal. Now consider the map $\mathcal{R} : \alg{N} \to \alg{B}(\mathcal{H}_S)$, defined by $\mathcal{R}(N) = \operatorname{diag}(N, N, N, N)$, which is a normal unital cp map.
  Then since $\alg{N} = \mathcal{R}_{AB}'$, it is clear from the description of $\pi$ above that $\mathcal{R}(\alg{N}) \subset \pi(\widehat{\mathcal{R}}_{AB})'$. Moreover, $\mathcal{E}^c \circ \mathcal{R} = \operatorname{id}_{\alg{N}}$, hence $\alg{N}$ is correctable for $\mathcal{E}$ (with respect to the identity projection).
  Similarly one can see that in this representation the twirl channel $\mc E_1$ from Section~\ref{sec:channels} is represented as $\mc E_1(A)=\frac14\sum_{g\in G}V^*\pi(V_gAV_g^*)V,\, A\in\mc R_{AB}$, with $G=\mathbb Z_2\times\mathbb Z_2$, and $\pi(\mc R_0)$ is given by matrices of the form $\mr{diag}(A,A,A,A)$ with $A\in\mc R_0$, where $\mc{R}_0$ is the fixed-point algebra as before.

  This example can be generalised to the abelian quantum double model in a straightforward way. The non-abelian model is more difficult, since there the symmetry is not described by $G \times \widehat{G}$ any more, and we do not expect to find a similar crossed product structure. However, the general setting of the quantum dimension being related to the Jones index still applies, and we expect a similar correctable subalgebra result to hold with respect to the canonical conditional expectation $\mathcal{E}$ one obtains from the index theory. Moreover, what is interesting is that the index gives us a measure of the amount of classical information that is private for Eve. This suggests that the Jones index might be a useful tool in the study of the capacity of quantum channels. We hope to return to this question in the future.

\section{Stability under perturbations}
  There are a few technical assumptions that we needed to make in our analysis of the systems in the thermodynamic limit. In particular, we assume that the superselection sectors associated to the anyons can be strictly localised in cone regions. Although any topological charge should certainly be localisable in such a region, strict localisation is likely a too strong condition in general. This generalisation is important when considering \emph{perturbations} of the system, which is necessary if one wants to show that the quantum dimension is truly an invariant of a topologically ordered quantum phase.

  This can be seen as follows. Because of the topological order condition, we expect that the properties of the anyonic excitations will be the same across the whole phase (indeed, they should be by the very definition of a phase). That is, if we perturb the dynamics of our model (without closing the spectral gap), the perturbed ground state should have the same superselection sectors. However, the selection criterion, equation~\eqref{eq:superselect}, as we have used it here, will generally no longer hold: in the thermodynamic limit the ground states of the deformed model can be obtained by composing the original ground states with an automorphism $\alpha$~\cite{2011arXiv1102.0842B}. This automorphism is however not \emph{strictly} local. Rather, $\alpha$ satisfies a Lieb-Robinson type of bound, such that for strictly local $A$, $\alpha(A)$ can in general only be approximated up to a small (exponentially decreasing) error by a strictly local observable. As a consequence, if $\pi$ satisfies the selection criterion, it is not guaranteed that $\pi \circ \alpha$ does so too, since we only know unitary equivalence of $\pi$ and $\pi_0$ for observables \emph{outside} any given cone. However, since $\alpha$ is not strictly local, it does not map cone algebras into cone algebras.

  As a result it is necessary to adapt the superselection criterion, and in turn the inclusion of the von Neumann algebras associated to the two cones. It should be noted that a similar phenomenon also appears in Ref.~\cite{KaFuMu:2015}: their results are only strict in the case of zero correlation length. In either case it is expected that in the thermodynamic limit (or in the operator-algebraic case we are interested in, the limit of growing cone size) the small corrections vanish. We believe the information theoretic interpretation here will be of use in studying this question: for example, instead of correctable algebras in the previous section, one should use $\varepsilon$-correctable algebras~\cite{Crann:2015wj}, which allow for (arbitrarily small) errors in the correction. We hope to come back to this issue in future work.

\section{Summary and discussion}
  We have reviewed the total quantum dimension of topologically ordered systems, in particular how in the thermodynamic limit it can be obtained as the Jones-Kosaki-Longo index of an inclusion of certain algebras of observables. It has been argued by other authors~\cite{PhysRevLett.96.110405,PhysRevLett.96.110404} that the quantum dimension can also be obtained via topological entanglement entropy in finite dimensional systems, a fundamentally different approach. Nevertheless, it turns out that both quantities have an interpretation in terms of a secret sharing scheme, although the implementation details are different in both cases. Even though our secret sharing scheme is not very practical (and is not intended as such), it provides new insight to the quantum dimension, and gives a completely different viewpoint (or approach) of what appears to be same underlying concept. We believe that this may be beneficial to gaining a better understanding of such systems.

  The operator-algebraic approach we advocate here provides a rigorous and elegant mathematical framework. It also has other advantages. For example, inclusions of subfactors are well studied, in particular in the context of the Jones-Kosaki-Longo index, and many mathematical results are available. This puts the theory on firm mathematical footing. Moreover, a lot of structure comes for free with a finite index inclusion: we mentioned the conditional expectation $\mathcal{E}$, which can be interpreted as a quantum channel.

  We also believe this operator-algebraic approach might be beneficial in the important question of stability of topological phases. Although we have only explicitly mentioned the toric code as a test case, we argued that these structures hold more general in topologically ordered models (with the caveat mentioned in the previous section). Generalisation to the abelian quantum double is straightforward, but also in non-abelian models we expect to have a similar structure. An explicit verification, however, will of course be much more involved. Finally, while we mainly have studied what is usually referred to as ``long-range entangled'' phases, an algebraic approach to symmetry protected phases appears to be reasonable; as a toy model one can consider the Kitaev wire, and divide the system into three parts, as we did in the example of the Fibonacci chain. We conjecture that this can be related to a notion of entanglement entropy for symmetry protected phases, cf.~\cite{2013arXiv1307.6617M}.

  Although the setting we discussed here is tied to the setting of charges belonging to different superselection sectors, the conditional expectation (and hence a quantum channel) always exists for subfactors of finite index. Moreover, the index is related to a relative entropy, which opens up connections to quantum information theory:
  the discussion in Section~\ref{sec:channels} is an example of that.
  We believe that the index theory may be useful to study, for example, capacities of quantum channels, in particular for systems with infinitely many degrees of freedom. Except for the case of gaussian states, there are comparatively few tools available to deal with such examples.

\paragraph*{Acknowledgements}\hspace{-1em}.---
  TJO was supported by the ERC grants QFTCMPS and SIQS, and by the cluster of excellence EXC201 Quantum Engineering and Space-Time Research.
  PN has received funding from the European Union's Horizon 2020 research and innovation program under the Marie Sklodowska-Curie grant agreement No 657004. LF is supported by the European Research Council (ERC) through the Discrete Quantum Simulator (DQSIM) project.
  Many thanks go to K. Abdelkhalek, C.\ B\'eny, C.\ Brell, D.\ Reeb, R.\ Schwonnek and R.F.\ Werner for plenty of insightful discussions.

\appendix
\section{Operator algebras}\label{app:cstar}
  Dealing with quantum systems with infinitely many degrees of freedom, such as the thermodynamic limit of the quantum spin systems we are interested in here, introduces complications that are not present when discussing finite dimensional systems.
We prefer to use an operator-algebraic approach to tackle these.
In this appendix we give some reasons for why we elect this perspective, and introduce the main definitions and concepts.

To see an example of the difficulties that arise, consider an infinite chain of qubits. Naively, one might expect that the Hilbert space of this system is given by $\mathcal{H} = \bigotimes_{n=-\infty}^\infty \mathbb{C}^2$. There is however a problem with the definition of the inner product: let $\psi, \eta \in \mathcal{H}$. Then the inner product should be defined as
  \[
  	\langle \psi, \eta \rangle_\mathcal{H} := \prod_{n=-\infty}^\infty \langle \psi_n, \eta_n \rangle_{\mathbb{C}^2},
  \]
  analogously to the tensor product of a finite number of Hilbert spaces. The problem is that the expression on the right generally does not converge, since it is an infinite product. A simple example is given by taking a unit vector $\Omega \in \mathbb{C}^2$ and setting $\psi_n = \Omega$ and $\eta_n = (-1)^n \Omega$.

  We can work around this by using von Neumann's construction of the infinite tensor product: we choose a reference unit vector $\Omega_n$ for each $n$, and only consider vectors $\psi \in \mathcal{H}$ for which $\psi_n \neq \Omega_n$ for only finitely many $n$. For such vectors the expression above converges and defines an inner product. By taking the completion with respect to the norm obtained from this inner product, we arrive at a Hilbert space $\mathcal{H}$.

  This definition is somewhat undesirable, since it depends on the choice of reference vector, and a canonical choice may or may not be available (and results might depend on the choice of vector). In addition, it is not entirely clear what the observables are. One could consider all bounded operators $\alg{B}(\mathcal{H})$ as in single-particle quantum mechanics (potentially considering unbounded observables as well), but this has the downside that one loses some of the locality structure that the chain clearly has. These are some of the reasons why we prefer to work in an operator-algebraic (or, if one wishes, observable-centric) approach, which does not have these problems. For the benefit of the reader we recall the main definitions and explain how they can be interpreted in the context of quantum mechanics (see also~\cite{KeylSchlingemannWerner} and~\cite{MR887100,MR1441540}).

\subsection{$C^*$-algebras}
  We want to consider quantum spin systems with infinitely many sites. For concreteness, consider the square lattice $\mathbb{Z}^2$, where at each site there is a quantum spin, with Hilbert space $\mathbb{C}^d$. As remarked above, we cannot just take the infinite tensor product of $\mathbb{C}^d$, and we will focus on the local observables of the system.

  Let $\Lambda \subset \mathbb{Z}^2$ be a \emph{finite} subset, consisting of $|\Lambda|$ spins. Since this is a finite quantum spin system, it is described by a Hilbert space $\mathcal{H}_\Lambda = \bigotimes_{x \in \Lambda} \mathbb{C}^d$. Hence the associated observables are the (self-adjoint) elements of
  \[
  	\alg{A}(\Lambda) := \alg{B}(\mathcal{H}_\Lambda) = \bigotimes_{x \in \Lambda} M_d(\mathbb{C}).
  \]
  We will find it convenient to call $\alg{A}(\Lambda)$ the \emph{local observables} with support in $\Lambda$ (or \emph{localised} in $\Lambda$), even for those elements that are not self-adjoint.

  Now suppose that $\Lambda_1 \subset \Lambda_2$ are both finite subsets of $\mathbb{Z}^2$. Then $\mathcal{H}_{\Lambda_2} \simeq \mathcal{H}_{\Lambda_1} \otimes \mathcal{H}_{\Lambda_2 \setminus \Lambda_1}$. Hence we can identify $A \in \alg{A}(\Lambda_1)$ with $A \otimes I_{\mathcal{H}_{\Lambda_2 \setminus \Lambda_1}}$ in $\alg{A}(\Lambda_2)$. In addition, if $A \in \alg{A}(\Lambda_1)$ and $B \in \alg{A}(\Lambda_2)$, with $\Lambda_1 \cap \Lambda_2 = \emptyset$ and both finite, it is clear that $[A,B] = 0$. This is known as \emph{locality}, and hence we have a local structure. We want to consider the algebra generated by all such local observables. To this end, define the \emph{(strictly) local observables} by $\alg{A}_{loc} = \bigcup_{\Lambda} \alg{A}(\Lambda)$, where the union is over all finite subsets of $\mathbb{Z}_2$, and we identify those operators that come from inclusions $\alg{A}(\Lambda_1) \subset \alg{A}(\Lambda_2)$ in the obvious way.

  The algebra $\alg{A}_{loc}$ has a natural norm, induced by the operator norm on $M_d(\mathbb{C})$. It is however not complete with respect to this norm: there are Cauchy sequences in $\alg{A}_{loc}$ that do not converge. This can be solved by taking the closure with respect to this norm, i.e., by adding limits of Cauchy sequences. This gives a complete normed $*$-algebra $\alg{A}$, whose norm satisfies $\|A^*A\| = \|A\|^2$ for all $A \in \alg{A}$. Such an algebra is called a \emph{$C^*$-algebra}. We call the elements of $\alg{A}$ \emph{quasi-local observables}, since they can be approximated arbitrarily well (in the operator norm) by strictly local observables.

  In this setting states are given by positive linear functionals $\omega$ of norm one on $\alg{A}$. That is, linear maps $\omega: \alg{A} \to \mathbb{C}$ such that $\omega(A^*A) \geq 0$ and $\omega(I) = 1$ (or, equivalently, $\| \omega \| = 1$). The value $\omega(A)$ for a positive operator $A$ has the same interpretation as in Hilbert space quantum mechanics: it is the expectation value of $A$. We note that states are not necessarily of the form $\operatorname{Tr}(\rho A)$ for some density matrix $\rho$.

  Finally, once we have the algebra of observables we can specify the dynamics by specifying local Hamiltonians. These Hamiltonians generate, under suitable conditions (e.g., the interactions should decay fast enough), a time evolution on the algebra, which is most conveniently described as a one-parameter group $t \mapsto \alpha_t$ of automorphisms. That is, this gives a time evolution of the observables in the Heisenberg picture. Once dynamics are defined it is possible to talk about ground states: these are essentially the states that minimise the energy. In our case we are usually interested in translationally invariant ground states, and in many of the models of interest they are in addition frustration free: they minimise the expectation values of each local Hamiltonians individually.

  The Hilbert space picture can be very useful, and fortunately it is not lost in this algebraic approach. Indeed, the Gel'fand-Naimark-Segal (GNS) construction gives a representation of $\alg{A}$ on a Hilbert space. More precisely, suppose that $\omega$ is a state on $\alg{A}$. Then the GNS construction gives a triple $(\pi, \Omega, \mathcal{H})$, where $\mathcal{H}$ is a Hilbert space, $\pi$ is a representation of $\alg{A}$ as bounded operators on $\mathcal{H}$, that is, a linear map $\pi: \alg{A} \to \alg{B}(\mathcal{H})$ that is compatible with the product and adjoint operation of $\alg{A}$. The state $\omega$ is implemented in the Hilbert space by $\Omega \in \mathcal{H}$, in the sense that $\omega(A) = \langle \Omega, \pi(A) \Omega \rangle$ for all $A \in \alg{A}$. Note that this does \emph{not} imply that $\omega$ is a pure state. In fact, this is true if and only if $\pi$ acts irreducibly on $\mathcal{H}$, or equivalently, only multiples of the identity commute with every $\pi(A)$.

\subsection{Von Neumann algebras}
  Now consider a Hilbert space $\mathcal{H}$. Then $\alg{B}(\mathcal{H})$, the algebra of bounded operators on $\mathcal{H}$, is a $C^*$-algebra. Besides convergence in the operator norm, the underlying Hilbert space gives additional notions of convergence. If $A_i \in \alg{B}(\mathcal{H})$ (or more generally, a net $A_\lambda$ of operators) is a sequence of operators, we say it converges \emph{strongly}, or in the \emph{strong operator topology}, to an operator $A \in \alg{B}(\mathcal{H})$ if for any $\psi \in \mathcal{H}$, we have that $\| (A_i-A)\psi \| \to 0$. In other words, when acting on a \emph{fixed} vector, we get a convergent sequence. In general the rate of convergence depends on the vector $\psi$, and if $\mathcal{H}$ is infinite dimensional one cannot conclude that $A_i \to A$ in the operator norm.

  There is another topology that has a clear physical interpretation. We say that a sequence $A_n$ of operators converges in the \emph{weak operator topology} to some operator $A$ if for each $\psi \in \mathcal{H}$, we have that $|\langle\psi, A_n \psi \rangle - \langle \psi, A \psi \rangle| \to 0$ if $n \to \infty$. That is, a sequence of observables converges in this topology if we cannot distinguish them (in the limit $n \to \infty$) by measuring in arbitrary vector states.

  Now consider a unital $*$-subalgebra $\alg{M} \subset \alg{B}(\mathcal{H})$. We say that $\alg{M}$ is a \emph{von Neumann algebra} if it is closed in the weak operator topology. This is equivalent to being closed in the strong operator topology, since one can show that both topologies coincide on bounded sets. A perhaps more surprising (and very useful) fact is that this is equivalent to the algebraic condition $\alg{M} = \alg{M}''$, where $\alg{M}'' := (\alg{M}')'$, and the prime denotes the commutant in $\alg{B}(\mathcal{H})$. That is, $\alg{M}' := \{ T \in \alg{B}(\mathcal{H}) : TX = XT\,\textrm{ for all } X \in \alg{M} \}$. This is known as the \emph{bicommutant theorem}. It is easy to check that $\alg{M}''' = \alg{M}'$ if $\alg{M}$ is closed under the $*$-operation, hence this gives an easy way to obtain von Neumann algebras from subsets of $\alg{B}(\mathcal{H})$.

  Finally we would like to mention another useful property of von Neumann algebras, which is not true for general $C^*$-algebras: they are generated by their projections. This has the following application. Suppose that $O \in \alg{M}$ is some self-adjoint observable that we would want to measure. It is often the case that we cannot (or do not want) the whole observable $O$, for example due to limitations on equipment, but are content with the following question: does the measured value of $O$ lie in some interval $I = [a,b]$? This yes/no question corresponds to measuring a projection $P_{[a,b]}$. Indeed, it is the spectral projection of $O$ on the interval $I$. It follows from spectral theory that this projection also is in $\alg{M}$, and hence an observable. This is even true for positive unbounded operators, such as the Hamiltonian $H$ of the system, under mild additional assumptions (in particular, it should be \emph{affiliated} with $\alg{M}$~\cite[Lemma 2.5.8]{MR887100}). These properties make it natural to look at von Neumann algebras.

\bibliography{refs}

\def\cprime{$'$}
\begin{thebibliography}{87}%
\makeatletter
\providecommand \@ifxundefined [1]{%
 \@ifx{#1\undefined}
}%
\providecommand \@ifnum [1]{%
 \ifnum #1\expandafter \@firstoftwo
 \else \expandafter \@secondoftwo
 \fi
}%
\providecommand \@ifx [1]{%
 \ifx #1\expandafter \@firstoftwo
 \else \expandafter \@secondoftwo
 \fi
}%
\providecommand \natexlab [1]{#1}%
\providecommand \enquote  [1]{``#1''}%
\providecommand \bibnamefont  [1]{#1}%
\providecommand \bibfnamefont [1]{#1}%
\providecommand \citenamefont [1]{#1}%
\providecommand \href@noop [0]{\@secondoftwo}%
\providecommand \href [0]{\begingroup \@sanitize@url \@href}%
\providecommand \@href[1]{\@@startlink{#1}\@@href}%
\providecommand \@@href[1]{\endgroup#1\@@endlink}%
\providecommand \@sanitize@url [0]{\catcode `\\12\catcode `\$12\catcode
  `\&12\catcode `\#12\catcode `\^12\catcode `\_12\catcode `\%12\relax}%
\providecommand \@@startlink[1]{}%
\providecommand \@@endlink[0]{}%
\providecommand \url  [0]{\begingroup\@sanitize@url \@url }%
\providecommand \@url [1]{\endgroup\@href {#1}{\urlprefix }}%
\providecommand \urlprefix  [0]{URL }%
\providecommand \Eprint [0]{\href }%
\providecommand \doibase [0]{http://dx.doi.org/}%
\providecommand \selectlanguage [0]{\@gobble}%
\providecommand \bibinfo  [0]{\@secondoftwo}%
\providecommand \bibfield  [0]{\@secondoftwo}%
\providecommand \translation [1]{[#1]}%
\providecommand \BibitemOpen [0]{}%
\providecommand \bibitemStop [0]{}%
\providecommand \bibitemNoStop [0]{.\EOS\space}%
\providecommand \EOS [0]{\spacefactor3000\relax}%
\providecommand \BibitemShut  [1]{\csname bibitem#1\endcsname}%
\let\auto@bib@innerbib\@empty
\bibitem [{\citenamefont {Chen}\ \emph {et~al.}(2010)\citenamefont {Chen},
  \citenamefont {Gu},\ and\ \citenamefont {Wen}}]{PhysRevB.82.155138}%
  \BibitemOpen
  \bibfield  {author} {\bibinfo {author} {\bibfnamefont {Xie}\ \bibnamefont
  {Chen}}, \bibinfo {author} {\bibfnamefont {Zheng-Cheng}\ \bibnamefont {Gu}},
  \ and\ \bibinfo {author} {\bibfnamefont {Xiao-Gang}\ \bibnamefont {Wen}},\
  }\bibfield  {title} {\enquote {\bibinfo {title} {Local unitary
  transformation, long-range quantum entanglement, wave function
  renormalization, and topological order},}\ }\href {\doibase
  10.1103/PhysRevB.82.155138} {\bibfield  {journal} {\bibinfo  {journal} {Phys.
  Rev. B}\ }\textbf {\bibinfo {volume} {82}},\ \bibinfo {pages} {155138}
  (\bibinfo {year} {2010})},\ \Eprint {http://arxiv.org/abs/cond-mat/0404617}
  {arXiv:cond-mat/0404617} \BibitemShut {NoStop}%
\bibitem [{\citenamefont {Hastings}\ and\ \citenamefont
  {Wen}(2005)}]{PhysRevB.72.045141}%
  \BibitemOpen
  \bibfield  {author} {\bibinfo {author} {\bibfnamefont {Matthew~B.}\
  \bibnamefont {Hastings}}\ and\ \bibinfo {author} {\bibfnamefont {Xiao-Gang}\
  \bibnamefont {Wen}},\ }\bibfield  {title} {\enquote {\bibinfo {title}
  {Quasiadiabatic continuation of quantum states: {T}he stability of
  topological ground-state degeneracy and emergent gauge invariance},}\ }\href
  {\doibase 10.1103/PhysRevB.72.045141} {\bibfield  {journal} {\bibinfo
  {journal} {Phys. Rev. B}\ }\textbf {\bibinfo {volume} {72}},\ \bibinfo
  {pages} {045141} (\bibinfo {year} {2005})},\ \Eprint
  {http://arxiv.org/abs/cond-mat/0503554} {arXiv:cond-mat/0503554} \BibitemShut
  {NoStop}%
\bibitem [{\citenamefont {{Bachmann}}\ \emph {et~al.}(2012)\citenamefont
  {{Bachmann}}, \citenamefont {{Michalakis}}, \citenamefont {{Nachtergaele}},\
  and\ \citenamefont {{Sims}}}]{2011arXiv1102.0842B}%
  \BibitemOpen
  \bibfield  {author} {\bibinfo {author} {\bibfnamefont {Sven}\ \bibnamefont
  {{Bachmann}}}, \bibinfo {author} {\bibfnamefont {Spyridon}\ \bibnamefont
  {{Michalakis}}}, \bibinfo {author} {\bibfnamefont {Bruno}\ \bibnamefont
  {{Nachtergaele}}}, \ and\ \bibinfo {author} {\bibfnamefont {Robert}\
  \bibnamefont {{Sims}}},\ }\bibfield  {title} {\enquote {\bibinfo {title}
  {{Automorphic Equivalence within Gapped Phases of Quantum Lattice
  Systems}},}\ }\href {\doibase 10.1007/s00220-011-1380-0} {\bibfield
  {journal} {\bibinfo  {journal} {Commun. Math. Phys.}\ }\textbf {\bibinfo
  {volume} {309}},\ \bibinfo {pages} {835--871} (\bibinfo {year} {2012})},\
  \Eprint {http://arxiv.org/abs/1102.0842} {arXiv:1102.0842} \BibitemShut
  {NoStop}%
\bibitem [{\citenamefont {Bravyi}\ \emph {et~al.}(2010)\citenamefont {Bravyi},
  \citenamefont {Hastings},\ and\ \citenamefont {Michalakis}}]{BraHaMi:2010}%
  \BibitemOpen
  \bibfield  {author} {\bibinfo {author} {\bibfnamefont {Sergey}\ \bibnamefont
  {Bravyi}}, \bibinfo {author} {\bibfnamefont {Matthew}\ \bibnamefont
  {Hastings}}, \ and\ \bibinfo {author} {\bibfnamefont {Spyridon}\ \bibnamefont
  {Michalakis}},\ }\bibfield  {title} {\enquote {\bibinfo {title} {Topological
  quantum order: stability under local perturbations},}\ }\href {\doibase
  10.1063/1.3490195} {\bibfield  {journal} {\bibinfo  {journal} {J. Math.
  Phys.}\ }\textbf {\bibinfo {volume} {51}},\ \bibinfo {pages} {093512}
  (\bibinfo {year} {2010})},\ \Eprint {http://arxiv.org/abs/1001.0344}
  {arXiv:1001.0344} \BibitemShut {NoStop}%
\bibitem [{\citenamefont {Bravyi}\ and\ \citenamefont
  {Hastings}(2011)}]{MR2842961}%
  \BibitemOpen
  \bibfield  {author} {\bibinfo {author} {\bibfnamefont {Sergey}\ \bibnamefont
  {Bravyi}}\ and\ \bibinfo {author} {\bibfnamefont {Matthew~B.}\ \bibnamefont
  {Hastings}},\ }\bibfield  {title} {\enquote {\bibinfo {title} {A short proof
  of stability of topological order under local perturbations},}\ }\href
  {\doibase 10.1007/s00220-011-1346-2} {\bibfield  {journal} {\bibinfo
  {journal} {Commun. Math. Phys.}\ }\textbf {\bibinfo {volume} {307}},\
  \bibinfo {pages} {609--627} (\bibinfo {year} {2011})},\ \Eprint
  {http://arxiv.org/abs/1001.4363} {arXiv:1001.4363} \BibitemShut {NoStop}%
\bibitem [{\citenamefont {Levin}\ and\ \citenamefont
  {Wen}(2006)}]{PhysRevLett.96.110405}%
  \BibitemOpen
  \bibfield  {author} {\bibinfo {author} {\bibfnamefont {Michael}\ \bibnamefont
  {Levin}}\ and\ \bibinfo {author} {\bibfnamefont {Xiao-Gang}\ \bibnamefont
  {Wen}},\ }\bibfield  {title} {\enquote {\bibinfo {title} {Detecting
  topological order in a ground state wave function},}\ }\href {\doibase
  10.1103/PhysRevLett.96.110405} {\bibfield  {journal} {\bibinfo  {journal}
  {Phys. Rev. Lett.}\ }\textbf {\bibinfo {volume} {96}},\ \bibinfo {pages}
  {110405} (\bibinfo {year} {2006})},\ \Eprint
  {http://arxiv.org/abs/cond-mat/0510613} {arXiv:cond-mat/0510613} \BibitemShut
  {NoStop}%
\bibitem [{\citenamefont {Kitaev}\ and\ \citenamefont
  {Preskill}(2006)}]{PhysRevLett.96.110404}%
  \BibitemOpen
  \bibfield  {author} {\bibinfo {author} {\bibfnamefont {Alexei}\ \bibnamefont
  {Kitaev}}\ and\ \bibinfo {author} {\bibfnamefont {John}\ \bibnamefont
  {Preskill}},\ }\bibfield  {title} {\enquote {\bibinfo {title} {Topological
  entanglement entropy},}\ }\href {\doibase 10.1103/PhysRevLett.96.110404}
  {\bibfield  {journal} {\bibinfo  {journal} {Phys. Rev. Lett.}\ }\textbf
  {\bibinfo {volume} {96}},\ \bibinfo {pages} {110404} (\bibinfo {year}
  {2006})},\ \Eprint {http://arxiv.org/abs/hep-th/0510092}
  {arXiv:hep-th/0510092} \BibitemShut {NoStop}%
\bibitem [{\citenamefont {Kitaev}(2003)}]{MR1951039}%
  \BibitemOpen
  \bibfield  {author} {\bibinfo {author} {\bibfnamefont {Alexei}\ \bibnamefont
  {Kitaev}},\ }\bibfield  {title} {\enquote {\bibinfo {title} {Fault-tolerant
  quantum computation by anyons},}\ }\href {\doibase
  10.1016/S0003-4916(02)00018-0} {\bibfield  {journal} {\bibinfo  {journal}
  {Ann. Phys.}\ }\textbf {\bibinfo {volume} {303}},\ \bibinfo {pages} {2--30}
  (\bibinfo {year} {2003})},\ \Eprint {http://arxiv.org/abs/quant-ph/9707021}
  {arXiv:quant-ph/9707021} \BibitemShut {NoStop}%
\bibitem [{\citenamefont {Levin}\ and\ \citenamefont {Wen}(2005)}]{LeWe:2005}%
  \BibitemOpen
  \bibfield  {author} {\bibinfo {author} {\bibfnamefont {Michael~A.}\
  \bibnamefont {Levin}}\ and\ \bibinfo {author} {\bibfnamefont {Xiao-Gang}\
  \bibnamefont {Wen}},\ }\bibfield  {title} {\enquote {\bibinfo {title}
  {String-net condensation: {A} physical mechanism for topological phases},}\
  }\href {\doibase 10.1103/PhysRevB.71.045110} {\bibfield  {journal} {\bibinfo
  {journal} {Phys. Rev. B}\ }\textbf {\bibinfo {volume} {71}},\ \bibinfo
  {pages} {045110} (\bibinfo {year} {2005})},\ \Eprint
  {http://arxiv.org/abs/cond-mat/0404617} {arXiv:cond-mat/0404617} \BibitemShut
  {NoStop}%
\bibitem [{\citenamefont {Preskill}(1999)}]{Preskill:1999}%
  \BibitemOpen
  \bibfield  {author} {\bibinfo {author} {\bibfnamefont {John}\ \bibnamefont
  {Preskill}},\ }\href@noop {} {\enquote {\bibinfo {title} {Lecture notes for
  {P}hysics 219: {Q}uantum computation},}\ } (\bibinfo {year} {1999}),\
  \bibinfo {note}
  {\url{http://www.theory.caltech.edu/people/preskill/ph219/}}\BibitemShut
  {NoStop}%
\bibitem [{\citenamefont {Rowell}\ and\ \citenamefont
  {Wang}(2016)}]{RowelWang:2015}%
  \BibitemOpen
  \bibfield  {author} {\bibinfo {author} {\bibfnamefont {Eric~C.}\ \bibnamefont
  {Rowell}}\ and\ \bibinfo {author} {\bibfnamefont {Zhenghan}\ \bibnamefont
  {Wang}},\ }\bibfield  {title} {\enquote {\bibinfo {title} {Degeneracy and
  non-abelian statistics},}\ }\href {\doibase 10.1103/PhysRevA.93.030102}
  {\bibfield  {journal} {\bibinfo  {journal} {Phys. Rev. A}\ }\textbf {\bibinfo
  {volume} {93}},\ \bibinfo {pages} {030102} (\bibinfo {year} {2016})},\
  \Eprint {http://arxiv.org/abs/1508.04793} {arXiv:1508.04793} \BibitemShut
  {NoStop}%
\bibitem [{\citenamefont {Naaijkens}(2013)}]{klindex}%
  \BibitemOpen
  \bibfield  {author} {\bibinfo {author} {\bibfnamefont {Pieter}\ \bibnamefont
  {Naaijkens}},\ }\bibfield  {title} {\enquote {\bibinfo {title}
  {{K}osaki-{L}ongo index and classifcation of charges in 2d quantum spin
  models},}\ }\href {\doibase 10.1063/1.4818272} {\bibfield  {journal}
  {\bibinfo  {journal} {J. Math. Phys.}\ }\textbf {\bibinfo {volume} {54}},\
  \bibinfo {pages} {081901} (\bibinfo {year} {2013})},\ \Eprint
  {http://arxiv.org/abs/1303.4420} {arXiv:1303.4420} \BibitemShut {NoStop}%
\bibitem [{\citenamefont {Kato}\ \emph {et~al.}(2016)\citenamefont {Kato},
  \citenamefont {Furrer},\ and\ \citenamefont {Murao}}]{KaFuMu:2015}%
  \BibitemOpen
  \bibfield  {author} {\bibinfo {author} {\bibfnamefont {Kohtaro}\ \bibnamefont
  {Kato}}, \bibinfo {author} {\bibfnamefont {Fabian}\ \bibnamefont {Furrer}}, \
  and\ \bibinfo {author} {\bibfnamefont {Mio}\ \bibnamefont {Murao}},\
  }\bibfield  {title} {\enquote {\bibinfo {title} {Information-theoretical
  analysis of topological entanglement entropy and multipartite
  correlations},}\ }\href {\doibase 10.1103/PhysRevA.93.022317} {\bibfield
  {journal} {\bibinfo  {journal} {Phys. Rev. A}\ }\textbf {\bibinfo {volume}
  {93}},\ \bibinfo {pages} {022317} (\bibinfo {year} {2016})},\ \Eprint
  {http://arxiv.org/abs/1505.01917} {arXiv:1505.01917} \BibitemShut {NoStop}%
\bibitem [{\citenamefont {Kitaev}(2006)}]{MR2200691}%
  \BibitemOpen
  \bibfield  {author} {\bibinfo {author} {\bibfnamefont {Alexei}\ \bibnamefont
  {Kitaev}},\ }\bibfield  {title} {\enquote {\bibinfo {title} {Anyons in an
  exactly solved model and beyond},}\ }\href {\doibase
  10.1016/j.aop.2005.10.005} {\bibfield  {journal} {\bibinfo  {journal} {Ann.
  Phys.}\ }\textbf {\bibinfo {volume} {321}},\ \bibinfo {pages} {2--111}
  (\bibinfo {year} {2006})},\ \Eprint {http://arxiv.org/abs/cond-mat/0506438}
  {arXiv:cond-mat/0506438} \BibitemShut {NoStop}%
\bibitem [{\citenamefont {Wang}(2010)}]{Wang}%
  \BibitemOpen
  \bibfield  {author} {\bibinfo {author} {\bibfnamefont {Zhenghan}\
  \bibnamefont {Wang}},\ }\href {\doibase 10.1090/cbms/112} {\emph {\bibinfo
  {title} {Topological Quantum Computation}}},\ \bibinfo {series} {CBMS
  Regional Conference Series in Mathematics}, Vol.\ \bibinfo {volume} {112}\
  (\bibinfo  {publisher} {CBMS, Washington, DC},\ \bibinfo {year} {2010})\ pp.\
  \bibinfo {pages} {xiv+115}\BibitemShut {NoStop}%
\bibitem [{\citenamefont {Wen}\ and\ \citenamefont
  {Niu}(1990)}]{PhysRevB.41.9377}%
  \BibitemOpen
  \bibfield  {author} {\bibinfo {author} {\bibfnamefont {Xiao-Gang}\
  \bibnamefont {Wen}}\ and\ \bibinfo {author} {\bibfnamefont {Qian}\
  \bibnamefont {Niu}},\ }\bibfield  {title} {\enquote {\bibinfo {title}
  {Ground-state degeneracy of the fractional quantum {H}all states in the
  presence of a random potential and on high-genus {R}iemann surfaces},}\
  }\href {\doibase 10.1103/PhysRevB.41.9377} {\bibfield  {journal} {\bibinfo
  {journal} {Phys. Rev. B}\ }\textbf {\bibinfo {volume} {41}},\ \bibinfo
  {pages} {9377--9396} (\bibinfo {year} {1990})}\BibitemShut {NoStop}%
\bibitem [{\citenamefont {Hu}\ \emph {et~al.}(2012)\citenamefont {Hu},
  \citenamefont {Stirling},\ and\ \citenamefont {Wu}}]{PhysRevB.85.075107}%
  \BibitemOpen
  \bibfield  {author} {\bibinfo {author} {\bibfnamefont {Yuting}\ \bibnamefont
  {Hu}}, \bibinfo {author} {\bibfnamefont {Spencer~D.}\ \bibnamefont
  {Stirling}}, \ and\ \bibinfo {author} {\bibfnamefont {Yong-Shi}\ \bibnamefont
  {Wu}},\ }\bibfield  {title} {\enquote {\bibinfo {title} {Ground-state
  degeneracy in the levin-wen model for topological phases},}\ }\href {\doibase
  10.1103/PhysRevB.85.075107} {\bibfield  {journal} {\bibinfo  {journal} {Phys.
  Rev. B}\ }\textbf {\bibinfo {volume} {85}},\ \bibinfo {pages} {075107}
  (\bibinfo {year} {2012})},\ \Eprint {http://arxiv.org/abs/1105.5771}
  {arXiv:1105.5771} \BibitemShut {NoStop}%
\bibitem [{\citenamefont {M{\"u}ger}(2003)}]{MR1966524}%
  \BibitemOpen
  \bibfield  {author} {\bibinfo {author} {\bibfnamefont {Michael}\ \bibnamefont
  {M{\"u}ger}},\ }\bibfield  {title} {\enquote {\bibinfo {title} {From
  subfactors to categories and topology. {I}. {F}robenius algebras in and
  {M}orita equivalence of tensor categories},}\ }\href {\doibase
  10.1016/S0022-4049(02)00247-5} {\bibfield  {journal} {\bibinfo  {journal} {J.
  Pure Appl. Algebra}\ }\textbf {\bibinfo {volume} {180}},\ \bibinfo {pages}
  {81--157} (\bibinfo {year} {2003})},\ \Eprint
  {http://arxiv.org/abs/math/0111204} {arXiv:math/0111204} \BibitemShut
  {NoStop}%
\bibitem [{\citenamefont {Isakov}\ \emph {et~al.}(2011)\citenamefont {Isakov},
  \citenamefont {Hastings},\ and\ \citenamefont {Melko}}]{Isakov:2011fk}%
  \BibitemOpen
  \bibfield  {author} {\bibinfo {author} {\bibfnamefont {Sergei~V.}\
  \bibnamefont {Isakov}}, \bibinfo {author} {\bibfnamefont {Matthew~B.}\
  \bibnamefont {Hastings}}, \ and\ \bibinfo {author} {\bibfnamefont {Roger~G.}\
  \bibnamefont {Melko}},\ }\bibfield  {title} {\enquote {\bibinfo {title}
  {Topological entanglement entropy of a {B}ose-{H}ubbard spin liquid},}\
  }\href {\doibase 10.1038/nphys2036} {\bibfield  {journal} {\bibinfo
  {journal} {Nat. Phys.}\ }\textbf {\bibinfo {volume} {7}},\ \bibinfo {pages}
  {772--775} (\bibinfo {year} {2011})},\ \Eprint
  {http://arxiv.org/abs/1102.1721} {arXiv:1102.1721} \BibitemShut {NoStop}%
\bibitem [{\citenamefont {Castelnovo}\ and\ \citenamefont
  {Chamon}(2007)}]{PhysRevB.76.184442}%
  \BibitemOpen
  \bibfield  {author} {\bibinfo {author} {\bibfnamefont {Claudio}\ \bibnamefont
  {Castelnovo}}\ and\ \bibinfo {author} {\bibfnamefont {Claudio}\ \bibnamefont
  {Chamon}},\ }\bibfield  {title} {\enquote {\bibinfo {title} {Entanglement and
  topological entropy of the toric code at finite temperature},}\ }\href
  {\doibase 10.1103/PhysRevB.76.184442} {\bibfield  {journal} {\bibinfo
  {journal} {Phys. Rev. B}\ }\textbf {\bibinfo {volume} {76}},\ \bibinfo
  {pages} {184442} (\bibinfo {year} {2007})},\ \Eprint
  {http://arxiv.org/abs/0704.3616} {arXiv:0704.3616} \BibitemShut {NoStop}%
\bibitem [{\citenamefont {Brown}\ \emph {et~al.}(2013)\citenamefont {Brown},
  \citenamefont {Bartlett}, \citenamefont {Doherty},\ and\ \citenamefont
  {Barrett}}]{2013arXiv1303.4455B}%
  \BibitemOpen
  \bibfield  {author} {\bibinfo {author} {\bibfnamefont {Benjamin~J.}\
  \bibnamefont {Brown}}, \bibinfo {author} {\bibfnamefont {Stephen~D.}\
  \bibnamefont {Bartlett}}, \bibinfo {author} {\bibfnamefont {Andrew~C.}\
  \bibnamefont {Doherty}}, \ and\ \bibinfo {author} {\bibfnamefont {Sean~D.}\
  \bibnamefont {Barrett}},\ }\bibfield  {title} {\enquote {\bibinfo {title}
  {Topological entanglement entropy with a twist},}\ }\href {\doibase
  10.1103/PhysRevLett.111.220402} {\bibfield  {journal} {\bibinfo  {journal}
  {Phys. Rev. Lett.}\ }\textbf {\bibinfo {volume} {111}},\ \bibinfo {pages}
  {220402} (\bibinfo {year} {2013})},\ \Eprint {http://arxiv.org/abs/1303.4455}
  {arXiv:1303.4455} \BibitemShut {NoStop}%
\bibitem [{\citenamefont {Grover}\ \emph {et~al.}(2011)\citenamefont {Grover},
  \citenamefont {Turner},\ and\ \citenamefont
  {Vishwanath}}]{PhysRevB.84.195120}%
  \BibitemOpen
  \bibfield  {author} {\bibinfo {author} {\bibfnamefont {Tarun}\ \bibnamefont
  {Grover}}, \bibinfo {author} {\bibfnamefont {Ari~M.}\ \bibnamefont {Turner}},
  \ and\ \bibinfo {author} {\bibfnamefont {Ashvin}\ \bibnamefont
  {Vishwanath}},\ }\bibfield  {title} {\enquote {\bibinfo {title} {Entanglement
  entropy of gapped phases and topological order in three dimensions},}\ }\href
  {\doibase 10.1103/PhysRevB.84.195120} {\bibfield  {journal} {\bibinfo
  {journal} {Phys. Rev. B}\ }\textbf {\bibinfo {volume} {84}},\ \bibinfo
  {pages} {195120} (\bibinfo {year} {2011})},\ \Eprint
  {http://arxiv.org/abs/1108.4038} {arXiv:1108.4038} \BibitemShut {NoStop}%
\bibitem [{\citenamefont {Huang}\ \emph {et~al.}(2013)\citenamefont {Huang},
  \citenamefont {Chen},\ and\ \citenamefont {Lin}}]{2013arXiv1303.4190H}%
  \BibitemOpen
  \bibfield  {author} {\bibinfo {author} {\bibfnamefont {Ching-Yu}\
  \bibnamefont {Huang}}, \bibinfo {author} {\bibfnamefont {Xie}\ \bibnamefont
  {Chen}}, \ and\ \bibinfo {author} {\bibfnamefont {Feng-Li}\ \bibnamefont
  {Lin}},\ }\bibfield  {title} {\enquote {\bibinfo {title} {Symmetry-protected
  quantum state renormalization},}\ }\href {\doibase
  10.1103/PhysRevB.88.205124} {\bibfield  {journal} {\bibinfo  {journal} {Phys.
  Rev. B}\ }\textbf {\bibinfo {volume} {88}},\ \bibinfo {pages} {205124}
  (\bibinfo {year} {2013})},\ \Eprint {http://arxiv.org/abs/1303.4190}
  {arXiv:1303.4190} \BibitemShut {NoStop}%
\bibitem [{\citenamefont {Kim}(2012)}]{PhysRevB.86.245116}%
  \BibitemOpen
  \bibfield  {author} {\bibinfo {author} {\bibfnamefont {Isaac~H.}\
  \bibnamefont {Kim}},\ }\bibfield  {title} {\enquote {\bibinfo {title}
  {Perturbative analysis of topological entanglement entropy from conditional
  independence},}\ }\href {\doibase 10.1103/PhysRevB.86.245116} {\bibfield
  {journal} {\bibinfo  {journal} {Phys. Rev. B}\ }\textbf {\bibinfo {volume}
  {86}},\ \bibinfo {pages} {245116} (\bibinfo {year} {2012})},\ \Eprint
  {http://arxiv.org/abs/1210.2360} {arXiv:1210.2360} \BibitemShut {NoStop}%
\bibitem [{\citenamefont {Zou}\ and\ \citenamefont
  {Haah}(2016)}]{ZouHaah:2016}%
  \BibitemOpen
  \bibfield  {author} {\bibinfo {author} {\bibfnamefont {Liujun}\ \bibnamefont
  {Zou}}\ and\ \bibinfo {author} {\bibfnamefont {Jeongwang}\ \bibnamefont
  {Haah}},\ }\href@noop {} {\enquote {\bibinfo {title} {Spurious long-range
  entanglement and replica correlation length},}\ } (\bibinfo {year} {2016}),\
  \bibinfo {note} {preprint},\ \Eprint {http://arxiv.org/abs/1604.06101}
  {arXiv:1604.06101} \BibitemShut {NoStop}%
\bibitem [{\citenamefont {Shamir}(1979)}]{Shamir:1979}%
  \BibitemOpen
  \bibfield  {author} {\bibinfo {author} {\bibfnamefont {Adi}\ \bibnamefont
  {Shamir}},\ }\bibfield  {title} {\enquote {\bibinfo {title} {How to share a
  secret},}\ }\href {\doibase 10.1145/359168.359176} {\bibfield  {journal}
  {\bibinfo  {journal} {Commun. ACM}\ }\textbf {\bibinfo {volume} {22}},\
  \bibinfo {pages} {612--613} (\bibinfo {year} {1979})}\BibitemShut {NoStop}%
\bibitem [{\citenamefont {Blakley}(1979)}]{Blakley:1979}%
  \BibitemOpen
  \bibfield  {author} {\bibinfo {author} {\bibfnamefont {George~R.}\
  \bibnamefont {Blakley}},\ }\bibfield  {title} {\enquote {\bibinfo {title}
  {Safeguarding cryptographic keys},}\ }in\ \href {\doibase
  10.1109/AFIPS.1979.98} {\emph {\bibinfo {booktitle} {Proceedings of the 1979
  AFIPS National Computer Conference}}}\ (\bibinfo  {publisher} {IEEE Computer
  Society},\ \bibinfo {address} {Los Alamitos, CA, USA},\ \bibinfo {year}
  {1979})\ pp.\ \bibinfo {pages} {313--317}\BibitemShut {NoStop}%
\bibitem [{\citenamefont {Cleve}\ \emph {et~al.}(1999)\citenamefont {Cleve},
  \citenamefont {Gottesman},\ and\ \citenamefont {Lo}}]{ClevGotLo:1999}%
  \BibitemOpen
  \bibfield  {author} {\bibinfo {author} {\bibfnamefont {Richard}\ \bibnamefont
  {Cleve}}, \bibinfo {author} {\bibfnamefont {Daniel}\ \bibnamefont
  {Gottesman}}, \ and\ \bibinfo {author} {\bibfnamefont {Hoi-Kwong}\
  \bibnamefont {Lo}},\ }\bibfield  {title} {\enquote {\bibinfo {title} {How to
  share a quantum secret},}\ }\href {\doibase 10.1103/physrevlett.83.648}
  {\bibfield  {journal} {\bibinfo  {journal} {Phys. Rev. Lett.}\ }\textbf
  {\bibinfo {volume} {83}},\ \bibinfo {pages} {648--651} (\bibinfo {year}
  {1999})},\ \Eprint {http://arxiv.org/abs/quant-ph/9901025}
  {arXiv:quant-ph/9901025} \BibitemShut {NoStop}%
\bibitem [{\citenamefont {Gottesman}(2000)}]{Gottesman:2000}%
  \BibitemOpen
  \bibfield  {author} {\bibinfo {author} {\bibfnamefont {Daniel}\ \bibnamefont
  {Gottesman}},\ }\bibfield  {title} {\enquote {\bibinfo {title} {Theory of
  quantum secret sharing},}\ }\href {\doibase 10.1103/physreva.61.042311}
  {\bibfield  {journal} {\bibinfo  {journal} {Phys. Rev. A}\ }\textbf {\bibinfo
  {volume} {61}},\ \bibinfo {pages} {042311} (\bibinfo {year} {2000})},\
  \Eprint {http://arxiv.org/abs/quant-ph/9910067} {arXiv:quant-ph/9910067}
  \BibitemShut {NoStop}%
\bibitem [{\citenamefont {Kretschmann}\ \emph {et~al.}(2008)\citenamefont
  {Kretschmann}, \citenamefont {Kribs},\ and\ \citenamefont
  {Spekkens}}]{KretKriSpe:2008}%
  \BibitemOpen
  \bibfield  {author} {\bibinfo {author} {\bibfnamefont {Dennis}\ \bibnamefont
  {Kretschmann}}, \bibinfo {author} {\bibfnamefont {David~W.}\ \bibnamefont
  {Kribs}}, \ and\ \bibinfo {author} {\bibfnamefont {Robert~W.}\ \bibnamefont
  {Spekkens}},\ }\bibfield  {title} {\enquote {\bibinfo {title}
  {Complementarity of private and correctable subsystems in quantum
  cryptography and error correction},}\ }\href {\doibase
  10.1103/physreva.78.032330} {\bibfield  {journal} {\bibinfo  {journal} {Phys.
  Rev. A}\ }\textbf {\bibinfo {volume} {78}},\ \bibinfo {pages} {032330}
  (\bibinfo {year} {2008})},\ \Eprint {http://arxiv.org/abs/0711.3438}
  {arXiv:0711.3438} \BibitemShut {NoStop}%
\bibitem [{Note1()}]{Note1}%
  \BibitemOpen
  \bibinfo {note} {Perhaps the best analogy here is that Alice and Bob control
  separate nation-size states that cannot be completely encircled by an
  antagonistic spying nation.}\BibitemShut {Stop}%
\bibitem [{Note2()}]{Note2}%
  \BibitemOpen
  \bibinfo {note} {Note that, in the finite case, the set $\protect \mathcal
  {O}_E$ of Eve's observables \protect \emph {does not} obviously form an
  algebra.}\BibitemShut {Stop}%
\bibitem [{\citenamefont {Knill}\ and\ \citenamefont
  {Laflamme}(1997)}]{KnillLafla:1995}%
  \BibitemOpen
  \bibfield  {author} {\bibinfo {author} {\bibfnamefont {Emanuel}\ \bibnamefont
  {Knill}}\ and\ \bibinfo {author} {\bibfnamefont {Raymond}\ \bibnamefont
  {Laflamme}},\ }\bibfield  {title} {\enquote {\bibinfo {title} {Theory of
  quantum error-correcting codes},}\ }\href {\doibase 10.1103/physreva.55.900}
  {\bibfield  {journal} {\bibinfo  {journal} {Phys. Rev. A}\ }\textbf {\bibinfo
  {volume} {55}},\ \bibinfo {pages} {900--911} (\bibinfo {year} {1997})},\
  \Eprint {http://arxiv.org/abs/quant-ph/9604034} {arXiv:quant-ph/9604034}
  \BibitemShut {NoStop}%
\bibitem [{\citenamefont {Knill}\ \emph {et~al.}(2000)\citenamefont {Knill},
  \citenamefont {Laflamme},\ and\ \citenamefont {Viola}}]{KnilLafVio:2000}%
  \BibitemOpen
  \bibfield  {author} {\bibinfo {author} {\bibfnamefont {Emanuel}\ \bibnamefont
  {Knill}}, \bibinfo {author} {\bibfnamefont {Raymond}\ \bibnamefont
  {Laflamme}}, \ and\ \bibinfo {author} {\bibfnamefont {Lorenza}\ \bibnamefont
  {Viola}},\ }\bibfield  {title} {\enquote {\bibinfo {title} {Theory of quantum
  error correction for general noise},}\ }\href {\doibase
  10.1103/physrevlett.84.2525} {\bibfield  {journal} {\bibinfo  {journal}
  {Phys. Rev. Lett.}\ }\textbf {\bibinfo {volume} {84}},\ \bibinfo {pages}
  {2525--2528} (\bibinfo {year} {2000})},\ \Eprint
  {http://arxiv.org/abs/quant-ph/9604034} {arXiv:quant-ph/9604034} \BibitemShut
  {NoStop}%
\bibitem [{\citenamefont {Nielsen}\ \emph {et~al.}(1998)\citenamefont
  {Nielsen}, \citenamefont {Caves}, \citenamefont {Schumacher},\ and\
  \citenamefont {Barnum}}]{NieCavScBa:1998}%
  \BibitemOpen
  \bibfield  {author} {\bibinfo {author} {\bibfnamefont {Michael~A.}\
  \bibnamefont {Nielsen}}, \bibinfo {author} {\bibfnamefont {Carlton~M.}\
  \bibnamefont {Caves}}, \bibinfo {author} {\bibfnamefont {Benjamin}\
  \bibnamefont {Schumacher}}, \ and\ \bibinfo {author} {\bibfnamefont {Howard}\
  \bibnamefont {Barnum}},\ }\bibfield  {title} {\enquote {\bibinfo {title}
  {Information-theoretic approach to quantum error correction and reversible
  measurement},}\ }\href {\doibase 10.1098/rspa.1998.0160} {\bibfield
  {journal} {\bibinfo  {journal} {Proc. R. Soc. A}\ }\textbf {\bibinfo {volume}
  {454}},\ \bibinfo {pages} {277--304} (\bibinfo {year} {1998})},\ \Eprint
  {http://arxiv.org/abs/quant-ph/9706064} {arXiv:quant-ph/9706064} \BibitemShut
  {NoStop}%
\bibitem [{\citenamefont {Gottesman}(2010)}]{Gottesman:2009}%
  \BibitemOpen
  \bibfield  {author} {\bibinfo {author} {\bibfnamefont {Daniel}\ \bibnamefont
  {Gottesman}},\ }\bibfield  {title} {\enquote {\bibinfo {title} {An
  introduction to quantum error correction and fault-tolerant quantum
  computation},}\ }in\ \href {\doibase 10.1090/psapm/068/2762145} {\emph
  {\bibinfo {booktitle} {Quantum Information Science and Its Contributions to
  Mathematics}}},\ \bibinfo {series} {Proceedings of Symposia in Applied
  Mathematics}, Vol.~\bibinfo {volume} {68},\ \bibinfo {editor} {edited by\
  \bibinfo {editor} {\bibfnamefont {Samuel J.~Lomonaco}\ \bibnamefont {Jr}}}\
  (\bibinfo  {publisher} {American Mathematical Society ({AMS})},\ \bibinfo
  {year} {2010})\ pp.\ \bibinfo {pages} {13--58},\ \Eprint
  {http://arxiv.org/abs/0904.2557} {arXiv:0904.2557} \BibitemShut {NoStop}%
\bibitem [{\citenamefont {Haah}(2014)}]{Haah:2014}%
  \BibitemOpen
  \bibfield  {author} {\bibinfo {author} {\bibfnamefont {Jeongwan}\
  \bibnamefont {Haah}},\ }\bibfield  {title} {\enquote {\bibinfo {title} {An
  invariant of topologically ordered states under local unitary
  transformations},}\ }\href {\doibase 10.1007/s00220-016-2594-y} {\bibfield
  {journal} {\bibinfo  {journal} {Commun. Math. Phys.}\ }\textbf {\bibinfo
  {volume} {342}},\ \bibinfo {pages} {771--801} (\bibinfo {year} {2014})},\
  \Eprint {http://arxiv.org/abs/1407.2926} {arXiv:1407.2926} \BibitemShut
  {NoStop}%
\bibitem [{\citenamefont {{Pfeifer}}(2014)}]{2013arXiv1310.0373P}%
  \BibitemOpen
  \bibfield  {author} {\bibinfo {author} {\bibfnamefont {Robert~N.~C.}\
  \bibnamefont {{Pfeifer}}},\ }\bibfield  {title} {\enquote {\bibinfo {title}
  {{Measures of entanglement in non-Abelian anyonic systems}},}\ }\href
  {\doibase 10.1103/physrevb.89.035105} {\bibfield  {journal} {\bibinfo
  {journal} {Phys. Rev. B}\ }\textbf {\bibinfo {volume} {89}},\ \bibinfo
  {pages} {035105} (\bibinfo {year} {2014})},\ \Eprint
  {http://arxiv.org/abs/1310.0373} {arXiv:1310.0373} \BibitemShut {NoStop}%
\bibitem [{\citenamefont {{Kato}}\ \emph {et~al.}(2014)\citenamefont {{Kato}},
  \citenamefont {{Furrer}},\ and\ \citenamefont
  {{Murao}}}]{2013arXiv1310.4140K}%
  \BibitemOpen
  \bibfield  {author} {\bibinfo {author} {\bibfnamefont {Kohtaro}\ \bibnamefont
  {{Kato}}}, \bibinfo {author} {\bibfnamefont {Fabian}\ \bibnamefont
  {{Furrer}}}, \ and\ \bibinfo {author} {\bibfnamefont {Mio}\ \bibnamefont
  {{Murao}}},\ }\bibfield  {title} {\enquote {\bibinfo {title}
  {{Information-theoretical formulation of anyonic entanglement}},}\ }\href
  {\doibase 10.1103/physreva.90.062325} {\bibfield  {journal} {\bibinfo
  {journal} {Phys. Rev. A}\ }\textbf {\bibinfo {volume} {90}},\ \bibinfo
  {pages} {062325} (\bibinfo {year} {2014})},\ \Eprint
  {http://arxiv.org/abs/1310.4140} {arXiv:1310.4140} \BibitemShut {NoStop}%
\bibitem [{\citenamefont {Trebst}\ \emph {et~al.}(2009)\citenamefont {Trebst},
  \citenamefont {Troyer}, \citenamefont {Wang},\ and\ \citenamefont
  {Ludwig}}]{TreTroWaLu:2009}%
  \BibitemOpen
  \bibfield  {author} {\bibinfo {author} {\bibfnamefont {Simon}\ \bibnamefont
  {Trebst}}, \bibinfo {author} {\bibfnamefont {Matthias}\ \bibnamefont
  {Troyer}}, \bibinfo {author} {\bibfnamefont {Zhenghan}\ \bibnamefont {Wang}},
  \ and\ \bibinfo {author} {\bibfnamefont {Andreas W.~W.}\ \bibnamefont
  {Ludwig}},\ }\bibfield  {title} {\enquote {\bibinfo {title} {A short
  introduction to fibonacci anyon models},}\ }\href {\doibase
  10.1143/PTPS.176.384} {\bibfield  {journal} {\bibinfo  {journal} {Prog.
  Theor. Phys. Supplement}\ }\textbf {\bibinfo {volume} {176}},\ \bibinfo
  {pages} {384} (\bibinfo {year} {2009})},\ \Eprint
  {http://arxiv.org/abs/0902.3275} {arXiv:0902.3275} \BibitemShut {NoStop}%
\bibitem [{\citenamefont {Bratteli}\ and\ \citenamefont
  {Robinson}(1987)}]{MR887100}%
  \BibitemOpen
  \bibfield  {author} {\bibinfo {author} {\bibfnamefont {Ola}\ \bibnamefont
  {Bratteli}}\ and\ \bibinfo {author} {\bibfnamefont {Derek~W.}\ \bibnamefont
  {Robinson}},\ }\href {\doibase 10.1007/978-3-662-02520-8} {\emph {\bibinfo
  {title} {Operator algebras and quantum statistical mechanics. 1}}},\ \bibinfo
  {edition} {2nd}\ ed.,\ Texts and Monographs in Physics\ (\bibinfo
  {publisher} {Springer-Verlag},\ \bibinfo {address} {New York},\ \bibinfo
  {year} {1987})\ pp.\ \bibinfo {pages} {xiv+505}\BibitemShut {NoStop}%
\bibitem [{\citenamefont {Bratteli}\ and\ \citenamefont
  {Robinson}(1997)}]{MR1441540}%
  \BibitemOpen
  \bibfield  {author} {\bibinfo {author} {\bibfnamefont {Ola}\ \bibnamefont
  {Bratteli}}\ and\ \bibinfo {author} {\bibfnamefont {Derek~W.}\ \bibnamefont
  {Robinson}},\ }\href {\doibase 10.1007/978-3-662-03444-6} {\emph {\bibinfo
  {title} {Operator algebras and quantum statistical mechanics. 2}}},\ \bibinfo
  {edition} {2nd}\ ed.,\ Texts and Monographs in Physics\ (\bibinfo
  {publisher} {Springer-Verlag},\ \bibinfo {address} {Berlin},\ \bibinfo {year}
  {1997})\ pp.\ \bibinfo {pages} {xiv+519}\BibitemShut {NoStop}%
\bibitem [{\citenamefont {Naaijkens}(2011)}]{toricendo}%
  \BibitemOpen
  \bibfield  {author} {\bibinfo {author} {\bibfnamefont {Pieter}\ \bibnamefont
  {Naaijkens}},\ }\bibfield  {title} {\enquote {\bibinfo {title} {Localized
  endomorphisms in {K}itaev's toric code on the plane},}\ }\href {\doibase
  10.1142/s0129055x1100431x} {\bibfield  {journal} {\bibinfo  {journal} {Rev.
  Math. Phys.}\ }\textbf {\bibinfo {volume} {23}},\ \bibinfo {pages} {347--373}
  (\bibinfo {year} {2011})},\ \Eprint {http://arxiv.org/abs/1012.3857}
  {arXiv:1012.3857} \BibitemShut {NoStop}%
\bibitem [{\citenamefont {Jones}(1983)}]{MR696688}%
  \BibitemOpen
  \bibfield  {author} {\bibinfo {author} {\bibfnamefont {Vaughan F.~R.}\
  \bibnamefont {Jones}},\ }\bibfield  {title} {\enquote {\bibinfo {title}
  {Index for subfactors},}\ }\href {\doibase 10.1007/BF01389127} {\bibfield
  {journal} {\bibinfo  {journal} {Invent. math.}\ }\textbf {\bibinfo {volume}
  {72}},\ \bibinfo {pages} {1--25} (\bibinfo {year} {1983})}\BibitemShut
  {NoStop}%
\bibitem [{\citenamefont {Kosaki}(1986)}]{MR829381}%
  \BibitemOpen
  \bibfield  {author} {\bibinfo {author} {\bibfnamefont {Hideki}\ \bibnamefont
  {Kosaki}},\ }\bibfield  {title} {\enquote {\bibinfo {title} {Extension of
  {J}ones' theory on index to arbitrary factors},}\ }\href {\doibase
  10.1016/0022-1236(86)90085-6} {\bibfield  {journal} {\bibinfo  {journal} {J.
  Funct. Anal.}\ }\textbf {\bibinfo {volume} {66}},\ \bibinfo {pages}
  {123--140} (\bibinfo {year} {1986})}\BibitemShut {NoStop}%
\bibitem [{\citenamefont {Longo}(1989)}]{MR1027496}%
  \BibitemOpen
  \bibfield  {author} {\bibinfo {author} {\bibfnamefont {Roberto}\ \bibnamefont
  {Longo}},\ }\bibfield  {title} {\enquote {\bibinfo {title} {Index of
  subfactors and statistics of quantum fields. {I}},}\ }\href {\doibase
  10.1007/bf02125124} {\bibfield  {journal} {\bibinfo  {journal} {Commun. Math.
  Phys.}\ }\textbf {\bibinfo {volume} {126}},\ \bibinfo {pages} {217--247}
  (\bibinfo {year} {1989})}\BibitemShut {NoStop}%
\bibitem [{\citenamefont {Naaijkens}(2012)}]{haagdtoric}%
  \BibitemOpen
  \bibfield  {author} {\bibinfo {author} {\bibfnamefont {Pieter}\ \bibnamefont
  {Naaijkens}},\ }\bibfield  {title} {\enquote {\bibinfo {title} {Haag duality
  and the distal split property for cones in the toric code},}\ }\href
  {\doibase 10.1007/s11005-012-0572-7} {\bibfield  {journal} {\bibinfo
  {journal} {Lett. Math. Phys.}\ }\textbf {\bibinfo {volume} {101}},\ \bibinfo
  {pages} {341--354} (\bibinfo {year} {2012})},\ \Eprint
  {http://arxiv.org/abs/1106.4171} {arXiv:1106.4171} \BibitemShut {NoStop}%
\bibitem [{\citenamefont {Fiedler}\ and\ \citenamefont
  {Naaijkens}(2015)}]{haagdabelian}%
  \BibitemOpen
  \bibfield  {author} {\bibinfo {author} {\bibfnamefont {Leander}\ \bibnamefont
  {Fiedler}}\ and\ \bibinfo {author} {\bibfnamefont {Pieter}\ \bibnamefont
  {Naaijkens}},\ }\bibfield  {title} {\enquote {\bibinfo {title} {Haag duality
  for {K}itaev’s quantum double model for abelian groups},}\ }\href {\doibase
  10.1142/S0129055X1550021X} {\bibfield  {journal} {\bibinfo  {journal} {Rev.
  Math. Phys.}\ }\textbf {\bibinfo {volume} {27}},\ \bibinfo {pages} {1550021}
  (\bibinfo {year} {2015})},\ \Eprint {http://arxiv.org/abs/1406.1084}
  {arXiv:1406.1084} \BibitemShut {NoStop}%
\bibitem [{\citenamefont {Doplicher}\ and\ \citenamefont
  {Longo}(1984)}]{DoLo:1984}%
  \BibitemOpen
  \bibfield  {author} {\bibinfo {author} {\bibfnamefont {Sergio}\ \bibnamefont
  {Doplicher}}\ and\ \bibinfo {author} {\bibfnamefont {Roberto}\ \bibnamefont
  {Longo}},\ }\bibfield  {title} {\enquote {\bibinfo {title} {Standard and
  split inclusions of von {N}eumann algebras},}\ }\href {\doibase
  10.1007/BF01388641} {\bibfield  {journal} {\bibinfo  {journal} {Invent.
  Math.}\ }\textbf {\bibinfo {volume} {75}},\ \bibinfo {pages} {493--536}
  (\bibinfo {year} {1984})}\BibitemShut {NoStop}%
\bibitem [{\citenamefont {Longo}(1984)}]{Longo:1984}%
  \BibitemOpen
  \bibfield  {author} {\bibinfo {author} {\bibfnamefont {R.}~\bibnamefont
  {Longo}},\ }\bibfield  {title} {\enquote {\bibinfo {title} {Solution of the
  factorial {S}tone-{W}eierstrass conjecture. an application of the theory of
  standard split {W}$^*$-inclusions},}\ }\href {\doibase 10.1007/bf01388497}
  {\bibfield  {journal} {\bibinfo  {journal} {Invent. Math.}\ }\textbf
  {\bibinfo {volume} {76}},\ \bibinfo {pages} {145--155} (\bibinfo {year}
  {1984})}\BibitemShut {NoStop}%
\bibitem [{\citenamefont {Werner}(1987)}]{Werner:1987:1}%
  \BibitemOpen
  \bibfield  {author} {\bibinfo {author} {\bibfnamefont {Reinhard}\
  \bibnamefont {Werner}},\ }\bibfield  {title} {\enquote {\bibinfo {title}
  {Local preparability of states and the split property in quantum field
  theory},}\ }\href {\doibase 10.1007/BF00401161} {\bibfield  {journal}
  {\bibinfo  {journal} {Lett. Math. Phys.}\ }\textbf {\bibinfo {volume} {13}},\
  \bibinfo {pages} {325--329} (\bibinfo {year} {1987})}\BibitemShut {NoStop}%
\bibitem [{\citenamefont {Summers}(1996)}]{Summers:1996}%
  \BibitemOpen
  \bibfield  {author} {\bibinfo {author} {\bibfnamefont {Stephen~J.}\
  \bibnamefont {Summers}},\ }\href@noop {} {\enquote {\bibinfo {title} {Bell's
  inequalities and algebraic structure},}\ } (\bibinfo {year} {1996}),\
  \bibinfo {note} {preprint},\ \Eprint {http://arxiv.org/abs/funct-an/9701003}
  {arXiv:funct-an/9701003} \BibitemShut {NoStop}%
\bibitem [{\citenamefont {Haag}(1996)}]{MR1405610}%
  \BibitemOpen
  \bibfield  {author} {\bibinfo {author} {\bibfnamefont {Rudolf}\ \bibnamefont
  {Haag}},\ }\href {\doibase 10.1007/978-3-642-61458-3} {\emph {\bibinfo
  {title} {Local quantum physics: Fields, particles, algebras}}},\ \bibinfo
  {edition} {2nd}\ ed.\ (\bibinfo  {publisher} {Springer-Verlag},\ \bibinfo
  {address} {Berlin},\ \bibinfo {year} {1996})\ pp.\ \bibinfo {pages}
  {xvi+390}\BibitemShut {NoStop}%
\bibitem [{Note3()}]{Note3}%
  \BibitemOpen
  \bibinfo {note} {If we speak of ``local'' we always mean that the observable
  acts on finitely many particles on the lattice. Furthermore, in this context
  ``local'' additionally means that the observable is localised in one of the
  cones.}\BibitemShut {Stop}%
\bibitem [{Note4()}]{Note4}%
  \BibitemOpen
  \bibinfo {note} {Although we do not claim that this is the case here, this
  touches upon a more fundamental property of infinite dimensional systems.
  Recently Slofstra has found a counterexample to Tsirelson's problem~\cite
  {Slofstra:2016}, by showing that there are commuting operator models for
  two-party correlations that are not equivalent to a tensor product
  model.}\BibitemShut {Stop}%
\bibitem [{\citenamefont {Kribs}\ \emph {et~al.}(2006)\citenamefont {Kribs},
  \citenamefont {Laflamme}, \citenamefont {Poulin},\ and\ \citenamefont
  {Lesosky}}]{KriLafPoLe:2006}%
  \BibitemOpen
  \bibfield  {author} {\bibinfo {author} {\bibfnamefont {David~W.}\
  \bibnamefont {Kribs}}, \bibinfo {author} {\bibfnamefont {Raymond}\
  \bibnamefont {Laflamme}}, \bibinfo {author} {\bibfnamefont {David}\
  \bibnamefont {Poulin}}, \ and\ \bibinfo {author} {\bibfnamefont {Maia}\
  \bibnamefont {Lesosky}},\ }\bibfield  {title} {\enquote {\bibinfo {title}
  {Operator quantum error correction},}\ }\href
  {http://www.rintonpress.com/journals/qiconline.html} {\bibfield  {journal}
  {\bibinfo  {journal} {Quantum Inf. Comput.}\ }\textbf {\bibinfo {volume}
  {6}},\ \bibinfo {pages} {382--399} (\bibinfo {year} {2006})},\ \Eprint
  {http://arxiv.org/abs/quant-ph/0504189} {arXiv:quant-ph/0504189} \BibitemShut
  {NoStop}%
\bibitem [{\citenamefont {B{\'e}ny}\ \emph {et~al.}(2007)\citenamefont
  {B{\'e}ny}, \citenamefont {Kempf},\ and\ \citenamefont
  {Kribs}}]{BenyKemKri:2007}%
  \BibitemOpen
  \bibfield  {author} {\bibinfo {author} {\bibfnamefont {C{\'e}dric}\
  \bibnamefont {B{\'e}ny}}, \bibinfo {author} {\bibfnamefont {Achim}\
  \bibnamefont {Kempf}}, \ and\ \bibinfo {author} {\bibfnamefont {David~W.}\
  \bibnamefont {Kribs}},\ }\bibfield  {title} {\enquote {\bibinfo {title}
  {Quantum error correction of observables},}\ }\href {\doibase
  10.1103/PhysRevA.76.042303} {\bibfield  {journal} {\bibinfo  {journal} {Phys.
  Rev. A}\ }\textbf {\bibinfo {volume} {76}},\ \bibinfo {pages} {042303}
  (\bibinfo {year} {2007})},\ \Eprint {http://arxiv.org/abs/0705.1574}
  {arXiv:0705.1574} \BibitemShut {NoStop}%
\bibitem [{\citenamefont {Crann}\ \emph {et~al.}(2016)\citenamefont {Crann},
  \citenamefont {Kribs}, \citenamefont {Levene},\ and\ \citenamefont
  {Todorov}}]{Crann:2015wj}%
  \BibitemOpen
  \bibfield  {author} {\bibinfo {author} {\bibfnamefont {Jason}\ \bibnamefont
  {Crann}}, \bibinfo {author} {\bibfnamefont {David~W.}\ \bibnamefont {Kribs}},
  \bibinfo {author} {\bibfnamefont {Rupert~H.}\ \bibnamefont {Levene}}, \ and\
  \bibinfo {author} {\bibfnamefont {Ivan~G.}\ \bibnamefont {Todorov}},\
  }\bibfield  {title} {\enquote {\bibinfo {title} {Private algebras in quantum
  information and infinite-dimensional complementarity},}\ }\href {\doibase
  10.1063/1.4935399} {\bibfield  {journal} {\bibinfo  {journal} {J. Math.
  Phys.}\ }\textbf {\bibinfo {volume} {57}},\ \bibinfo {pages} {015208}
  (\bibinfo {year} {2016})},\ \bibinfo {note} {preprint},\ \Eprint
  {http://arxiv.org/abs/1510.06672} {arXiv:1510.06672} \BibitemShut {NoStop}%
\bibitem [{Note5()}]{Note5}%
  \BibitemOpen
  \bibinfo {note} {In \cite {haagdtoric} this is defined rigorously for the
  toric code, and in \cite {klindex} this is extended to more general
  models.}\BibitemShut {Stop}%
\bibitem [{\citenamefont {{Bombin}}\ and\ \citenamefont
  {{Martin-Delgado}}(2008)}]{BoMa:2008}%
  \BibitemOpen
  \bibfield  {author} {\bibinfo {author} {\bibfnamefont {Hector}\ \bibnamefont
  {{Bombin}}}\ and\ \bibinfo {author} {\bibfnamefont {Miguel~A.}\ \bibnamefont
  {{Martin-Delgado}}},\ }\bibfield  {title} {\enquote {\bibinfo {title} {Family
  of non-abelian {K}itaev models on a lattice: {T}opological condensation and
  confinement},}\ }\href {\doibase 10.1103/PhysRevB.78.115421} {\bibfield
  {journal} {\bibinfo  {journal} {Phys. Rev. B}\ }\textbf {\bibinfo {volume}
  {78}},\ \bibinfo {eid} {115421} (\bibinfo {year} {2008})},\ \Eprint
  {http://arxiv.org/abs/0712.0190} {arXiv:0712.0190} \BibitemShut {NoStop}%
\bibitem [{Note6()}]{Note6}%
  \BibitemOpen
  \bibinfo {note} {The choice of taking $\protect \mathcal {C}=\protect
  \{\Omega ,V_X\Omega ,V_Z\Omega ,V_XV_Z\Omega \protect \}$ from the spaces
  $\protect \mathcal {R}_{AB} V_i \Omega $ can be interpreted as choosing
  different implementations of secret sharing schemes with same information to
  be hidden and same access structure. Additionally Alice and Bob can act
  locally on the state within their cones without the corresponding other being
  able to notice.}\BibitemShut {Stop}%
\bibitem [{\citenamefont {Beigi}\ \emph {et~al.}(2011)\citenamefont {Beigi},
  \citenamefont {Shor},\ and\ \citenamefont {Whalen}}]{BeShoWha:2011}%
  \BibitemOpen
  \bibfield  {author} {\bibinfo {author} {\bibfnamefont {Salman}\ \bibnamefont
  {Beigi}}, \bibinfo {author} {\bibfnamefont {Peter~W.}\ \bibnamefont {Shor}},
  \ and\ \bibinfo {author} {\bibfnamefont {Daniel}\ \bibnamefont {Whalen}},\
  }\bibfield  {title} {\enquote {\bibinfo {title} {The quantum double model
  with boundary: Condensations and symmetries},}\ }\href {\doibase
  10.1007/s00220-011-1294-x} {\bibfield  {journal} {\bibinfo  {journal}
  {Commun. Math. Phys.}\ }\textbf {\bibinfo {volume} {306}},\ \bibinfo {pages}
  {663--694} (\bibinfo {year} {2011})},\ \Eprint
  {http://arxiv.org/abs/1006.5479} {arXiv:1006.5479} \BibitemShut {NoStop}%
\bibitem [{\citenamefont {Zhou}\ and\ \citenamefont
  {You}(2007)}]{ZhouYou:2007}%
  \BibitemOpen
  \bibfield  {author} {\bibinfo {author} {\bibfnamefont {Duanlu}\ \bibnamefont
  {Zhou}}\ and\ \bibinfo {author} {\bibfnamefont {Li}~\bibnamefont {You}},\
  }\href@noop {} {\enquote {\bibinfo {title} {Characterizing the complete
  hierarchy of correlations in an $n$-party system},}\ } (\bibinfo {year}
  {2007}),\ \bibinfo {note} {preprint},\ \Eprint
  {http://arxiv.org/abs/quant-ph/0701029} {arXiv:quant-ph/0701029} \BibitemShut
  {NoStop}%
\bibitem [{Note7()}]{Note7}%
  \BibitemOpen
  \bibinfo {note} {The optimal rate determines how many bits can at most be
  encoded in the state $\rho _{ABC}$ that it there exists a decoding channel
  that reliably can recover the information in asymptotic many uses of the
  scheme.}\BibitemShut {Stop}%
\bibitem [{\citenamefont {Verstraete}\ and\ \citenamefont
  {Cirac}(2003)}]{VerstraeteCirac}%
  \BibitemOpen
  \bibfield  {author} {\bibinfo {author} {\bibfnamefont {Frank}\ \bibnamefont
  {Verstraete}}\ and\ \bibinfo {author} {\bibfnamefont {Ignacio}\ \bibnamefont
  {Cirac}},\ }\bibfield  {title} {\enquote {\bibinfo {title} {Quantum
  nonlocality in the presence of superselection rules and data hiding
  protocols},}\ }\href {\doibase 10.1103/PhysRevLett.91.010404} {\bibfield
  {journal} {\bibinfo  {journal} {Phys. Rev. Lett.}\ }\textbf {\bibinfo
  {volume} {91}},\ \bibinfo {pages} {010404} (\bibinfo {year} {2003})},\
  \Eprint {http://arxiv.org/abs/quant-ph/0302039} {arXiv:quant-ph/0302039}
  \BibitemShut {NoStop}%
\bibitem [{\citenamefont {Kitaev}\ \emph {et~al.}(2004)\citenamefont {Kitaev},
  \citenamefont {Mayers},\ and\ \citenamefont {Preskill}}]{Kitaev:2004p5175}%
  \BibitemOpen
  \bibfield  {author} {\bibinfo {author} {\bibfnamefont {Alexei}\ \bibnamefont
  {Kitaev}}, \bibinfo {author} {\bibfnamefont {Dominic}\ \bibnamefont
  {Mayers}}, \ and\ \bibinfo {author} {\bibfnamefont {John}\ \bibnamefont
  {Preskill}},\ }\bibfield  {title} {\enquote {\bibinfo {title}
  {{Superselection rules and quantum protocols}},}\ }\href {\doibase
  10.1103/physreva.69.052326} {\bibfield  {journal} {\bibinfo  {journal} {Phys.
  Rev. A}\ }\textbf {\bibinfo {volume} {69}},\ \bibinfo {pages} {052326}
  (\bibinfo {year} {2004})},\ \Eprint {http://arxiv.org/abs/quant-ph/0310088}
  {arXiv:quant-ph/0310088} \BibitemShut {NoStop}%
\bibitem [{\citenamefont {Pimsner}\ and\ \citenamefont
  {Popa}(1986)}]{MR860811}%
  \BibitemOpen
  \bibfield  {author} {\bibinfo {author} {\bibfnamefont {Mihai}\ \bibnamefont
  {Pimsner}}\ and\ \bibinfo {author} {\bibfnamefont {Sorin}\ \bibnamefont
  {Popa}},\ }\bibfield  {title} {\enquote {\bibinfo {title} {Entropy and index
  for subfactors},}\ }\href
  {http://www.numdam.org/item?id=ASENS_1986_4_19_1_57_0} {\bibfield  {journal}
  {\bibinfo  {journal} {Ann. Sci. École Norm. Sup.}\ }\textbf {\bibinfo
  {volume} {19}},\ \bibinfo {pages} {57--106} (\bibinfo {year}
  {1986})}\BibitemShut {NoStop}%
\bibitem [{\citenamefont {Hiai}(1991)}]{MR1096438}%
  \BibitemOpen
  \bibfield  {author} {\bibinfo {author} {\bibfnamefont {Fumio}\ \bibnamefont
  {Hiai}},\ }\bibfield  {title} {\enquote {\bibinfo {title} {Minimum index for
  subfactors and entropy. {II}},}\ }\href {\doibase 10.2969/jmsj/04320347}
  {\bibfield  {journal} {\bibinfo  {journal} {J. Math. Soc. Japan}\ }\textbf
  {\bibinfo {volume} {43}},\ \bibinfo {pages} {347--379} (\bibinfo {year}
  {1991})}\BibitemShut {NoStop}%
\bibitem [{\citenamefont {Hiai}(1990)}]{MR1150623}%
  \BibitemOpen
  \bibfield  {author} {\bibinfo {author} {\bibfnamefont {Fumio}\ \bibnamefont
  {Hiai}},\ }\bibfield  {title} {\enquote {\bibinfo {title} {Minimum index for
  subfactors and entropy},}\ }\href {\doibase 10.2969/jmsj/04320347} {\bibfield
   {journal} {\bibinfo  {journal} {J. Operator Theory}\ }\textbf {\bibinfo
  {volume} {24}},\ \bibinfo {pages} {301--336} (\bibinfo {year}
  {1990})}\BibitemShut {NoStop}%
\bibitem [{Note8()}]{Note8}%
  \BibitemOpen
  \bibinfo {note} {Von Neumann algebras which have trivial centres (in other
  words, factors), can be classified in types I, II$_1$, II$_\infty $ and Type
  III\spacefactor \@m {}. Type I factors are precisely those that are
  isomorphic to $\protect \mathfrak {B}(\protect \mathcal {H})$ for some
  Hilbert space $\protect \mathcal {H}$. The type of the factors has important
  implications for the technical parts of the index theory, but the qualitative
  features are largely the same.}\BibitemShut {Stop}%
\bibitem [{\citenamefont {Araki}(1975/76)}]{MR0425631}%
  \BibitemOpen
  \bibfield  {author} {\bibinfo {author} {\bibfnamefont {Huzihiro}\
  \bibnamefont {Araki}},\ }\bibfield  {title} {\enquote {\bibinfo {title}
  {Relative entropy of states of von {N}eumann algebras},}\ }\href {\doibase
  10.2977/prims/1195191148} {\bibfield  {journal} {\bibinfo  {journal} {Publ.
  Res. Inst. Math. Sci.}\ }\textbf {\bibinfo {volume} {11}},\ \bibinfo {pages}
  {809--833} (\bibinfo {year} {1975/76})}\BibitemShut {NoStop}%
\bibitem [{\citenamefont {Araki}(1977/78)}]{MR0454656}%
  \BibitemOpen
  \bibfield  {author} {\bibinfo {author} {\bibfnamefont {Huzihiro}\
  \bibnamefont {Araki}},\ }\bibfield  {title} {\enquote {\bibinfo {title}
  {Relative entropy for states of von {N}eumann algebras. {II}},}\ }\href
  {\doibase 10.2977/prims/1195190105} {\bibfield  {journal} {\bibinfo
  {journal} {Publ. Res. Inst. Math. Sci.}\ }\textbf {\bibinfo {volume} {13}},\
  \bibinfo {pages} {173--192} (\bibinfo {year} {1977/78})}\BibitemShut
  {NoStop}%
\bibitem [{\citenamefont {Ohya}\ and\ \citenamefont {Petz}(1993)}]{MR1230389}%
  \BibitemOpen
  \bibfield  {author} {\bibinfo {author} {\bibfnamefont {Masanori}\
  \bibnamefont {Ohya}}\ and\ \bibinfo {author} {\bibfnamefont {D{\'e}nes}\
  \bibnamefont {Petz}},\ }\href {\doibase 10.1007/978-3-642-57997-4} {\emph
  {\bibinfo {title} {Quantum entropy and its use}}},\ Texts and Monographs in
  Physics\ (\bibinfo  {publisher} {Springer-Verlag},\ \bibinfo {address}
  {Berlin},\ \bibinfo {year} {1993})\ pp.\ \bibinfo {pages}
  {viii+335}\BibitemShut {NoStop}%
\bibitem [{\citenamefont {Holevo}(2012)}]{MR2986302}%
  \BibitemOpen
  \bibfield  {author} {\bibinfo {author} {\bibfnamefont {Alexander~S.}\
  \bibnamefont {Holevo}},\ }\href {\doibase 10.1515/9783110273403} {\emph
  {\bibinfo {title} {Quantum systems, channels, information}}},\ \bibinfo
  {series} {De Gruyter Studies in Mathematical Physics}, Vol.~\bibinfo {volume}
  {16}\ (\bibinfo  {publisher} {De Gruyter, Berlin},\ \bibinfo {year} {2012})\
  pp.\ \bibinfo {pages} {xiv+349}\BibitemShut {NoStop}%
\bibitem [{\citenamefont {Wilde}(2013)}]{MR3088659}%
  \BibitemOpen
  \bibfield  {author} {\bibinfo {author} {\bibfnamefont {Mark~M.}\ \bibnamefont
  {Wilde}},\ }\href {\doibase 10.1017/CBO9781139525343} {\emph {\bibinfo
  {title} {Quantum information theory}}}\ (\bibinfo  {publisher} {Cambridge
  University Press, Cambridge},\ \bibinfo {year} {2013})\ pp.\ \bibinfo {pages}
  {xvi+655},\ \Eprint {http://arxiv.org/abs/1106.1445} {arXiv:1106.1445}
  \BibitemShut {NoStop}%
\bibitem [{\citenamefont {Holevo}(2011)}]{Holevo.ChoiJamiolkowski}%
  \BibitemOpen
  \bibfield  {author} {\bibinfo {author} {\bibfnamefont {Alexander~S.}\
  \bibnamefont {Holevo}},\ }\bibfield  {title} {\enquote {\bibinfo {title}
  {Entropy gain and the {C}hoi-{J}amiolkowski correspondence for
  infinite-dimensional quantum evolutions},}\ }\href {\doibase
  10.1007/s11232-011-0010-5} {\bibfield  {journal} {\bibinfo  {journal} {Theor.
  Math. Phys.}\ }\textbf {\bibinfo {volume} {166}},\ \bibinfo {pages}
  {123--138} (\bibinfo {year} {2011})}\BibitemShut {NoStop}%
\bibitem [{\citenamefont {Holevo}(1979)}]{holevo1979capacity}%
  \BibitemOpen
  \bibfield  {author} {\bibinfo {author} {\bibfnamefont {Alexander~S.}\
  \bibnamefont {Holevo}},\ }\bibfield  {title} {\enquote {\bibinfo {title}
  {Capacity of a quantum communications channel},}\ }\href@noop {} {\bibfield
  {journal} {\bibinfo  {journal} {Probl. Inform. Transm.}\ }\textbf {\bibinfo
  {volume} {15}},\ \bibinfo {pages} {247--253} (\bibinfo {year}
  {1979})}\BibitemShut {NoStop}%
\bibitem [{\citenamefont {Holevo}(1998)}]{HolevoCapacity}%
  \BibitemOpen
  \bibfield  {author} {\bibinfo {author} {\bibfnamefont {Alexander~S.}\
  \bibnamefont {Holevo}},\ }\bibfield  {title} {\enquote {\bibinfo {title} {The
  capacity of the quantum channel with general signal states},}\ }\href
  {\doibase 10.1109/18.651037} {\bibfield  {journal} {\bibinfo  {journal}
  {{IEEE} Trans. Inform. Theory}\ }\textbf {\bibinfo {volume} {44}},\ \bibinfo
  {pages} {269--273} (\bibinfo {year} {1998})},\ \Eprint
  {http://arxiv.org/abs/quant-ph/9611023} {arXiv:quant-ph/9611023} \BibitemShut
  {NoStop}%
\bibitem [{\citenamefont {Schumacher}\ and\ \citenamefont
  {Westmoreland}(1998)}]{PhysRevLett.80.5695}%
  \BibitemOpen
  \bibfield  {author} {\bibinfo {author} {\bibfnamefont {Benjamin}\
  \bibnamefont {Schumacher}}\ and\ \bibinfo {author} {\bibfnamefont
  {Michael~D.}\ \bibnamefont {Westmoreland}},\ }\bibfield  {title} {\enquote
  {\bibinfo {title} {Quantum privacy and quantum coherence},}\ }\href {\doibase
  10.1103/PhysRevLett.80.5695} {\bibfield  {journal} {\bibinfo  {journal}
  {Phys. Rev. Lett.}\ }\textbf {\bibinfo {volume} {80}},\ \bibinfo {pages}
  {5695--5697} (\bibinfo {year} {1998})}\BibitemShut {NoStop}%
\bibitem [{\citenamefont {Cai}\ \emph {et~al.}(2004)\citenamefont {Cai},
  \citenamefont {Winter},\ and\ \citenamefont {Yeung}}]{Cai2004}%
  \BibitemOpen
  \bibfield  {author} {\bibinfo {author} {\bibfnamefont {Ning}\ \bibnamefont
  {Cai}}, \bibinfo {author} {\bibfnamefont {Andreas}\ \bibnamefont {Winter}}, \
  and\ \bibinfo {author} {\bibfnamefont {Raymond~W.}\ \bibnamefont {Yeung}},\
  }\bibfield  {title} {\enquote {\bibinfo {title} {Quantum privacy and quantum
  wiretap channels},}\ }\href {\doibase 10.1007/s11122-005-0002-x} {\bibfield
  {journal} {\bibinfo  {journal} {Problems of Information Transmission}\
  }\textbf {\bibinfo {volume} {40}},\ \bibinfo {pages} {318--336} (\bibinfo
  {year} {2004})}\BibitemShut {NoStop}%
\bibitem [{\citenamefont {Devetak}(2005)}]{Devetak2005}%
  \BibitemOpen
  \bibfield  {author} {\bibinfo {author} {\bibfnamefont {Igor}\ \bibnamefont
  {Devetak}},\ }\bibfield  {title} {\enquote {\bibinfo {title} {The private
  classical capacity and quantum capacity of a quantum channel},}\ }\href
  {\doibase 10.1109/TIT.2004.839515} {\bibfield  {journal} {\bibinfo  {journal}
  {IEEE Transactions on Information Theory}\ }\textbf {\bibinfo {volume}
  {51}},\ \bibinfo {pages} {44--55} (\bibinfo {year} {2005})},\ \Eprint
  {http://arxiv.org/abs/quant-ph/0304127} {arXiv:quant-ph/0304127} \BibitemShut
  {NoStop}%
\bibitem [{\citenamefont {Kawahigashi}\ \emph {et~al.}(2001)\citenamefont
  {Kawahigashi}, \citenamefont {Longo},\ and\ \citenamefont
  {M{\"u}ger}}]{MR1838752}%
  \BibitemOpen
  \bibfield  {author} {\bibinfo {author} {\bibfnamefont {Yasuyuki}\
  \bibnamefont {Kawahigashi}}, \bibinfo {author} {\bibfnamefont {Roberto}\
  \bibnamefont {Longo}}, \ and\ \bibinfo {author} {\bibfnamefont {Michael}\
  \bibnamefont {M{\"u}ger}},\ }\bibfield  {title} {\enquote {\bibinfo {title}
  {Multi-interval subfactors and modularity of representations in conformal
  field theory},}\ }\href {\doibase 10.1007/pl00005565} {\bibfield  {journal}
  {\bibinfo  {journal} {Commun. Math. Phys.}\ }\textbf {\bibinfo {volume}
  {219}},\ \bibinfo {pages} {631--669} (\bibinfo {year} {2001})},\ \Eprint
  {http://arxiv.org/abs/math/9903104} {arXiv:math/9903104} \BibitemShut
  {NoStop}%
\bibitem [{\citenamefont {Jochym-O'Connor}\ \emph {et~al.}(2014)\citenamefont
  {Jochym-O'Connor}, \citenamefont {Kribs}, \citenamefont {Laflamme},\ and\
  \citenamefont {Plosker}}]{PhysRevA.90.032305}%
  \BibitemOpen
  \bibfield  {author} {\bibinfo {author} {\bibfnamefont {Tomas}\ \bibnamefont
  {Jochym-O'Connor}}, \bibinfo {author} {\bibfnamefont {David~W.}\ \bibnamefont
  {Kribs}}, \bibinfo {author} {\bibfnamefont {Raymond}\ \bibnamefont
  {Laflamme}}, \ and\ \bibinfo {author} {\bibfnamefont {Sarah}\ \bibnamefont
  {Plosker}},\ }\bibfield  {title} {\enquote {\bibinfo {title} {Quantum
  subsystems: {E}xploring the complementarity of quantum privacy and error
  correction},}\ }\href {\doibase 10.1103/PhysRevA.90.032305} {\bibfield
  {journal} {\bibinfo  {journal} {Phys. Rev. A}\ }\textbf {\bibinfo {volume}
  {90}},\ \bibinfo {pages} {032305} (\bibinfo {year} {2014})},\ \Eprint
  {http://arxiv.org/abs/1405.1798} {arXiv:1405.1798} \BibitemShut {NoStop}%
\bibitem [{\citenamefont {B{\'e}ny}\ \emph {et~al.}(2009)\citenamefont
  {B{\'e}ny}, \citenamefont {Kempf},\ and\ \citenamefont {Kribs}}]{MR2541204}%
  \BibitemOpen
  \bibfield  {author} {\bibinfo {author} {\bibfnamefont {C{\'e}dric}\
  \bibnamefont {B{\'e}ny}}, \bibinfo {author} {\bibfnamefont {Achim}\
  \bibnamefont {Kempf}}, \ and\ \bibinfo {author} {\bibfnamefont {David~W.}\
  \bibnamefont {Kribs}},\ }\bibfield  {title} {\enquote {\bibinfo {title}
  {Quantum error correction on infinite-dimensional {H}ilbert spaces},}\ }\href
  {\doibase 10.1063/1.3155783} {\bibfield  {journal} {\bibinfo  {journal} {J.
  Math. Phys.}\ }\textbf {\bibinfo {volume} {50}},\ \bibinfo {pages} {062108}
  (\bibinfo {year} {2009})},\ \Eprint {http://arxiv.org/abs/0811.0421}
  {arXiv:0811.0421} \BibitemShut {NoStop}%
\bibitem [{\citenamefont {{Marvian}}(2013)}]{2013arXiv1307.6617M}%
  \BibitemOpen
  \bibfield  {author} {\bibinfo {author} {\bibfnamefont {Ian}\ \bibnamefont
  {{Marvian}}},\ }\href@noop {} {\enquote {\bibinfo {title} {{Symmetry
  Protected Topological Entanglement}},}\ } (\bibinfo {year} {2013}),\ \bibinfo
  {note} {preprint},\ \Eprint {http://arxiv.org/abs/1307.6617}
  {arXiv:1307.6617} \BibitemShut {NoStop}%
\bibitem [{\citenamefont {Keyl}\ \emph {et~al.}(2002)\citenamefont {Keyl},
  \citenamefont {Schlingemann},\ and\ \citenamefont
  {Werner}}]{KeylSchlingemannWerner}%
  \BibitemOpen
  \bibfield  {author} {\bibinfo {author} {\bibfnamefont {Michael}\ \bibnamefont
  {Keyl}}, \bibinfo {author} {\bibfnamefont {Dirk}\ \bibnamefont
  {Schlingemann}}, \ and\ \bibinfo {author} {\bibfnamefont {Reinhard~F.}\
  \bibnamefont {Werner}},\ }\href@noop {} {\enquote {\bibinfo {title}
  {Infinitely entangled states},}\ } (\bibinfo {year} {2002}),\ \bibinfo {note}
  {preprint},\ \Eprint {http://arxiv.org/abs/quant-ph/0212014}
  {arXiv:quant-ph/0212014} \BibitemShut {NoStop}%
\bibitem [{\citenamefont {Slofstra}(2016)}]{Slofstra:2016}%
  \BibitemOpen
  \bibfield  {author} {\bibinfo {author} {\bibfnamefont {William}\ \bibnamefont
  {Slofstra}},\ }\href@noop {} {\enquote {\bibinfo {title} {Tsirelson's problem
  and an embedding theorem for groups arising from non-local games},}\ }
  (\bibinfo {year} {2016}),\ \bibinfo {note} {preprint},\ \Eprint
  {http://arxiv.org/abs/1606.03140} {arXiv:1606.03140} \BibitemShut {NoStop}%
\end{thebibliography}%
\end{document}